\documentclass[%
 reprint,
 amsmath,amssymb,
 aps,
]{revtex4-2}

\usepackage{xspace}
\usepackage{subfig}
\usepackage{relsize}
\usepackage{multirow}
\usepackage{braket}
\DeclareMathAlphabet{\mathpzc}{OT1}{pzc}{m}{it}
\newcommand{\mathpzcB}[1]{\mathlarger{\mathlarger{\mathpzc{#1}}}}

\newcommand{\HEPfit}{\texttt{HEPfit}\xspace}

\usepackage{color}
\usepackage[usenames,dvipsnames,svgnames,table]{xcolor}

\usepackage{graphicx}
\usepackage{dcolumn}
\usepackage{bm}
\usepackage{hyperref}

\begin{document}


\title{On {\boldmath $SU(3)_{F}$} Breaking through Final State Interactions and CP Asymmetries in {\boldmath $D\to P P$} Decays}

\author{Franco Buccella}
\email{franco.buccella@na.infn.it}
\affiliation{%
 INFN, Sezione di Napoli,  via Cintia, 80126 Napoli, Italy.
 }
 \author{Ayan Paul}
\email{ayan.paul@desy.de}
\affiliation{DESY, Notkestrasse 85, D-22607 Hamburg, Germany.
}%
\affiliation{Institut f\"ur Physik, Humboldt-Universit\"at zu Berlin, D-12489 Berlin, Germany.
}%
\author{Pietro Santorelli}%
 \email{pietro.santorelli@na.infn.it}
 \affiliation
{
 Dipartimento di Fisica "Ettore Pancini", Universit\`{a} di Napoli Federico II, 
Complesso Universitario di Monte S. Angelo,
Via Cintia Edificio 6, 80126 Napoli, Italy.\\
}%
\affiliation{%
 INFN, Sezione di Napoli,  via Cintia, 80126 Napoli, Italy.
 }

\date{\today}

\begin{abstract}
We analyse $D$ decays to two
  pseudoscalars $(\pi,K)$ assuming the dominant source of $\mathrm{SU}(3)_F$
  breaking lies in final state interactions. We obtain an excellent
  agreement with experimental data and are able to predict CP
  violation in several channels based on current data on branching
  ratios and $\Delta \mathrm{A_{CP}}$. We also make predictions for
  $\delta_{K\pi}$ and the branching fraction for the decay
  $D_s^+\to K^+K_L$.
\end{abstract}

\maketitle


\section{Introduction}
\label{sec:intro}
The search for physics beyond the Standard Model (SM) has brought
about a search for CP violation beyond what is generated by the phase
in the CKM matrix
\cite{Cabbibo:1964zsa,Cabibbo:1963yz,Kobayashi:1973fv}. The
challenging measurement and conclusive evidence of a non-vanishing
value for the ratio
$\epsilon'/\epsilon$~\cite{Batley:2002gn,AlaviHarati:2002ye,Abouzaid:2010ny}
has excluded the superweak hypothesis of Wolfenstein
\cite{Wolfenstein:1964ks}. However, $\epsilon'/\epsilon$ has not yet
facilitated a severe test of the SM. This is primarily due to the poor
predictive power plaguing this ratio in the presence of cancellations
between QCD and electroweak penguins. The possibility of a
quantitative evaluation of these matrix elements often transforms into
a theological debate, in spite of the encouraging and pioneering
lattice results by the RBC and UKQCD collaborations
\cite{Bai:2015nea}. On the other hand, the triumphant measurement of
$O(1)$ CP violation in the golden decay channel
$J/\psi
K_S$~\cite{Chen:2006nk,Aubert:2007hm,Aubert:2009aw,Adachi:2012et} of
the neutral $B$ meson, where the measured time-dependent asymmetry
depends, to an excellent approximation, only on the CP violating phase
of the CKM matrix~\cite{Bander:1979px,Carter:1980tk,Bigi:1981qs}, has been a
striking confirmation of the SM. More recently, LHCb has also confirmed the
validity of the SM through a measurement of CP violation in $B_s$ physics \cite{Aaij:2014zsa}. LHCb has also played a key role in bringing the pioneering first
results on neutral $D^0$ meson mixing by previous experiments
\cite{Aitala:1996vz,Cawlfield:2005ze,Aubert:2007aa,Bitenc:2008bk,Aitala:1996fg,Godang:1999yd,Link:2004vk,Zhang:2006dp,Aubert:2007wf,Aaltonen:2013pja,Ko:2014qvu,Aubert:2008zh,Aitala:1999dt,Link:2000cu,Csorna:2001ww,Abe:2001ed,Lees:2012qh,Staric:2012ta,Aaltonen:2014efa,Ko:2012jh,delAmoSanchez:2010xz,Peng:2014oda,TheBABAR:2016gom,Asner:2012xb,Aubert:2007if,DiCanto:2012zw}
to a mature stage, along with an impressive progress in the
measurement of CP violation \cite{Aaij:2015yda,Aaij:2014gsa,Aaij:2016cfh,Aaij:2016roz,Aaij:2016rhq,Aaij:2011ad,LHCb-CONF-2016-009,Aaij:2015xoa}.

In the recent past, the controversial
measurements~\cite{Aubert:2007if,Collaboration:2012qw,Aaij:2011in,Ko:2012px,DiCanto:2012zw,Aaij:2013bra,Aaij:2014gsa,Aaij:2015yda,Aaij:2016cfh}
of the CP violating asymmetry found in the decay of the neutral $D^0$
meson to pairs of charged kaons and pions, had effectively stirred the
question of whether such rather high values found in the first
experimental results could be accommodated within the Standard
Model. While many arguments were placed in favour of contributions
coming from beyond the
SM~\cite{Bigi:2011re,Grossman:2006jg,Isidori:2011qw,Wang:2011uu,Hiller:2012wf,Giudice:2012qq,Altmannshofer:2012ur,Chen:2012am,Gedalia:2012pi,Mannel:2012hb,KerenZur:2012fr,Barbieri:2012bh,Dolgov:2012ez,Delaunay:2012cz},
concrete arguments were also made for the presence of large phases
coming from Final State Interactions (FSI) allowing for the
accommodation of the asymmetry within the
SM~\cite{Golden:1989qx,Pirtskhalava:2011va,Bhattacharya:2012ah,Feldmann:2012js,Brod:2011re,Brod:2012ud,Cheng:2012xb,Hiller:2012xm}. In
fact, both the isospin relations for the Cabibbo allowed (CA) decays
into $\bar{K} \pi$ and singly Cabibbo suppressed (SCS) decays into
$ \pi \pi$ of $D$ mesons are characterized by large angles in the
complex plane for the corresponding
triangles~\cite{LeYaouanc:1992iq,Franco:2012ck}. In the case of the
$\pi \pi$ final states the phase difference between the $I=2$ and
$I=0$ amplitudes is about $\pi/2$. These large phases have been, for a
long time, advocated as the main cause for the large $SU(3)_{F}$
violations in exclusive $D$
decays~\cite{Buccella:1994nf,Buccella:1996uy}. Indeed, in the $D$ mass
region there is a nonet of scalar resonances and their mass splittings
imply large $\mathrm{SU}(3)_F$ violations generated by FSI.

Identifying the dominant source of $\mathrm{SU}(3)_F$ violation is of
crucial phenomenological importance in $D$ decays, since on one
hand the imposition of exact $\mathrm{SU}(3)_F$ completely fails in reproducing
experimental data, while on the other hand introducing
$\mathrm{SU}(3)_F$ breaking in a general manner leads to a complete
loss of predictivity due to the proliferation of independent
parameters (see \textit{e.g.}  ref.~\cite{Hiller:2012xm}). Several
interesting attempts at reducing the number of parameters have been
made. The authors of
refs.~\cite{Muller:2015lua,Muller:2015rna,Nierste:2015zra} advocated
the use of $1/N_c$ counting to reduce the size of the parameter set to a tractable
number. However, relying on the $1/N_c^2$ suppression of formally
divergent corrections seems questionable. In
ref. \cite{Pirtskhalava:2011va}, the dominance of lower rank
representations was argued for, and only $\mathrm{SU}(3)_F$ triplets
were considered as additional operators in the effective
Hamiltonian. However, there is no compelling reason to truncate the
effective Hamiltonian in such a drastic manner.

In contrast to the above-mentioned approaches, the assumption that FSI
is the dominant source of $\mathrm{SU}(3)_F$ breaking rests solidly on
the large observed strong phases, it provides a very good description
of available experimental data, it allows to predict several CP
asymmetries which are currently poorly measured and it can be
tested against independent determinations of the relevant rescattering
matrices. It also allows us to predict the relative strong phase between the doubly Cabibbo suppressed (DCS) and CA
charged $K\pi$ decays, $\delta_{K\pi}$.

The paper is organized as follows. In the next section we write the
amplitudes for all the decays considered. In section~\ref{sec:CPamp}
we discuss the parameterization of the $\Delta U =0$ part of the amplitudes proportional to $V_{cb}^*V_{ub}^{}$. In section \ref{sec:exp} we give
a brief overview of the current status of experimental measurements of CP asymmetries
in SCS decays. Using the experimental branching ratios of $D^0$, $D^+$ and $D_s^+$ into
final states with kaons and/or pions and measurements of CP
asymmetries we fit the values of the parameters in
section~\ref{sec:fit}. In this section we also take a critical look at our
parameterization and its consequences on correlations between CP
asymmetries followed by an estimate of how future measurements will improve the
constraints on penguin amplitudes. We finally summarize our results in
section~\ref{sec:sum}. The details of the fit results are given in the
appendix.

\section{Amplitudes and Parametrization}
\label{sec:Amp}

It is customary to describe weak decay amplitudes in terms of
topologies of Wick contractions (or renormalization-group-invariant
combinations thereof). Notice that any Wick contraction, as defined in
refs.~\cite{hep-ph/9703353,Buras:1998ra}, can be seen as an emission
followed by rescattering \cite{hep-ph/9703353,hep-ph/9601265,Cheng:2002wu}. Thus,
rescattering establishes a link between emissions and long-distance
contributions to other subleading topologies such as penguins,
annihilations, weak exchange, etc. The large phases observed in
two-body nonleptonic $D$ decays imply the importance of FSI, leading
to an effective description of decay amplitudes in terms of emissions
followed by rescattering. This description was developed in
refs. \cite{Buccella:1994nf,Buccella:1996uy}, where FSI effects were
parameterized in terms of resonances\footnote{For recent alternate approaches see~\cite{Li:2012cfa,Biswas:2015aaa}}.

In a previous study~\cite{Buccella:2013tya} of the SCS decays of the $D^0$
into a pair of pseudoscalars, exact
$\mathrm{SU}(3)_F$ symmetry was assumed amongst the emission matrix
elements of the non-leptonic Hamiltonian. The necessary
$\mathrm{SU}(3)_F$ breaking was determined by FSI, described as the
effect of resonances in the scattering of the final
particles. Assuming no exotic resonances belonging to the 27
representation, the possible resonances have $\mathrm{SU}(3)_F$ and
isospin quantum numbers $(8, I=1)$, $(8,I=0)$ and $(1,I=0)$.
Moreover, the two states with $I=0$ can mix, yielding two
resonances:
\begin{eqnarray}
  \ket{f_0} &=& \sin \phi  \ket{8,I=0} + \cos \phi  \ket{1,I=0} , \\
  \ket{f'_0} &=& - \cos \phi   \ket{8,I=0} + \sin  \phi  \ket{1,I=0}.
\end{eqnarray}
 The main contribution from the Hamiltonian,
$H(|\Delta C|=1, \Delta S=0)$ transforms as a $U-$spin triplet and
therefore relates the $D^0$, which is a $U-$spin singlet, to the
$U-$spin triplets of the 8 and 27 representation of $\mathrm{SU}(3)_F$. So in
ref.~\cite{Buccella:2013tya} two parameters were introduced for the
matrix elements of the weak Hamiltonian, namely $T$ and $C$. The phase of the $I=1$ octet amplitude, $\delta_1$, and the two phases and mixing angle between the $I=0$ singlet and octet amplitude,
$\delta_0$, $\delta_0^\prime$ and $\phi$, were taken as free
parameters. The strong phases should be related to the mass and width of the resonances. However, the lack of complete experimental information
on the scalar resonances do not allow for the determination of the strong phases and so we determine them from the fit.
It should also be noted that notwithstanding the lack of exotic resonances belonging to the 27 representation, there is a small phase associated with this amplitude which is compatible with 0~\cite{Franco:2012ck}. We set this phase to 0 and hence all the other phases should be interpreted as a difference with respect to this phase.

Other attempts have been made previously to study $D\to P P$ decays in the $\mathrm{SU}(3)_F$
framework with perturbative breaking of the
symmetry~\cite{Bhattacharya:2009ps,Brod:2011re,Hiller:2012xm,Muller:2015lua,Nierste:2015zra,Muller:2015rna}. The
spotlight has always been on prescriptions for estimating the penguin
amplitudes by formulating a reasonable parametrization in the
$\mathrm{SU}(3)_F$ framework and then using available data on branching fractions and CP
asymmetries. In this work we extend the formalism that was developed in ref.~\cite{Buccella:2013tya} by
including more decay modes of the $D$ meson system and introducing new parameters to aptly parametrize
the additional decay amplitudes.

The $D^+$ and $D^+_s$ form a $U-$spin doublet and the
matrix elements of the weak Hamiltonian, which relate $D^0$ to the
$Q=0$, $U=1$ and $D^+$ to the $Q=1$, $U=\frac{1}{2}$ of the octet, are
independent. This requires the introduction of  a $\mathrm{SU}(3)_F$ invariant
parameter $\Delta$ in the $\Delta U = 1$ part of the amplitude. The terms proportional to $\Delta$ vanish in the factorization ansatz
and appears only in the $D^+_{(s)}$ decay amplitudes. As we will explain in section~\ref{sec:CPamp},
$\Delta$ is related by $\mathrm{SU}(3)_F$ to a vanishingly small contribution in the $\Delta U=0$ part of the amplitude suppressed by an
approximate selection rule.

To expand the FSI description we need a phase, $\delta_\frac{1}{2}$, for the FSI of
the $I = \frac{1}{2}$ member of the octet. The phases $\delta_0$, $\delta_0^\prime$,
$\delta_1$ and $\delta_\frac{1}{2}$ and the mixing angle $\phi$ are defined such that in the the limit of $\mathrm{SU}(3)_{F}$ conservation
$\delta_0 = \delta_1 = \delta_\frac{1}{2}$, the amplitudes are independent of $\delta_0^\prime$ and $\phi=\pi/2$. The
phase for the decay modes with $D^+_s$ in the initial state is expected to be different from those in the $D^0$ and
$D^+$ decay modes as an effect of $\mathrm{SU}(3)_{F}$ breaking and the consequent shift in the mass of the $D^+_s$. Keeping
in mind that both the phases shift in the same direction, we parameterize the phases with $\epsilon_\delta$ as
\begin{equation}
\delta^\prime_{1} = \delta_{1} (1-\epsilon_\delta)\textrm{  and  } \delta^\prime_{\frac{1}{2}} = \delta_{\frac{1}{2}}(1-\epsilon_\delta).
\label{eq:split}
\end{equation}

The extension to the CA and DCS final states requires the introduction of additional
sources of $\mathrm{SU}(3)_{F}$ violation in addition to $\delta_\frac{1}{2}$
in the $I = \frac{1}{2}$ octet channel. To understand this better one must note that the $\mathrm{SU}(3)_{F}$ relationship for $D^+$
decays:
\begin{equation}
\tan{\theta_C} A ( D^+ \rightarrow \bar{K^0} \pi^+ ) =
\sqrt{2} A ( D^+ \rightarrow \pi^0 \pi^+) ,
\label{eq:kappa}
\end{equation}
which implies (neglecting the interference with the DCS final state)
the ratio of the decay amplitudes into two pions and into
$K_S \pi^+$ being equal to $\tan{\theta_C}$ is in disagreement with data. To correct for this discrepancy we allow for a breaking of the 27 amplitudes
through the introductions of a parameters $\kappa$ and $\kappa^\prime$ which, respectively, split the 27 matrix element in the CA and DCS channels from the 27 matrix element  in the SCS channel.

Next, we observe that the ratio of
the branching fractions of the DCS to the CA decays of $D^0$ into a
kaon and a pion with opposite electric charge given by
\begin{equation}
\frac{{\rm BR}(D^0 \rightarrow K^+ \pi^-)}{{\rm BR}(D^0 \rightarrow K^- \pi^+)} = \tan^4\theta_C.
\label{eq:LNG}
\end{equation}
is violated and the ratio is actually larger than the value $\tan^4\theta_C$. To accommodate for this
we allow for the $\mathrm{SU}(3)_F$ breaking parameter $K$ which contributes with opposite signs to the octet part for the CA and DCS
channels to correct the prediction in eq.~\ref{eq:LNG}. This parameter represents the non-conservation of the strangeness
changing current which, in the factorization ansatz, corresponds to the axial current that destroys the initial $D$
meson state and the divergence in the vector current proportional to the mass difference of the strange quark and
that of the lighter quarks $u$ and $d$. This also generates a term proportional to $K^\prime$ in the DCS
decays of the $D^+$ and the CA decays of the $D^+_s$ meson which comes from the fact that for annihilating the charged mesons a charged current is necessary.

With this parametrization we arrive at the following amplitudes for the SCS, CA and DCS amplitudes. Although we present the amplitudes with $\eta_8$ in the final state, we do not make any attempt to include
$\eta-\eta^\prime$ mixing in this work and hence do not use the experimental measurements of these channels for the fits. While considering the singlet state, $\eta_1$, would increase the number of measurements
that we could fit the parameters to, including the singlet state would also require additional parameters since it is
has a significant gluonic content \cite{Ball:1995zv,DiDonato:2011kr}. A discussion of the complexities of addressing
$\eta-\eta^\prime$ mixing can be found in~\cite{Feldmann:1999uf} and references therein. Hence, we shall postpone
this exercise to a future work.

In summary, the $\mathrm{SU}(3)_{F}$ breaking parameters that we need to introduce to fit to the $\Delta U=1$ part of the amplitudes that are sensitive to the measurements of the branching fraction are:
\begin{itemize}
    \item For the phases generated by FSI in the different isospin amplitudes:
    \begin{itemize}
        \item $\delta_0$: The FSI phase of the singlet component of the $I=0$ amplitude.
        \item $\delta_0^\prime$: The FSI phase of the octet component of the $I=0$ amplitude.
        \item $\delta_\frac{1}{2}$: The FSI phace of the $I=1/2$ amplitude.
        \item $\epsilon_\delta$: Defined in equation~\eqref{eq:split} as the splitting in the phase of $I=0$ amplitudes for the heavier $D^+_s$ meson from the lighter $D^0$ and $D^+$ mesons.
    \end{itemize}
    \item A mixing angle $\phi$ that characterizes the mixing between the singlet and octet components in the $I=0$ amplitude.
    \item $K$ and $K^\prime$: Comes from the non-conservation of the strangeness changing neutral and charged currents respectively and accommodates for the deviation of data from equation~\eqref{eq:LNG}.
    \item $\kappa$ and $\kappa^\prime$: Introduced to allow for $\mathrm{SU}(3)_{F}$ breaking in the 27 matrix element to alleviate the discrepancy in equation~\eqref{eq:kappa} for the CA and DCS amplitudes respectively.
\end{itemize}

The $\Delta U=1$ amplitudes in terms of these parameters can be written as:

\begin{widetext}
{\allowdisplaybreaks
\begin{center}
{\bf SCS modes} (to be multiplied by $\frac{1}{2}(V^{}_{cs}V^*_{us} - V^{}_{cd}V^*_{ud})$)
\end{center}
\begin{eqnarray}
A(D^0\to\pi^+\pi^-) &=& \left(T - \frac{2}{3} C\right) \left[-\frac{3}{10} \left(e^{i \delta_0} + e^{i \delta_0^\prime}\right) + \left(-\frac{3}{10}  \cos(2 \phi) + \frac{3}{4\sqrt{10}} \sin(2 \phi)\right)  \left(e^{i \delta_0^\prime} - e^{i \delta_0}\right)\right]\nonumber\\&& -  \frac{2}{5}\left(T + C\right)\nonumber\\
A(D^0\to\pi^0\pi^0) &=& \left(T - \frac{2}{3} C\right) \left[-\frac{3}{10} \left(e^{i \delta_0} + e^{i \delta_0^\prime}\right) + \left(-\frac{3}{10}  \cos(2 \phi) + \frac{3}{4\sqrt{10}} \sin(2 \phi)\right)  \left(e^{i \delta_0^\prime} - e^{i \delta_0}\right)\right]\nonumber\\&& +  \frac{3}{5}\left(T + C\right)\nonumber\\
A(D^0\to K^+K^-) &=& \left(T - \frac{2}{3} C\right) \left[\frac{3}{20} \left(e^{i \delta_0} + e^{i \delta_0^\prime}\right) + \left(\frac{3}{20}  \cos(2 \phi) + \frac{3}{4\sqrt{10}} \sin(2 \phi)\right)  \left(e^{i \delta_0^\prime} - e^{i \delta_0}\right)\right.\nonumber\\&&\left. + \frac{3}{10} e^{i \delta_1}\right] +  \frac{2}{5}\left(T + C\right)\nonumber\\
A(D^0\to K^0\bar{K}^0) &=& \left(T - \frac{2}{3} C\right) \left[\frac{3}{20} \left(e^{i \delta_0} + e^{i \delta_0^\prime}\right) + \left(\frac{3}{20}  \cos(2 \phi) + \frac{3}{4\sqrt{10}} \sin(2 \phi)\right)  \left(e^{i \delta_0^\prime} - e^{i \delta_0}\right)\right.\nonumber\\&& \left.-  \frac{3}{10} e^{i \delta_1}\right]\nonumber\\
 A(D^0\to \eta_8\eta_8) &=& \left(T - \frac{2}{3} C\right) \left[\frac{3}{10} \left(e^{i \delta_0} + e^{i \delta_0^\prime}\right) + \left(\frac{3}{10}  \cos(2 \phi) + \frac{3}{4\sqrt{10}} \sin(2 \phi)\right)  \left(e^{i \delta_0^\prime} - e^{i \delta_0}\right)\right]\nonumber\\&& -  \frac{3}{5}\left(T + C\right)\nonumber\\
A(D^0\to \pi^0\eta_8) &=& \frac{\sqrt{3}}{5} \left[\left(T - \frac{2}{3} C\right)  e^{i\delta_1} - \left(T + C\right)\right]\nonumber\\
A(D^+\to K^+\bar{K}^0) &=& \frac{1}{5}  (2  T - 3 C + \Delta)  e^{i \delta_1} + \frac{3}{5} (T + C)\nonumber\\
A(D^+\to \pi^+\pi^0) &=& \frac{1}{\sqrt{2}} (T + C)\nonumber\\
A(D^+\to \pi^+\eta_8) &=& \frac{\sqrt{2}}{5\sqrt{3}}  (2  T - 3 C + \Delta)  e^{i \delta_1} - \frac{3\sqrt{3}}{5\sqrt{2}} (T + C)\nonumber\\
A(D_s^+\to \pi^+K^0) &=& -\frac{1}{5}  (2  T - 3 C + \Delta - K^\prime)  e^{i \delta^\prime_{\frac{1}{2}}} - \frac{3}{5} (T + C)\nonumber\\
A(D_s^+\to \pi^0K^+) &=& -\frac{1}{5\sqrt{2}}  (2  T - 3 C + \Delta - K^\prime)  e^{i \delta^\prime_{\frac{1}{2}}} + \frac{2}{5\sqrt{2}} (T + C)\nonumber\\
A(D_s^+\to K^+\eta_8) &=& \frac{1}{5\sqrt{6}}  (2  T - 3 C + \Delta - K^\prime)  e^{i \delta^\prime_{\frac{1}{2}}} - \frac{2\sqrt{6}}{5} (T + C)
\label{eq:SCS_P}
\end{eqnarray}
\begin{center}
{\bf CA modes} (to be multiplied by $V^{}_{cs}V^*_{ud}$)
\end{center}
\begin{eqnarray}
A(D^0\to \pi^+K^-) &=& \frac{1}{5}  (3  T - 2 C - K)  e^{i \delta_{\frac{1}{2}}} + \frac{2}{5} (T + C + \kappa)\nonumber\\
A(D^0\to \pi^0\bar{K}^0) &=& -\frac{1}{5\sqrt{2}}  (3  T - 2 C - K)  e^{i \delta_{\frac{1}{2}}} + \frac{3}{5\sqrt{2}} (T + C + \kappa)\nonumber\\
A(D^0\to \bar{K}^0 \eta_8) &=& -\frac{1}{5\sqrt{6}}  (3  T - 2 C - K)  e^{i \delta_{\frac{1}{2}}} + \frac{3}{5\sqrt{6}} (T + C + \kappa)\nonumber\\
A(D^+ \to \pi^+\bar{K}^0)  &=& (T + C+ \kappa)\nonumber\\
A(D_s^+\to K^+\bar{K}^0) &=& -\frac{1}{5}  (2  T - 3 C + \Delta)  e^{i \delta^\prime_{1}} + \frac{2}{5} (T + C + \kappa)\nonumber\\
A(D_s^+\to \pi^+\eta_8) &=& -\frac{\sqrt{2}}{5\sqrt{3}}  (2  T - 3 C + \Delta)  e^{i \delta^\prime_{1}} - \frac{\sqrt{6}}{5} (T + C + \kappa)
\label{eq:CA_P}
\end{eqnarray}
\begin{center}
{\bf DCS modes} (to be multiplied by $-V^{}_{cd}V^*_{us}$)
\end{center}
\begin{eqnarray}
A(D^0\to \pi^0 K^0) &=& \frac{1}{5\sqrt{2}}  (3  T - 2 C + K)  e^{i \delta_{\frac{1}{2}}} - \frac{3}{5\sqrt{2}} (T + C + \kappa^\prime)\nonumber\\
A(D^0\to K^0 \eta_8) &=& \frac{1}{5\sqrt{6}}  (3  T - 2 C + K)  e^{i \delta_{\frac{1}{2}}} - \frac{3}{5\sqrt{6}} (T + C + \kappa^\prime)\nonumber\\
A(D^+ \to \pi^+K^0) &=& \frac{1}{5}  (2  T - 3 C + \Delta - K^\prime)  e^{i \delta_{\frac{1}{2}}} - \frac{2}{5} (T + C + \kappa^\prime)\nonumber\\
A(D^0\to \pi^- K^+) &=& -\frac{1}{5}  (3  T - 2 C + K)  e^{i \delta_{\frac{1}{2}}} - \frac{2}{5} (T + C + \kappa^\prime)\nonumber\\
A(D^+ \to \pi^0K^+) &=& \frac{1}{5\sqrt{2}}  (2  T - 3 C + \Delta - K^\prime)  e^{i \delta_{\frac{1}{2}}} + \frac{3}{5\sqrt{2}} (T + C + \kappa^\prime)\nonumber\\
 A(D^+ \to K^+\eta_8) &=& -\frac{1}{5\sqrt{6}}  (2  T - 3 C + \Delta - K^\prime)  e^{i \delta_{\frac{1}{2}}} - \frac{3}{5\sqrt{6}} (T + C + \kappa^\prime)\nonumber\\
A(D_s^+ \to K^+K^0)  &=& -(T + C + \kappa^\prime)
\label{eq:DCS_P}
\end{eqnarray}
}
\end{widetext}

The CKM factors are to be kept explicit and hence the amplitudes given in equations (\ref{eq:SCS_P})--(\ref{eq:DCS_P}) should be multiplied by $\frac{1}{2}(V^{}_{us}V^*_{cs} - V^{}_{ud}V^*_{cd})$, $V^{}_{ud}V^*_{cs}$ and $-V^{}_{us}V^*_{cd}$ for SCS, CA and DCS modes respectively. The branching fraction is then defined as
\begin{widetext}
\begin{eqnarray}
\textrm{BR}(D\to P_1 P_2)=
\frac{\tau_DG_F^2}{16\pi m_D^2}\frac{\sqrt{ (m_D^2 - (m_{P_1} + m_{P_2})^2)(m_D^2 - (m_{P_1} - m_{P_2})^2) }}{2m_D}
\times\left|\mathcal{A}(D\to P_1 P_2)\right|^2,
\end{eqnarray}
\end{widetext}
where $m_D$, $m_{P_1}$ and $m_{P_2}$ are the masses of the $D$ meson in the initial state and the pseudoscalars in the final state respectively, $G_F$ is the Fermi constant and $\tau_D$ is the relevant $D$ meson lifetime.

\section{The \texorpdfstring{\boldmath $\Delta U =0$}{DeltaU=0} amplitudes}
\label{sec:CPamp}

The $\Delta U = 0$ contributions to the SCS decays
proportional to $V^{}_{cb} V^*_{ub}$ need to be considered both for the
amplitudes related to the penguin operator:
\begin{eqnarray}
\bar{u}_L(x) \gamma^{\mu} \lambda_a c_L(x)
[(\bar{u}(x) \gamma_{\mu} \lambda_a u(x) \nonumber\\
+\; \bar{d}(x) \gamma^{\mu} \lambda_a d(x)+
\bar{s}(x) \gamma_{\mu} \lambda_a s(x)]
\end{eqnarray}
and to the operator:
\begin{equation}
\bar{u_L} \gamma_{\mu} s_L \bar{s_L} \gamma^{\mu} c_L (x) +
\bar{u_L} \gamma_{\mu} d_L \bar{d_L} \gamma^{\mu} c_L (x).
\label{eq:delta}
\end{equation}
The latter have to be considered as a consequence of the unitarity of the
CKM matrix and are referred to as the pseudo-penguin operators. Parameterizing this part of the amplitude requires the introduction of three additional real parameters $P$, $\Delta_3$ and $\Delta_4$. The penguin contributions are encapsulated in $P$. The matrix elements of the operator defined in equation~(\ref{eq:delta}) depend on four reduced matrix elements, $\left<27|15|\bar{3}\right>$, $\left<8|15|\bar{3}\right>$, $\left<8|3|\bar{3}\right>$ and $\left<1|3|\bar{3}\right>$. The first two are related to the ones for the $\Delta U = 1$ part. We introduce two parameters, $\Delta_3$ and $\Delta_4$, which are combinations of the four reduced matrix elements which are defined in such a way that, by neglecting final state interactions, one has:
\begin{eqnarray}
B(D^0 \rightarrow K^+ K^-) &=& P +T + \Delta_3 \equiv \mathpzcB{P},\nonumber\\
B(D^0 \rightarrow K^0 \bar{K}^0) &=& \Delta_4.
\end{eqnarray}

The asymmetries consist of three contributions. The first contribution comes from the terms proportional to $\mathpzcB{P}$. While $T$ can be extracted from the branching fraction data neither $P$ nor $\Delta_3$ can be estimated from first principles. The second contribution is proportional to $T + C$ and it can be completely determined from the branching fraction data. The third contribution is proportional to $\Delta_4$. This contribution, which is a sum of penguin and pseudo-penguin contributions, is vanishingly small due to an approximate selections rule which disfavours the simultaneous creation of $d\bar{d}$ and $s\bar{s}$ pairs\footnote{The approximate selection rule that leads to the suppression of the simultaneous creation of $d\bar{d}$ and $s\bar{s}$ pairs in our framework is analogous to a Zweig suppression (or OZI rule)~\cite{Okubo:1963fa,Zweig:570209,Iizuka:1966fk}. We refrain from calling it the Zweig suppression since the process $\phi \to \rho \pi$, forbidden by the Zweig rule, might occur by allowing the conversion of the initial $s \bar{s}$ into a pair of light quarks accompanied by the creation of another pair of light quarks. To exclude the formation of a $K^0 \bar{K}^0$ pair through the  $\Delta U = 0$ decay amplitude we do not allow the conversion of the $u \bar{u}$ pair into a $d \bar{d}$ or $s \bar{s}$ pair and assume that the constituents of the final mesons are just the three quarks produced in the $c$ decay and the spectator $\bar{q}$. This has a consequence also for the $\Delta U = 1$ part related by $\mathrm{SU}(3)_{F}$ to the $\Delta U = 0$. Our selection rule is better motivated, since asymptotic freedom implies that the strong coupling constant is a decreasing function of the scale.}. Moreover, $\mathrm{SU}(3)_{F}$ relates the $\Delta$ in the $\Delta U=1$ part to the $\Delta_4$ in the $\Delta U =0$ part. It is interesting to note here that indeed the $\Delta U = 0$ contributions of the 15 would allow a contribution to $D^0 \to K^0 \bar{K}^0$. To forbid it, according to the selection rule, one should put $\Delta \sim 0$. The fact that the fit to the branching ratios implies for $\Delta$ a small value consistent with $0$ lies in favor of the selection rule. However $\Delta_3$ is not affected by this approximate selection rule and hence does not need to be vanishingly small. There is a contribution proportional to $\Delta$ in the $\Delta U =0$ part of the amplitudes of the $D^+$ modes. However, as we shall see, $\Delta$ is very small and hence this contribution turns out to be insignificant. The explicit form of the $\Delta U =0$ part of the amplitude are as follows:

\begin{widetext}
{\allowdisplaybreaks
\begin{eqnarray}
B(D^0\to\pi^+\pi^-) &=& \mathpzcB{P} \left(\frac{1}{2} \left(e^{i\delta_0^\prime} + e^{i\delta_0}\right) + \left(e^{i\delta_0^\prime} - e^{i\delta_0}\right)  \left(-\frac{1}{6} \cos(2\phi) - \frac{7}{4 \sqrt{10}} \sin(2\phi)\right)\right) \nonumber\\
&&+\left(T + C\right) \left(-\frac{3}{20}  \left(e^{i\delta_0^\prime} + e^{i\delta_0}\right) + \frac{3}{10} + \left(\frac{1}{60}  \cos(2\phi) + \frac{1}{2\sqrt{10.}}  \sin(2\phi) \right) \left(e^{i\delta_0^\prime} - e^{i\delta_0} \right)\right)  \nonumber\\
&&+\Delta_4  \left(e^{i\delta_0^\prime} - e^{i\delta_0}\right)  \left(-\frac{1}{3} \cos(2\phi) - \frac{1}{4\sqrt{10}} \sin(2\phi)\right),\nonumber\\
B(D^0\to\pi^0\pi^0) &=& B(D^0\to\pi^+\pi^-) - (T + C),\nonumber\\
B(D^0\to K^+K^-) &=& \mathpzcB{P}\left(\frac{1}{4}  \left(e^{i\delta_0^\prime} + e^{i\delta_0}\right) + \left(e^{i\delta_0^\prime} - e^{i\delta_0}\right) \left(-\frac{5}{12}  \cos(2\phi) + \frac{1}{4\sqrt{10}} \sin(2\phi)\right) + \frac{1}{2} e^{i\delta_1}\right) \nonumber\\
&& +\left(T + C\right)  \left(-\frac{1}{20}  \left(e^{i\delta_0^\prime} + e^{i\delta_0}\right) + \frac{3}{10} + \frac{7}{60}  \cos(2\phi) \left(e^{i\delta_0^\prime} - e^{i\delta_0}\right) - \frac{1}{5} e^{i\delta_1}\right) \nonumber\\
&& +\Delta_4  \left(\frac{1}{4}  \left(e^{i\delta_0^\prime} + e^{i\delta_0}\right) + \left(e^{i\delta_0^\prime} - e^{i\delta_0}\right) \left(-\frac{1}{12}  \sin(2\phi) + \frac{3}{4\sqrt{10}}  \sin(2\phi)\right) - \frac{1}{2}  e^{i\delta_1}\right),\nonumber\\
B(D^0\to K^0\bar{K}^0) &=& \mathpzcB{P}\left(\frac{1}{4} \left(e^{i\delta_0^\prime} + e^{i\delta_0}\right) + \left(e^{i\delta_0^\prime} - e^{i\delta_0}\right)  \left(-\frac{5}{12} \cos(2\phi) + \frac{1}{4\sqrt{10}} \sin(2\phi)\right) - \frac{1}{2} e^{i\delta_1}\right) \nonumber\\
&&+ \left(T + C\right) \left(-\frac{1}{20} \left(e^{i\delta_0^\prime} + e^{i\delta_0}\right) - \frac{1}{10} + \frac{7}{60} \cos(2\phi) \left(e^{i\delta_0^\prime} - e^{i\delta_0}\right) + \frac{1}{5}  e^{i\delta_1}\right) \nonumber\\
&&+\Delta_4 \left(\frac{1}{4}  \left(e^{i\delta_0^\prime} + e^{i\delta_0}\right) + \left(e^{i\delta_0^\prime} - e^{i\delta_0}\right)  \left(-\frac{1}{12} \cos(2\phi) + \frac{3}{4\sqrt{10}} \sin(2\phi)\right) + \frac{1}{2}  e^{i\delta_1}\right),\nonumber\\
B(D^0\to \pi^0\eta_8) &=& \frac{1}{\sqrt{3}}  \left(\left(\mathpzcB{P} - \Delta_4\right)  e^{i\delta_1} - \left(T + C\right) \left(\frac{2}{5}  e^{i\delta_1} + \frac{3}{5}\right)\right),\nonumber\\
B(D^+\to K^+\bar{K}^0) &=& \left(\mathpzcB{P} - \Delta_4 - \frac{1}{5} \left(\Delta + T + C \right)\right) e^{i\delta_1} + \frac{1}{5}  \left(T + C\right),\nonumber\\
B(D^+\to \pi^+\eta_8) &=& \sqrt{\frac{2}{3}}  \left(\mathpzcB{P} - \Delta_4  - \frac{1}{5} \left(\Delta + T + C\right)\right)  e^{i\delta_1} - \frac{\sqrt{6}}{10}  \left(T + C\right),\nonumber\\
B(D^+_s \to K^0 \pi^+) &=& -\left(\mathpzcB{P} - \Delta_4  - \frac{1}{5}\left(\Delta + T + C\right)\right) e^{i \delta^\prime_{\frac{1}{2}}}
-\frac{1}{5}\left(T + C\right),\nonumber\\
B(D^+_s \to K^+ \pi^0) &=& -\sqrt{\frac{1}{2}} \left(\mathpzcB{P} - \Delta_4 - \frac{1}{5}\left(\Delta + T + C\right)\right)
e^{i \delta^\prime_{\frac{1}{2}}} + \frac{2 \sqrt{2}}{5} \left(T + C\right),\nonumber\\
B(D^+_s \to K^+ \eta_8) &=& \sqrt{\frac{1}{6}} \left(\mathpzcB{P} - \Delta_4 -\frac{1}{5}\left(\Delta + T + C\right)\right)
e^{i \delta^\prime_{\frac{1}{2}}} + \frac{\sqrt{6}}{5} \left(T + C\right).
\label{eq:SCS_CPV}
\end{eqnarray}
}
\end{widetext}

The total amplitude for the SCS decays where we consider CP violation can now be written as
\begin{eqnarray}
\mathcal{A}&&(D\to P_1 P_2)=\nonumber\\
&&\frac{1}{2}\Big[(V^{}_{us}V^*_{cs} - V^{}_{ud}V^*_{cd})A(D\to P_1 P_2)\nonumber\\
&&+(V^{}_{us}V^*_{cs} + V^{}_{ud}V^*_{cd})B(D\to P_1 P_2)\Big]
\end{eqnarray}
with $P_1$ and $P_2$ as $\pi$, $K$ or $\eta_8$.

The $\Delta U=0$ part of the $D^0\to K_SK_S$ decay amplitude includes a weak exchange topology. This contribution can be potentially large and lead to the
enhancement of the CP asymmetry in this channel~\cite{Atwood:2012ac,Nierste:2015zra} due to the Zweig suppression of the $\Delta U =1$ part of the
amplitude. It has been pointed out in~\cite{Nierste:2015zra} that the possibly large weak exchange contribution and the suppression of the branching fraction decorrelates the CP asymmetry in this channel from the other SCS channels where the weak exchange topology does not contribute. In our
framework, this weak exchange topology is generated by rescattering and hence related to the parameters in the $\Delta U =1$ amplitude. The generation of the weak exchange topology by rescattering was also discussed in~\cite{Cheng:2010ry}. This characterization of the weak exchange topology leads to the parametric correlation between all the SCS $\Delta U =0$ amplitudes leading to a correlation amongst the CP violation in these channels.

\section{Measurements of CP asymmetries}
\label{sec:exp}

\begin{table}[h!]
\begin{center}
\begin{tabular}{|l|l|l|}
\hline
channel				& mean $\pm$ rms (\%)			& reference											\\
\hline				                                        																
$D^0\to K^+K^-$	    	&$-0.16\pm0.12$			        & HFLAV~\cite{Amhis:2016xyh}	                                				\\
$D^0\to \pi^+\pi^-$		&$\phantom{-}0.00\pm0.15$		& HFLAV~\cite{Amhis:2016xyh}	                               			\\
$D^0\to \pi^0\pi^0$		&$-0.03  \pm 0.64$			        & HFLAV~\cite{Bonvicini:2000qm,Nisar:2014fkc}	               		\\
$D^+\to K^+K_S$	    	&$-0.11  \pm 0.25$			        & HFLAV~\cite{Link:2001zj,Mendez:2009aa,Ko:2012uh,Lees:2012jv}	\\
\hline
$D^+_s\to K_S\pi^+$		&$\phantom{-}0.38\pm0.48$		& HFLAV~\cite{Amhis:2016xyh}	                             			\\
$D^+_s\to K^+\pi^0$		&$-0.266\pm0.238\pm0.009$	        & CLEO~\cite{Mendez:2009aa}	                            				\\
\hline
$D^0\to K_SK_S$	    	&$-2.9\pm5.2\pm2.2$			& LHCb~\cite{Aaij:2015fua} 		                                			\\
$D^0\to K_SK_S$	    	&$-0.2\pm1.53\pm0.17$		        & Belle~\cite{Abdesselam:2016gqq}                              			\\
\hline
\end{tabular}
\end{center}
\caption{Measurements of CP asymmetries in various channels.}
\label{tab:ACPexp}
\end{table}%

Much progress has been made in the measurement of CP asymmetries, a compendium
of which can be found on the HFLAV~\cite{Amhis:2016xyh} website. It is important
to note here that the CP asymmetries measured by the experiments in the neutral
$D^0$ channel is the sum of direct and indirect CP asymmetries (time integrated), while the HFLAV averages
are direct CP asymmetries only. For the decay of the charged
$D$ mesons the direct CP asymmetry is measured. The most notable of the CP
asymmetry measurements is the very precise measurement of $\Delta{\rm A}_{\rm
CP} = {\rm A}_{\rm CP}(D^0\to K^+K^-)-{\rm A}_{\rm CP}(D^0\to\pi^+\pi^-)$ by
LHCb~\cite{Aaij:2016cfh} with their 7 TeV and 8 TeV data. Combining this result
with the previous LHCb measurement~\cite{Aaij:2014gsa} and the LHCb measurement
of indirect asymmetry~\cite{Aaij:2013ria,Aaij:2015yda} using ${\rm A_\Gamma}\sim
-\Delta{\rm A}_{\rm CP}^{\rm ind}$ and the measurement of $y_{\rm
CP}$~\cite{Aaij:2011ad} they extracted the difference of the direct CP asymmetry
as
\begin{equation}
\Delta{\rm A}_{\rm CP}^{\rm dir}=(-0.061\pm0.076)\%,
\label{eq:DACP}
\end{equation}
while the HFLAV world average stood
at~\cite{Amhis:2016xyh}:
\begin{equation}
\Delta{\rm A}_{\rm CP}^{\rm dir}=(-0.137\pm0.070)\%.
\label{eq:DACP_H}
\end{equation}

Recently, LHCb has released the analysis of $\Delta{\rm A}_{\rm CP}$ with combined 9fb$^{-1}$ of data collected over Run I and II~\cite{Aaij:2019kcg}. With this analysis, the collaboration has measured $\Delta{\rm A}_{\rm CP}$ with more than 5$\sigma$ significance. This is the first measurement of CP asymmetry in the up-type quark sector and the only significant measurement of CP asymmetry in charm mesons. Considering the relevance of this measurement, we include this in our analysis even though it was released after the first version of our work was made public. From the LHCb measurements we have:
\begin{eqnarray}
\Delta{\rm A}_{\rm CP}&=&(-0.154\pm0.029)\%,\\
\Delta{\rm A}_{\rm CP}^{\rm dir}&=&(-0.156\pm0.029)\%,
\end{eqnarray}
which leades to a world average of:
\begin{eqnarray}
\Delta{\rm A}_{\rm CP}^{\rm dir}=(-0.164\pm0.028)\%,
\end{eqnarray}

We do not use the measurement of the individual asymmetries ${\rm A}_{\rm CP}(D^0\to\pi^+\pi^-)$
and ${\rm A}_{\rm CP}(D^0\to K^+K^-)$ since the LHCb results for
these~\cite{Aaij:2014gsa} are used in the estimation of $\Delta{\rm A}_{\rm
CP}^{\rm dir}$. The results from the other experiments on these individual
asymmetries do not improve the fit in any manner since they are much less
precise. We have numerically checked the validity of this statement. For the sake
of completeness we list some of the relevant CP asymmetries in table~\ref{tab:ACPexp}
that have been measured till date. We do not use these measurements in the fit but predict them
from a fit to the branching fractions and the HFLAV average of $\Delta{\rm A}_{\rm CP}^{\rm dir}$.

In the recent past some theoretical effort has been put on estimating CP
asymmetry in $D^0\to K_SK_S$~\cite{Hiller:2012xm,Brod:2011re,Nierste:2015zra}
along with experimental measurements being performed at LHCb~\cite{Aaij:2015fua}
and Belle~\cite{Abdesselam:2016gqq} as listed in table~\ref{tab:ACPexp}.
There is an
older measurement by CLEO~\cite{Bonvicini:2000qm} which we do not quote here
since it is much less precise. In~\cite{Brod:2011re}, ${\rm A}_{\rm CP}(D^0\to
K_SK_S)$ was estimated to be about 0.6\% in magnitude. In~\cite{Hiller:2012xm}
the CP asymmetry in $D^0\to K_SK_S$ was related to $\Delta {\rm A}^{\rm
dir}_{\rm CP}$ and an estimation of about 0.4\% was made for the former. In~\cite{Nierste:2015zra} it was shown
that this asymmetry can be of $O(1\%)$ due to possibly large contributions from the weak exchange
diagrams to the $\Delta U = 0$ part of the amplitude. In the following section we present our
results for $A_{\rm CP}(D^0\to K_SK_S)$ using data on both branching fractions
and $\Delta {\rm A}^{\rm dir}_{\rm CP}$ to constrain the parameters along with the prediction
of CP asymmetries of several other SCS modes.

The only SCS channel for which the CP asymmetry is predictably 0 in the SM is that in $D^+\to\pi^+\pi^0$~\cite{Buccella:1992sg,Grossman:2012eb} since it is driven by a single isospin amplitude and hence lacks the two separate strong {\it and} weak phases necessary for a non-zero CP asymmetry. Recently Belle has measured a CP asymmetry in this channel consistent with the null SM value~\cite{Babu:2017bjn}:
\begin{equation}
{\rm A}_{\rm CP}(D^+\to \pi^+\pi^0)=(2.31\pm1.24\pm0.23)\%
\end{equation}
which has a much better precision than the previous CLEO measurement~\cite{Mendez:2009aa} which has a error of 2.9\% and is consistent with the null SM value.

The BESIII Collaboration has performed the first measurements of CP asymmetry in $D^+\to K^+K_S$ and $D^+\to K^+K_L$~\cite{Ablikim:2018inm} which should be exactly equal since both are driven by $D^+\to K^+\bar{K}^0$ only. The two measurements are in good agreement with each other and consistent with 0:
\begin{eqnarray}
{\rm A}_{\rm CP}(D^+\to K^+K_S)=(-1.8 \pm 2.7 \pm 1.6)\%\;,\nonumber\\
{\rm A}_{\rm CP}(D^+\to K^+K_L)=(-4.2 \pm 3.2 \pm 1.2)\%\;.
\end{eqnarray}

\begin{table*}
\begin{center}
\begin{tabular}[b]{|l|c|c|c|}
\hline
Channel					& Fit	($\times 10^{-3}$)	        & PDG ($\times 10^{-3}$)			& BESIII ($\times 10^{-3}$)\\
\hline
\multicolumn{4}{|c|}{\bf SCS} \\
\hline
$D^0\to\pi^+\pi^-$		& \phantom{0}1.448 $\pm$ 0.019	& 1.407 $\pm$ 0.025				&1.508 $\pm$ 0.028		\\
$D_0^+\to \pi^0\pi^0$	& \phantom{0}0.816 $\pm$ 0.025	& 0.822 $\pm$ 0.025				&--					\\
$D^+\to \pi^+\pi^0$		& \phantom{0}1.235 $\pm$ 0.033	& 1.17 $\pm$ 0.06				&1.259 $\pm$ 0.040		\\
$D^0\to K^+K^-$		& \phantom{0}4.064 $\pm$ 0.044	& 3.97 $\pm$ 0.07				&4.233 $\pm$ 0.067		\\
$D^0\to K_SK_S$		& \phantom{0}0.168 $\pm$ 0.012	& \phantom{0}0.17 $\pm$ 0.012	&--					\\
$D^+\to K^+K_S$		& \phantom{0}3.164 $\pm$ 0.056	& 2.83 $\pm$ 0.16				&3.183 $\pm$ 0.067		\\
$D^+\to K^+K_L$		& \phantom{0}3.164 $\pm$ 0.056	& 		--					&3.21 $\pm$   0.16		\\
$D_s^+\to \pi^0K^+$		& \phantom{0}1.41 $\pm$ 0.15		& 0.63 $\pm$ 0.21 				&--					\\
$D_s^+\to\pi^+K_S$		& \phantom{0}1.24 $\pm$ 0.06		& 1.22 $\pm$ 0.06				&--					\\
\hline
\hline
\multicolumn{4}{|c|}{\bf CA \& DCS}\\
\hline
$D^+\to \pi^+K_S$		& 15.80 $\pm$ 0.29				& 14.7 $\pm$ 0.8				&15.91 $\pm$ 0.31		\\
$D^+\to \pi^+K_L$		& 14.37 $\pm$ 0.52				& 14.6 $\pm$ 0.5				&--					\\
$D^0\to\pi^+K^-$		& 38.96 $\pm$ 0.32 				& 38.9 $\pm$ 0.4				&--					\\
$D^0\to\pi^0 K_S$		& 12.29 $\pm$ 0.21				& 11.9 $\pm$ 0.4				&12.39 $\pm$ 0.28		\\
$D^0\to\pi^0 K_L$ 		& \phantom{0}9.73 $\pm$ 0.21		& 10.0 $\pm$ 0.7				&--					\\
$D_s^+\to K^+K_S$	    	& 14.67  $\pm$ 0.41		 		& 15.0 $\pm$ 0.5				&--					\\
$D^+\to\pi^0 K^+$		& \phantom{0}0.151 $\pm$ 0.013	& \phantom{0}0.181 $\pm$ 0.027	&0.231 $\pm$ 0.022		\\
$D^0\to\pi^-K^+$		&  \phantom{0}0.141 $\pm$ 0.003	&\phantom{0}0.1385 $\pm$ 0.0027	&--					\\
$D^0\to\pi^\pm K^\mp$	& 39.1 $\pm$ 0.32 				&--							&38.98 $\pm$ 0.52		\\
\hline
\end{tabular}

\caption{The branching fractions that were used in the fit~\cite{Patrignani:2016xqp,Ablikim:2018ydy,Ablikim:2018inm}. We also use $  {\rm BR}(D_s^+\to K^+K_{S,L})  = (29.5\pm1.4)\times 10^{-3}$~\cite{Zupanc:2013byn} measured by Belle. The fit results are the same for both the negative and the positive solutions for the phases, $\delta_i$.}
\label{tab:BR-fit}
\end{center}
\end{table*}

\section{Results and their consequences}
\label{sec:fit}
We use \HEPfit~\cite{HEPfit} to perform a fit in the Bayesian framework. The 7
amplitudes ($T$, $C$, $\Delta$, $K$, $K^\prime$, $\kappa$ and $\kappa^\prime$), the
$\mathrm{SU}(3)_{F}$ breaking parameter quantifying the shift in the $D^+_s$ phase
($\epsilon_\delta$), 4 phases ($\delta_0$, $\delta_0^\prime$,
$\delta_{\frac{1}{2}}$, $\delta_1$) and a mixing angle $\phi$ are constrained
using 17 branching fractions. The $\Delta U =0$ part of the amplitudes require three
additional parameters $P$, $\Delta_3$ and $\Delta_4$. The first two, $P$ and
$\Delta_3$ always appear as a sum in the $\Delta U =0$ part of all the decay amplitudes
and hence it is not possible to disentangle them individually from data.
Moreover, it is not possible to estimate the sizes of these parameters from
first principles and hence we work with the ratio $(P + \Delta_3)/T$ in our fit
and use $\Delta {\rm A}^{\rm dir}_{\rm CP}$ to constrain this combination. As
discussed before, $\Delta_4$ is expected to be tiny due to the
approximate selection rule and so we set it to 0 in our fit. The experimental
numbers used for the fit are listed in table~\ref{tab:BR-fit} and in
equation~(\ref{eq:DACP}). For the branching fractions we use the $D^0$ and the
$D_{(s)}^+$ decays with only $\pi$ and $K$ in the final state. In addition to the PDG averages listed in table~\ref{tab:BR-fit} we also use the recent measurements made by the BESIII Collaboration~\cite{Ablikim:2018ydy,Ablikim:2018inm} which are comparable or better than the PDG averages.

\subsection[Fit to branching fractions and \texorpdfstring{$\Delta {\rm A}^{\rm dir}_{\rm CP}$}{DeltaACP}]{Fit to branching fractions and \boldmath $\Delta {\rm A}^{\rm dir}_{\rm CP}$}
\label{sec:fit2}

The fit results are presented in table~\ref{tab:fit}
along with the correlation matrix for the fitted parameters. The parameter $(P + \Delta_3)/T$ is excluded from the
correlation matrix because it is essentially uncorrelated with the other
parameters being determined by $\Delta {\rm A}^{\rm dir}_{\rm CP}$ while the
other parameters are determined by the branching ratio data. As a cross-check we
also performed a fit using MINUIT routines and have verified that the results are
exactly the same. The error analysis was done in \HEPfit and is taken as the RMS
of the posterior distributions of the parameters and observables.

We find two equivalent solutions for the parameters in the $\Delta U =1$
part of the amplitude from the
branching fractions alone. The solutions are distinct only for the posterior
distributions of the 4 phases ($\delta_0$, $\delta_0^\prime$,
$\delta_{\frac{1}{2}}$, $\delta_1$) with $\delta_i \to -\delta_i$ relating these
two solutions. The rest of the parameters have identical solutions.
However, these two solutions lead to very different fits for $(P +
\Delta_3)/T$ from the ${\rm A}^{\rm dir}_{\rm CP}$ data and hence we present
both the solutions.

\begin{table*}
\begin{center}
\subfloat[Fit values of the parameters. $T$, $C$, $\Delta$, $\kappa^{(\prime)}$ and $K^{(\prime)}$ are in units of GeV$^3$. The angle $\phi$ and the phases $\delta_0,\delta_0^\prime,\delta_\frac{1}{2}$ and $\delta_1$ are in radians. The parameter $\epsilon_\delta$ is dimensionless. The phases $\delta_i$ have two solutions, positive and negative. The negative solution is better motivated as explained in the text. The solution for $(P+\Delta_3)/T$ changes accordingly while the solutions for all other parameters remain the same.]{
\begin{tabular}{|l|c||c|c|c|}
\hline
\multirow{1}{*}{}	& \multicolumn{1}{c||}{($\mu\pm\sigma$)}      & \multirow{2}{*}{}		& \multicolumn{2}{c|}{($\mu\pm\sigma$)}\\
\hline
$T$			      	& $\phantom{-}0.424 \pm 0.003$	 	&$\delta_0$		            	& $-2.373 \pm 0.062$      	& $\phantom{-}2.373 \pm 0.062$\\
$C$			        & $-0.211 \pm 0.003$				&$\delta_0^\prime$	    		& $-0.840 \pm 0.046$    	& $\phantom{-}0.840 \pm 0.046$\\
$\kappa$		    	& $-0.036 \pm 0.004$	 		 	&$\delta_\frac{1}{2}$			& $-1.632 \pm 0.020$      	& $\phantom{-}1.632 \pm 0.020$\\
$\kappa^\prime$       	& $-0.063 \pm 0.088$				&$\delta_1$		            	& $-1.085 \pm 0.038$     	& $\phantom{-}1.085 \pm 0.039$\\
$K$		          	& $\phantom{-}0.100 \pm 0.012$		&&&\\
$K^\prime$          	& $-0.153 \pm 0.072$				&&&\\
$\Delta$            	& $-0.026 \pm 0.019$        			&&&\\
$\phi$		        & $\phantom{-}0.435 \pm 0.025$		&$(P+\Delta_3)/T$	    		& $-1.897 \pm 0.211$	& $-0.465 \pm 0.211$\\
$\epsilon_\delta$   &	 $\phantom{-}0.067 \pm 0.061$			&&&\\
\hline
\end{tabular}
}\\
\subfloat[Correlation matrix of the parameters, $\lambda$ being the Wolfenstein parameter in the CKM matrix.]{
\setlength\tabcolsep{3pt}
\begin{tabular}{|c|cccccccccccccc|}
\hline
 & T & C & $\kappa$ & $\kappa^\prime$ & $\Delta$ & $K$ & $K^\prime$ & $\epsilon_\delta$ & $\delta_0$ & $\delta_0^\prime$ & $\delta_{1/2}$ & $\delta_1$ & $\phi$  & $\lambda$ \\
 \hline
T & $1$ & $-0.39$ & $-0.37$ & $-0.20$ & $-0.20$ & $\phantom{-}0.70$ & $\phantom{-}0.14$ & $\phantom{-}0.28$ & $-0.40$ & $-0.36$ & $-0.24$ & $-0.40$ & $-0.32$ & $-0.56$ \\
C && $\phantom{-}1.00$ & $-0.58$ & $\phantom{-}0.25$ & $\phantom{-}0.26$ & $-0.57$ & $-0.09$ & $-0.20$ & $\phantom{-}0.19$ & $\phantom{-}0.18$ & $\phantom{-}0.27$ & $\phantom{-}0.14$ & $\phantom{-}0.12$ & $\phantom{-}0.22$ \\
$\kappa$ &&& $\phantom{-}1.00$ & $-0.33$ & $-0.10$ & $\phantom{-}0.03$ & $-0.05$ & $-0.04$ & $\phantom{-}0.11$ & $\phantom{-}0.09$ & $-0.04$ & $\phantom{-}0.15$ & $\phantom{-}0.12$ &  $\phantom{-}0.26$ \\
$\kappa^\prime$ &&&& $\phantom{-}1.00$ & $\phantom{-}0.19$ & $-0.17$ & $\phantom{-}0.11$ & $-0.03$ & $\phantom{-}0.12$ & $\phantom{-}0.10$ & $\phantom{-}0.26$ & $\phantom{-}0.09$ & $\phantom{-}0.08$ & $-0.08$ \\
$\Delta$ &&&&& $\phantom{-}1.00$ & $-0.25$ & $\phantom{-}0.62$ & $\phantom{-}0.54$ & $-0.19$ & $-0.21$ & $-0.00$ & $-0.30$ & $-0.21$  & $\phantom{-}0.11$ \\
$K$ &&&&&& $\phantom{-}1.00$ & $\phantom{-}0.07$ & $\phantom{-}0.14$ & $-0.20$ & $-0.17$ & $\phantom{-}0.11$ & $-0.17$ & $-0.14$  & $-0.54$ \\
$K^\prime$ &&&&&&& $\phantom{-}1.00$ & $\phantom{-}0.81$ & $-0.40$ & $-0.42$ & $-0.06$ & $-0.52$ & $-0.37$ &  $\phantom{-}0.02$ \\
$\epsilon_d$ &&&&&&&& $\phantom{-}1.00$ & $-0.52$ & $-0.52$ & $-0.27$ & $-0.64$ & $-0.47$  & $-0.03$ \\
$\delta_0$ &&&&&&&&& $\phantom{-}1.00$ & $\phantom{-}0.94$ & $\phantom{-}0.24$ & $\phantom{-}0.81$ & $\phantom{-}0.95$  & $-0.11$ \\
$\delta_0^\prime$ &&&&&&&&&& $\phantom{-}1.00$ & $\phantom{-}0.22$ & $\phantom{-}0.81$ & $\phantom{-}0.86$ &  $-0.11$ \\
$\delta_{1/2}$ &&&&&&&&&&& $\phantom{-}1.00$ & $\phantom{-}0.25$ & $\phantom{-}0.19$ &  $-0.04$ \\
$\delta_1$ &&&&&&&&&&&& $\phantom{-}1.00$ & $\phantom{-}0.74$  & $-0.14$ \\
$\phi$ &&&&&&&&&&&&& $\phantom{-}1.00$ &  $-0.10$ \\
$\lambda$ &&&&&&&&&&&&&& $\phantom{-}1.00$ \\
\hline
\end{tabular}
}
\caption{The fit value of the parameters and their correlations}
\label{tab:fit}
\end{center}
\end{table*}

\begin{table*}
\begin{center}
\begin{tabular}[b]{|l|c|c||l|c|c|}
\hline
\multirow{2}{*}{${\rm A}_{\rm CP}$ ($D^0$)}	& \multicolumn{2}{c||}{($\mu\pm\sigma$) (\%)}      & \multirow{2}{*}{${\rm A}_{\rm CP}$ ($D^+_{(s)}$)}		& \multicolumn{2}{c|}{($\mu\pm\sigma$) (\%)}\\
\cline{2-3}\cline{5-6}
&$\delta_i\to$ -ve&$\delta_i\to$ +ve&&$\delta_i\to$ -ve&$\delta_i\to$ +ve\\
\hline
$D^0\to\pi^+\pi^-$           & $\phantom{-}0.117 \pm 0.020$    	&$\phantom{-}0.118 \pm 0.020$       		& $D^+\to K^+K_S$                   &$-0.028 \pm 0.005$        			&$-0.026  \pm 0.005$\\
$D^0\to\pi^0\pi^0$          &$\phantom{-}0.004 \pm 0.009$     	&$\phantom{-}0.079 \pm 0.010$        	& $D^+_s\to \pi^+K_S$               &$-0.040 \pm 0.007$      &$-0.036 \pm 0.007$\\
$D^0\to K^+K^-$             &$-0.047 \pm 0.008$      			&$-0.046 \pm 0.008$			         	& $D^+_s\to \pi^0K^+$               &$\phantom{-}0.048 \pm 0.006$        		&$-0.003 \pm 0.004$\\
$D^0\to K_SK_S$           & $\phantom{-}0.043 \pm 0.007$    	& $\phantom{-}0.038 \pm 0.007$&&&\\
\hline
\end{tabular}
\caption{Predictions of CP asymmetries using the branching fraction data and the HFLAV average of $\Delta{\rm A}_{\rm CP}^{\rm dir}$.}
\label{tab:ACP_HFLAV}
\end{center}
\end{table*}
 \begin{figure*}
\begin{center}
\subfloat{\includegraphics[width=.2\textwidth]{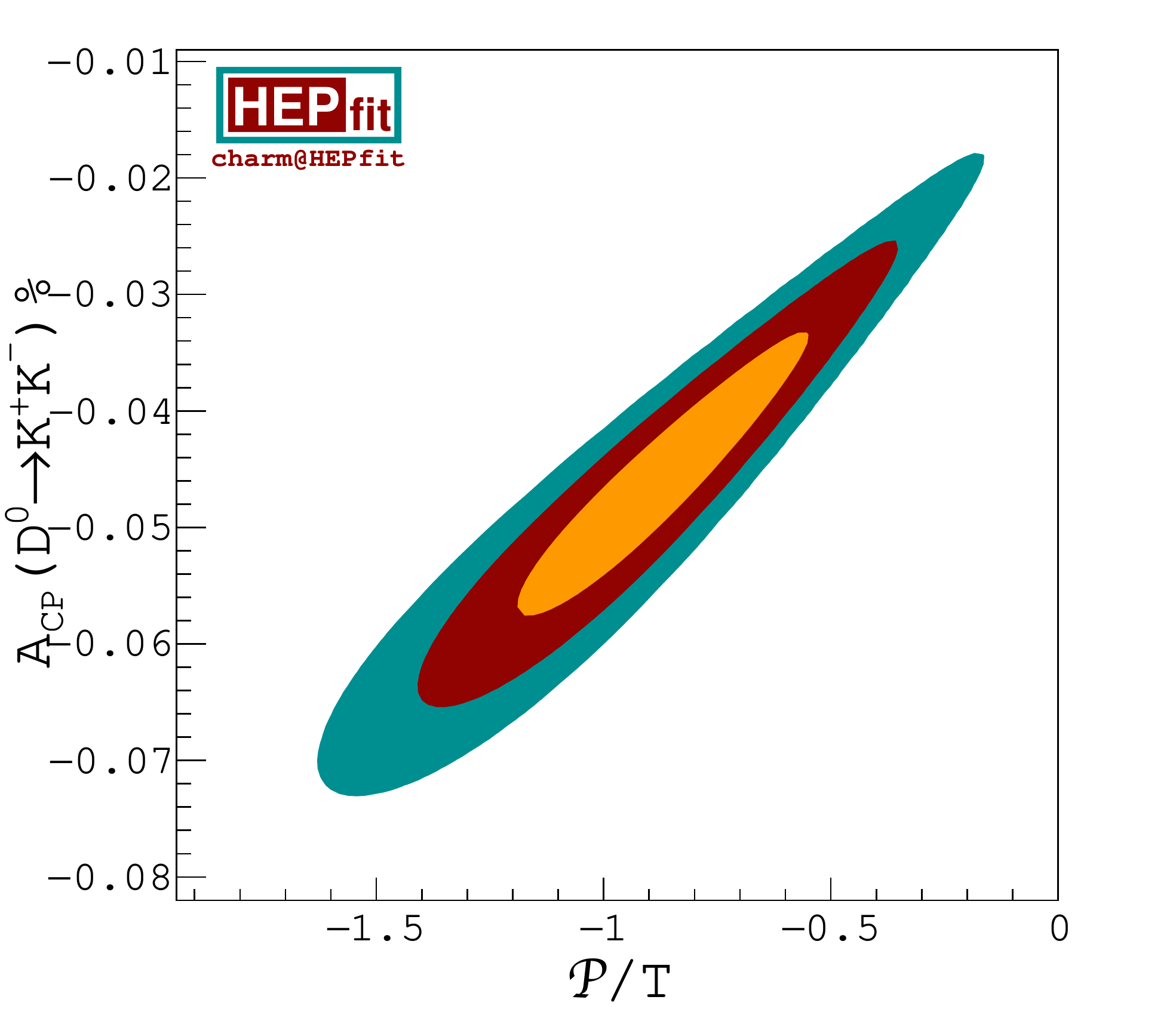}}
\subfloat{\includegraphics[width=.2\textwidth]{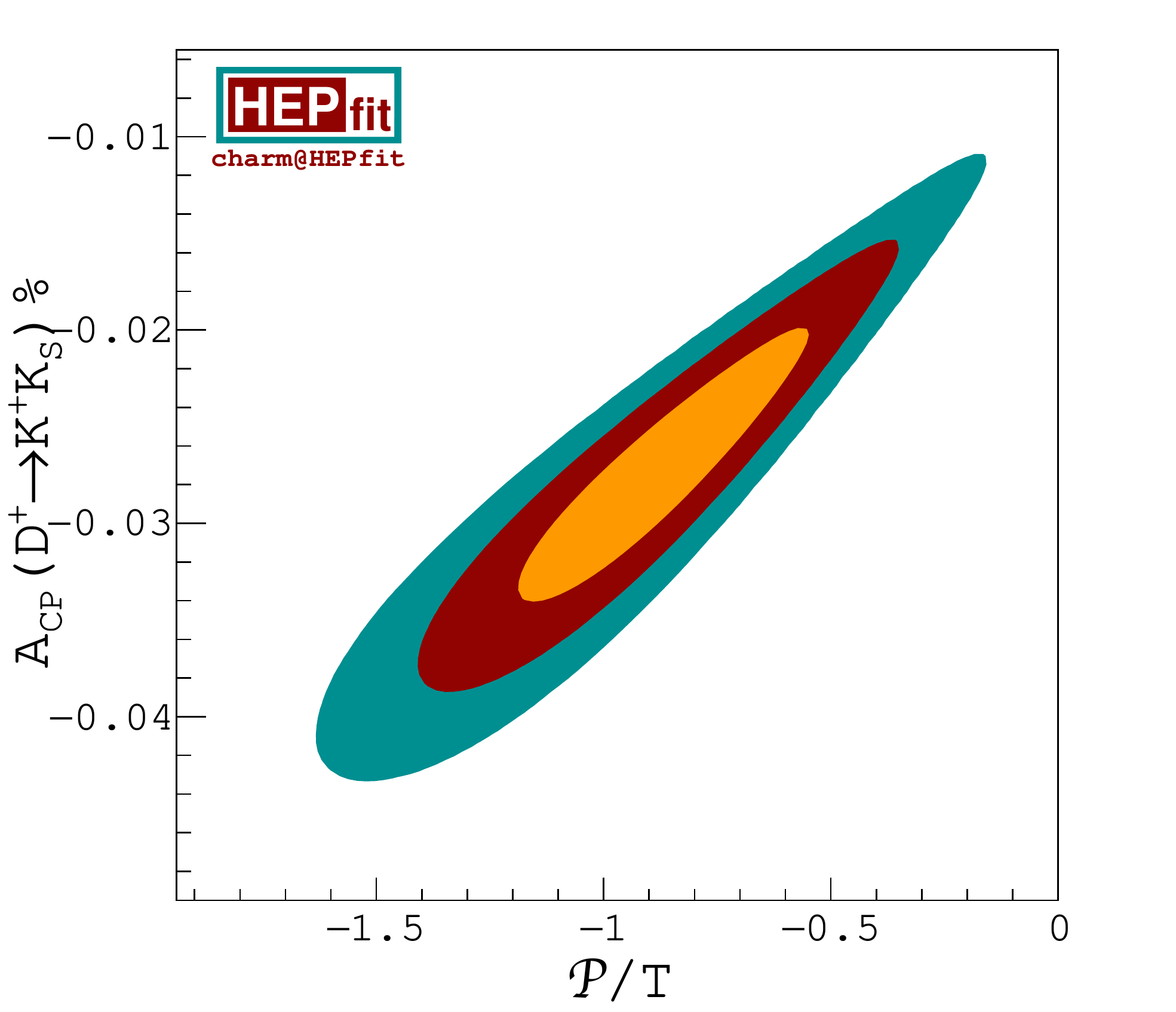}}
\subfloat{\includegraphics[width=.2\textwidth]{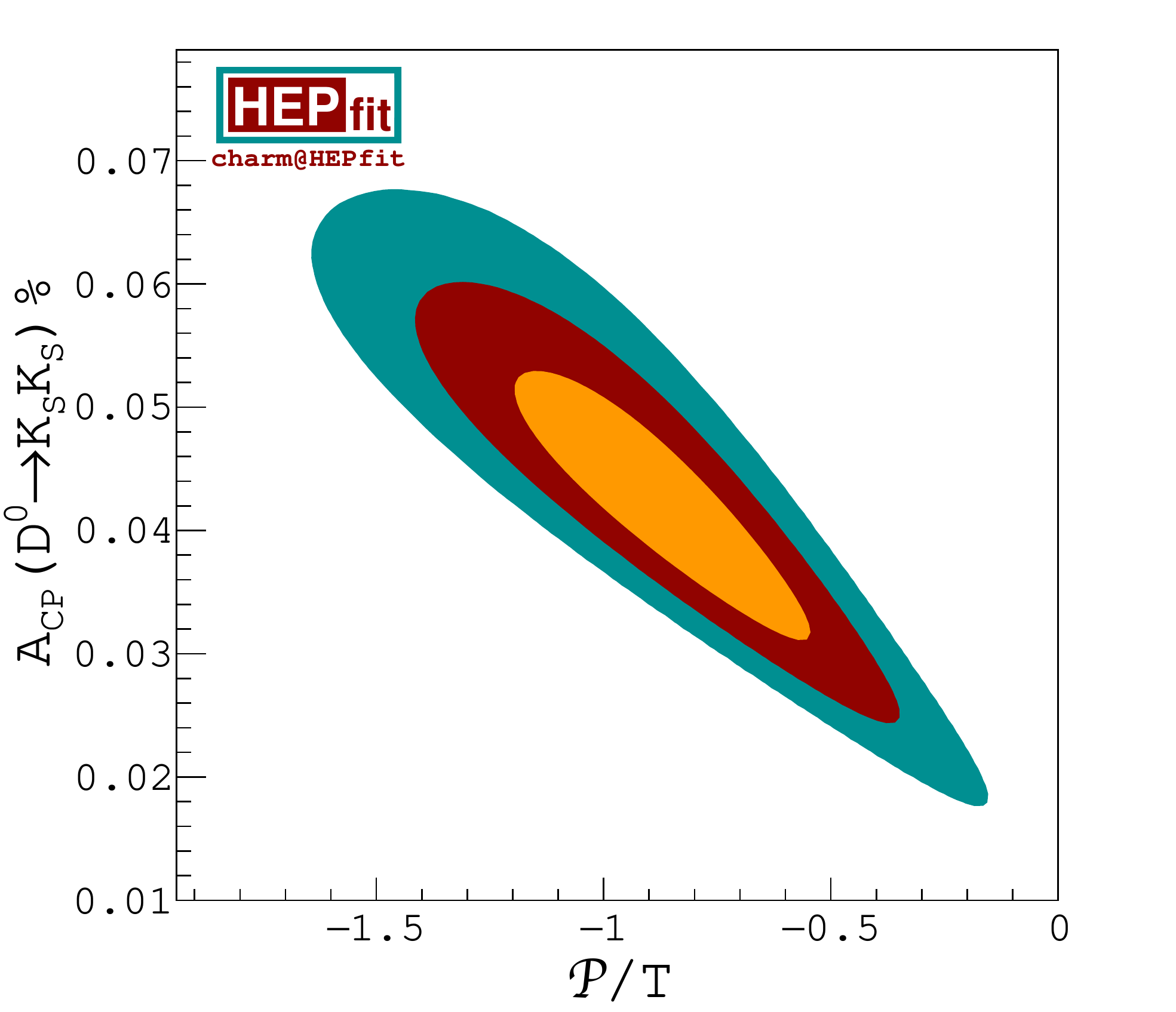}}
\subfloat{\includegraphics[width=.2\textwidth]{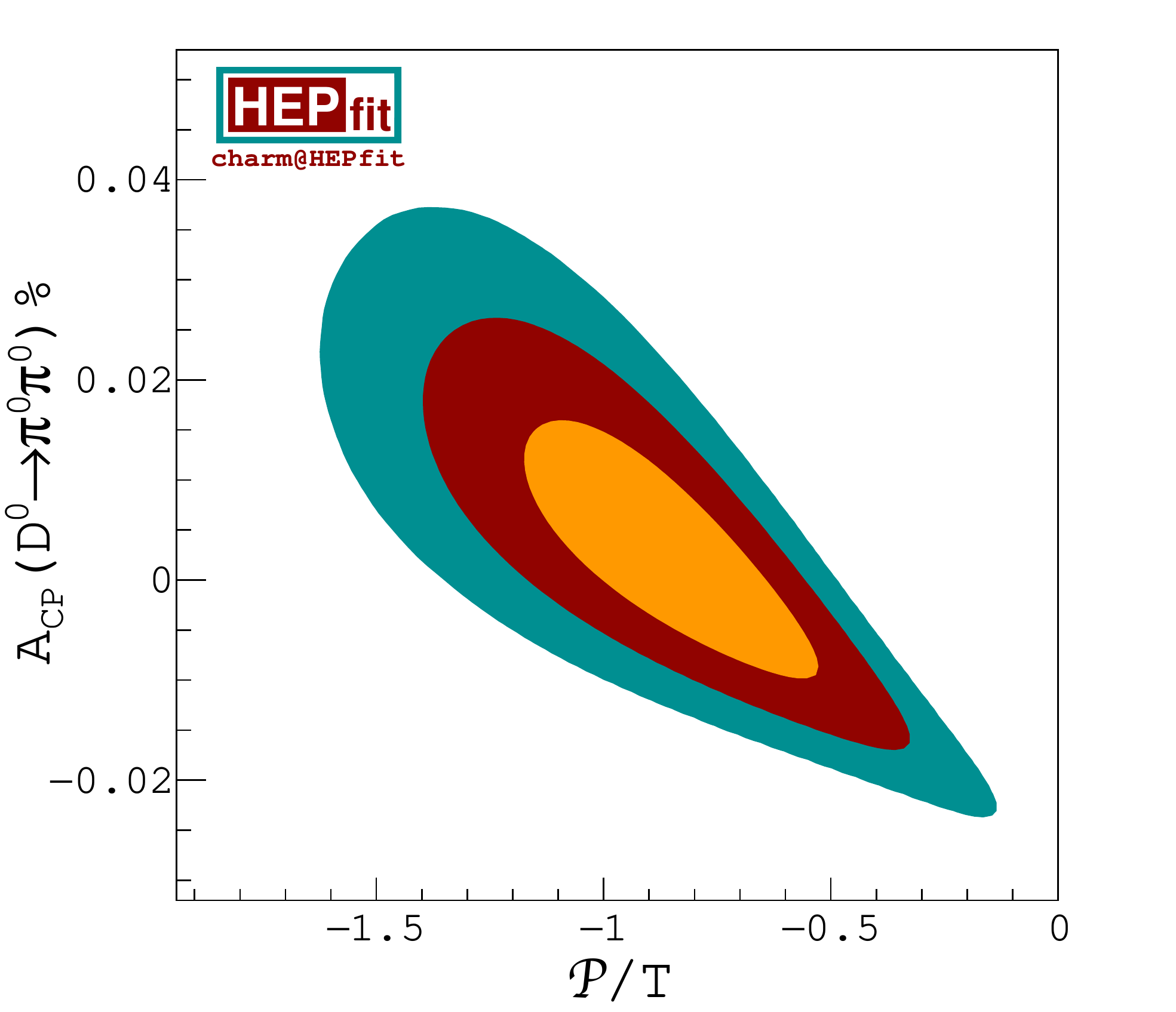}}\\
\subfloat{\includegraphics[width=.2\textwidth]{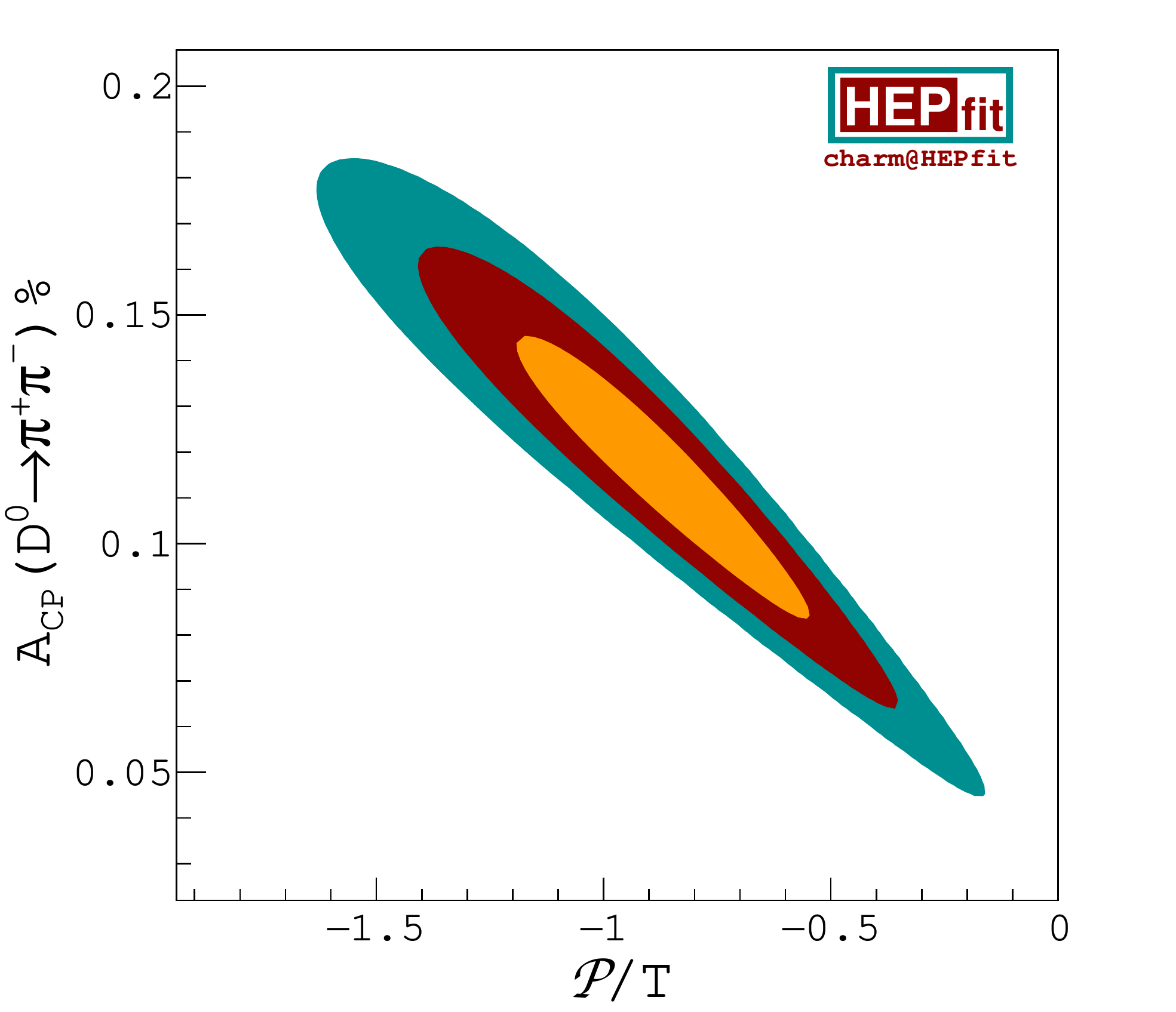}}
\subfloat{\includegraphics[width=.2\textwidth]{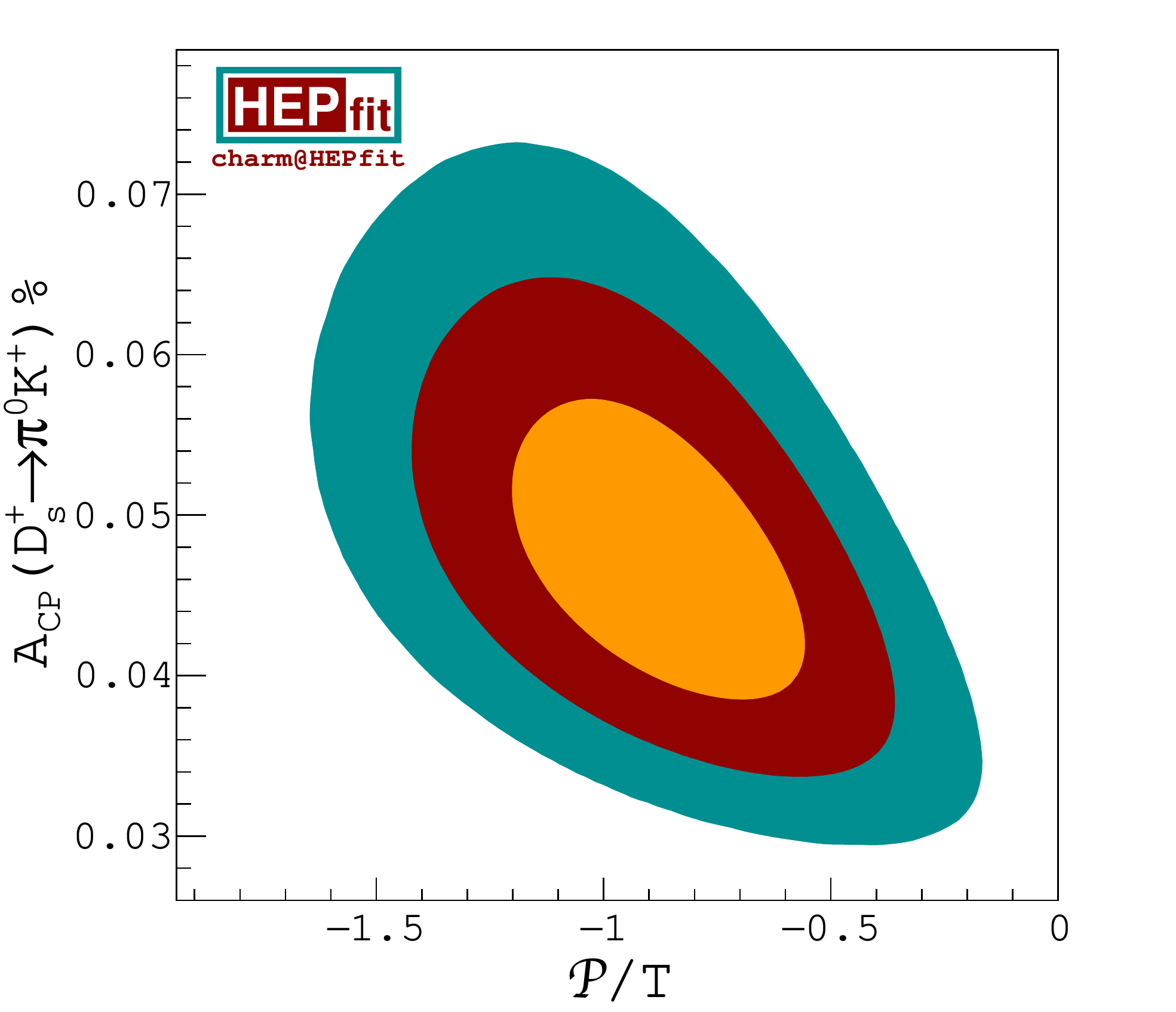}}
\subfloat{\includegraphics[width=.2\textwidth]{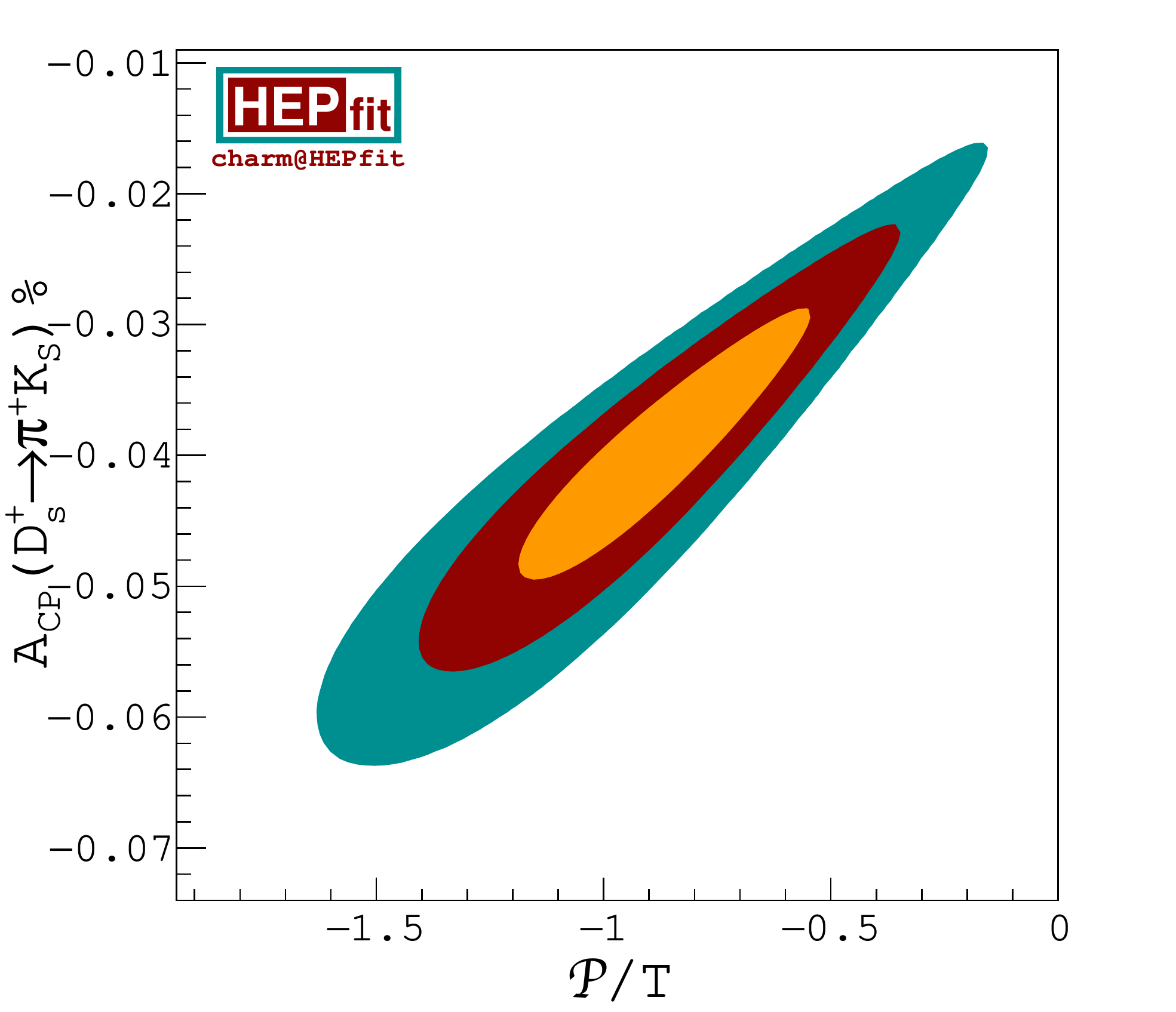}}
\subfloat{\includegraphics[width=.2\textwidth]{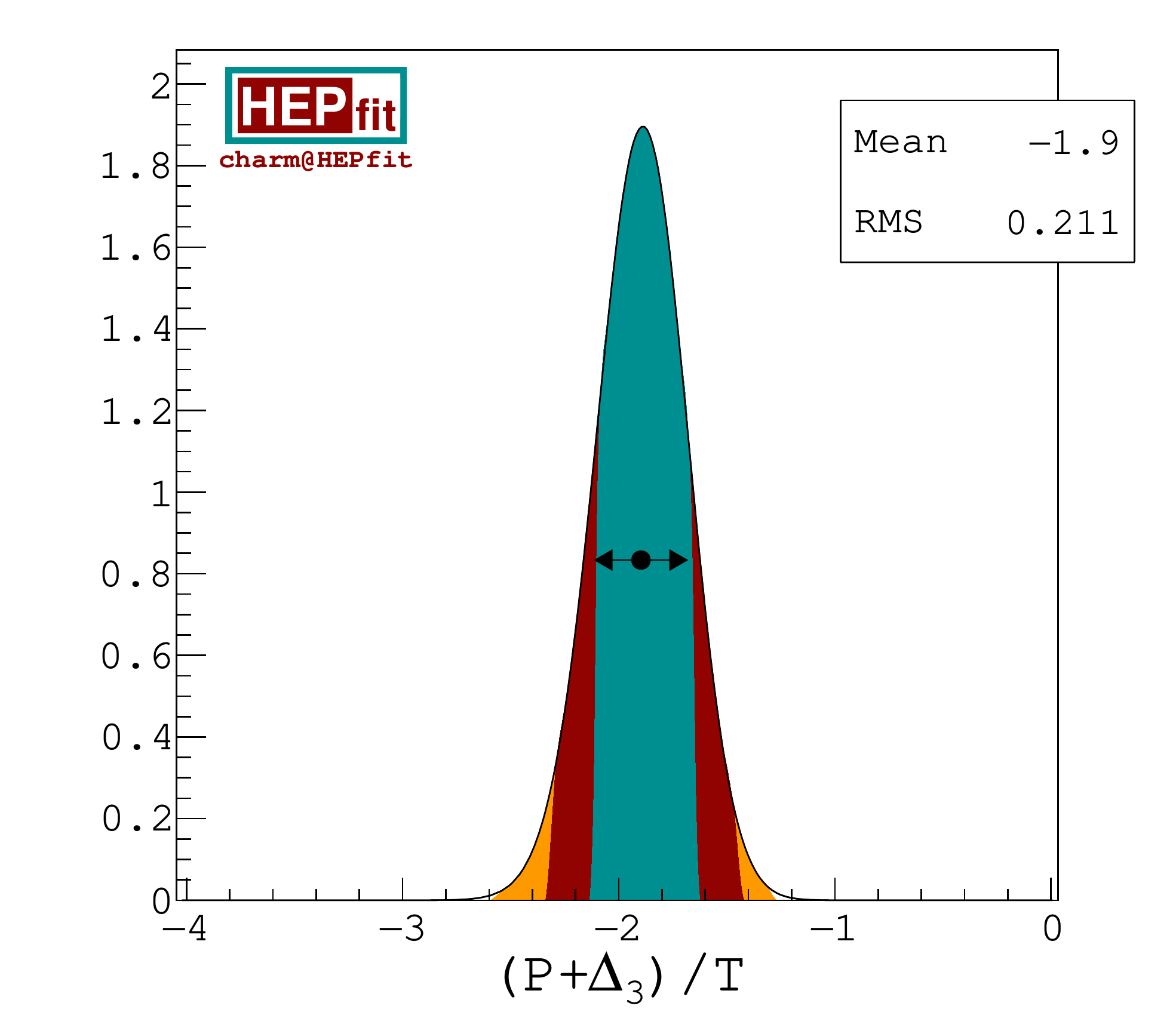}}
\caption{\it The correlations between $\mathpzcB{P}/T$ and the CP asymmetries (given in \%). HFLAV world average of $\Delta{\rm A}_{\rm CP}$ has been used for the fit and these CP asymmetries correspond to the negative solution for the phases. The orange, red and green regions are the 68\%, 95\% and 99\% probability regions respectively. The bottom right-most panel shows the fit to  $(P+\Delta_3)/T=\mathpzcB{P}/T-1$. The green, red and orange regions are the 68\%, 95\% and 99\% probability regions respectively.}
\label{fig:ACP_corr}
\end{center}
\end{figure*}

Since the phases coming from final state interactions should be interpreted as being generated by rescattering due to the presence of resonances, the phases should follow a distinct pattern determined by the masses of the resonances corresponding to particular isospin quantum numbers. The spectrum of the masses of these scalar resonances can be inferred upon by using the Gell-Mann-Okubo mass formula~\cite{Okubo:1961jc,Neeman:1961jhl,Okubo:1962zzc,GellMann:1962xb}. From the latter it can be seen that the mass of the resonance corresponding to the $I=1$ channel is smaller than the mass of the resonance corresponding to the $I=1/2$ channel. This implies that the strong phase shift due to resonance rescattering in the $I=1$ channel should be larger than that in the $I=1/2$ channel. As a consequence, the solution with negative phases seem to be the favourable solution. In the following discussion we shall focus more on the results with the negative solution keeping in mind that it is better motivated.

 The strongest constraint, by far, on $(P + \Delta_3)/T$ comes from $\Delta{\rm
 A}_{\rm CP}^{\rm dir}$. In figure~\ref{fig:ACP_corr} we show the  posterior
 distribution of $(P + \Delta_3)/T$. It was pointed out earlier that the CP asymmetries
 in the different channels are parametrically correlated. Hence
 the constraints from $\Delta{\rm A}_{\rm CP}^{\rm dir}$ also put constraints on
 CP asymmetries in the other decay modes. We use this to make predictions for the CP
 asymmetries in the other SCS channels which we present in table~\ref{tab:ACP_HFLAV}
 including the single channels ${\rm A}_{\rm CP}(D^0\to\pi^+\pi^-)$ and ${\rm
 A}_{\rm CP}(D^0\to K^+K^-)$. The errors in the prediction of the asymmetries
 vindicate our deduction that amongst the CP asymmetries $\Delta{\rm A}_{\rm
 CP}^{\rm dir}$ puts the strongest constraint on $(P + \Delta_3)/T$ by far.

 It is important to note here that the $(P + \Delta_3)/T\sim 2$ from the fit making $(P + \Delta_3)/T$
  comparable in size to the tree amplitudes parametrized by $T$. This implies that the penguin amplitudes,
  which appear in the terms proportional to $(P + T + \Delta_3)$, is the same size as
  the tree amplitudes. As a result, the penguin amplitudes
 can no longer be considered the dominant contribution in CP asymmetries of
 these channels. While it is still sizable compared to the contribution proportional
 to $T + C$, it can only bring about a factor of few, and not an order
 of magnitude, enhancement contrary to what was previously expected. This can be clearly gauged from
 figure~\ref{fig:ACP_corr}. The recent measurement of $\Delta{\rm A}_{\rm CP}$ which is over 5$\sigma$ in significance also implies that all the asymmetries listed in table.~\ref{tab:ACP_HFLAV} are non-zero with a significance of greater than 5$\sigma$ with the important exception of ${\rm A}_{\rm CP}(D^0\to\pi^0\pi^0)$ that can still be 0 for the negative solution of the phases but not for the positive solution. This parametric correlation between $\Delta{\rm A}_{\rm CP}$ and the other CP asymmetries also implies that the 
 predictions of CP asymmetries in these SCS modes using only the $\Delta {\rm
 A}_{\rm CP}$ data has errors bars that are much smaller than what is projected as the sensitivity at Belle II with 50 ab$^{-1}$
 of data as shown in table~\ref{tab:ACPF}. While on one hand this means that many of the CP asymmetries might be beyond the reach of the Belle II experiment, on the other hand it also means that if a CP asymmetry is measured in any of these channels that are far larger than what is predicted here, it will imply a possible necessity for sources of larger $\mathrm{SU}(3)_{F}$ beyond just large phases from FSI. An important test of this will be the measurement of ${\rm A}_{\rm CP}(D^0 \to K_S K_S)$ since its size is quite sensitive to how $\mathrm{SU}(3)_{F}$ is broken.

 Finally, we make some predictions from our fit. The branching fraction of the decay mode $D_s^+\to K^+K_L$ is yet unmeasured. However, the sum of the branching fractions for $D_s^+\to K^+K_S$ and $D_s^+\to K^+K_L$ has been measured by Belle yielding~\cite{Zupanc:2013byn}:
 \begin{eqnarray}
  {\rm BR}(D_s^+\to K^+K_S)&&+ {\rm BR}(D_s^+\to K^+K_L) = \nonumber\\
  &&(29.5\pm1.1\pm0.9)\times 10^{-3}
  \label{res:Sumkpk0}
 \end{eqnarray}
 while the branching fraction $ {\rm BR}(D_s^+\to K^+K_S)=(15.0\pm0.5)\times 10^{-3}$. Several predictions have been made in the past for the rate asymmetry between $D_s^+\to K^+K_L$ and $D_s^+\to K^+K_S$ which is tantamount to predicting the branching fraction of the former since the branching fraction of the latter mode is measured to a very good precision. In~\cite{Muller:2015lua,Wang:2017ksn} the branching fraction of $D_s^+\to K^+K_L$ is predicted to be smaller than the branching fraction of $D_s^+\to K^+K_S$. In contrast, we predict:
 \begin{equation}
 {\rm BR}(D_s^+\to K^+K_L)= (14.98\pm0.39)\times 10^{-3},
 \end{equation}
which is almost equal to the branching fraction of $D_s^+\to K^+K_S$ with the central value of the former being greater than the latter. The discrepancy is discussed in the next section where we also discuss the rate asymmetry and compare with results in the literature. If we do not use the result in eq.~(\ref{res:Sumkpk0}) we get
\begin{equation}
{\rm BR}(D_s^+\to K^+K_L)= (15.01\pm0.47)\times 10^{-3}.
\end{equation}

We also predict the relative strong phase between the amplitudes of the modes $D^0\to K^+\pi^-$ and $D^0\to K^-\pi^+$. The world average of the measured value of this phase is $(9.3^{+8.3}_{-9.2})^\circ$~\cite{Amhis:2016xyh}\footnote{The average corresponds to the one presented by HFLAV post-Moriond 2019.} when one assumes that there is no CP violation in the DCS decays. From our fit we get\footnote{The predicted value has the opposite sign for the positive solutions of the strong phases.}:
\begin{equation}
\delta_{K\pi}=\delta_{K^-\pi^+}-\delta_{K^+\pi^-} = 3.14^\circ \pm 5.69^\circ
\end{equation}
which is compatible with the measured value. Various other estimates of $\delta_{K\pi}$ can be found in~\cite{Chau:1993ec,Buccella:1994nf,Browder:1995ay,Falk:1999ts,Gao:2006nb}. In the exact $\mathrm{SU}(3)_{F}$ limit $\delta_{K\pi}$ should be 0~\cite{Kingsley:1975fe,Voloshin:1975yx,Wolfenstein:1995kv}, the deviation from which, as indicated by the fit result, underscores the significance of $\mathrm{SU}(3)_{F}$ breaking through strong phases in the framework that we use.
\subsection{Rate asymmetries}
\label{sec:RA}

\begin{figure*}
\begin{center}
\subfloat{\includegraphics[width=.25\textwidth]{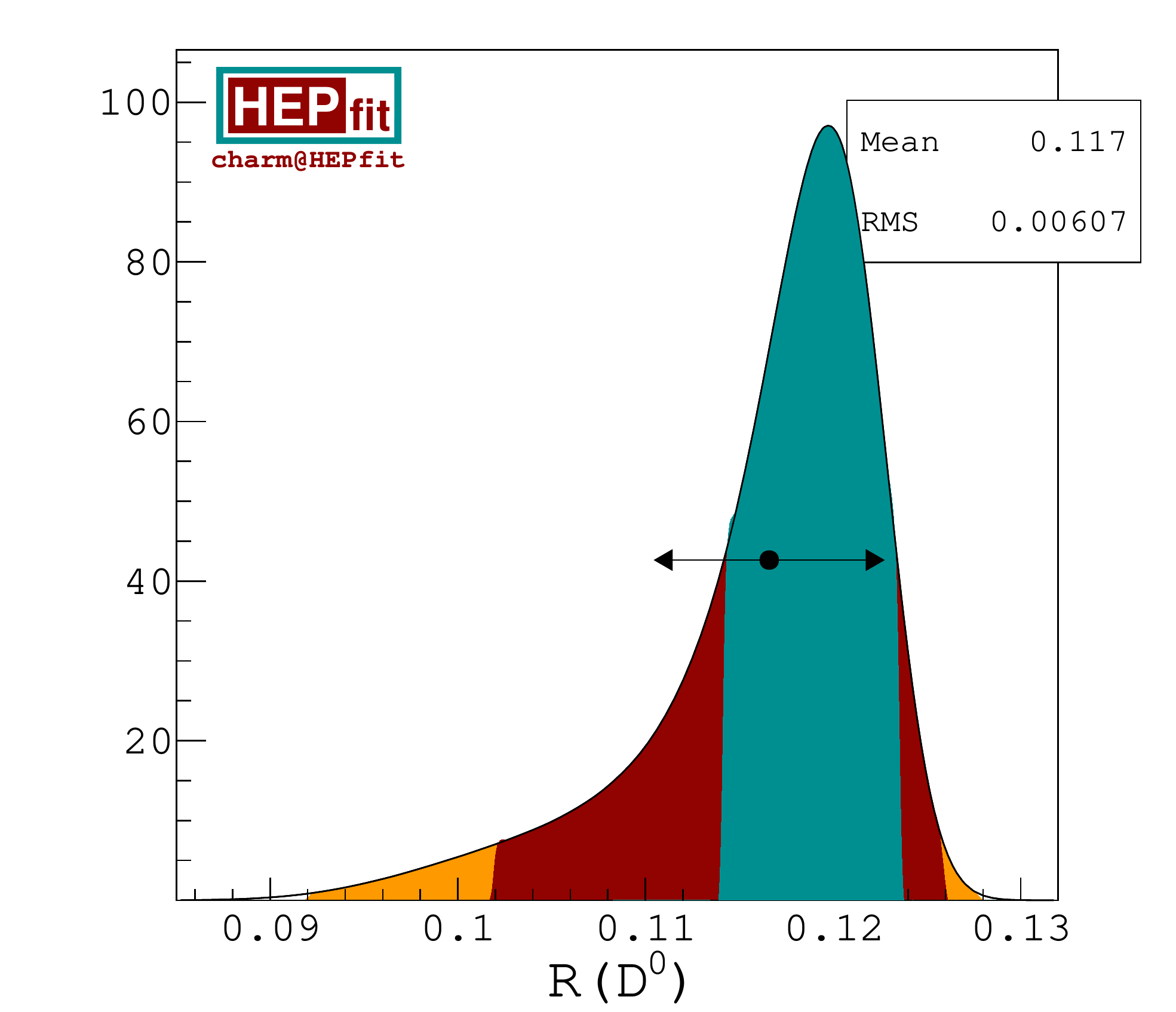}}
\subfloat{\includegraphics[width=.25\textwidth]{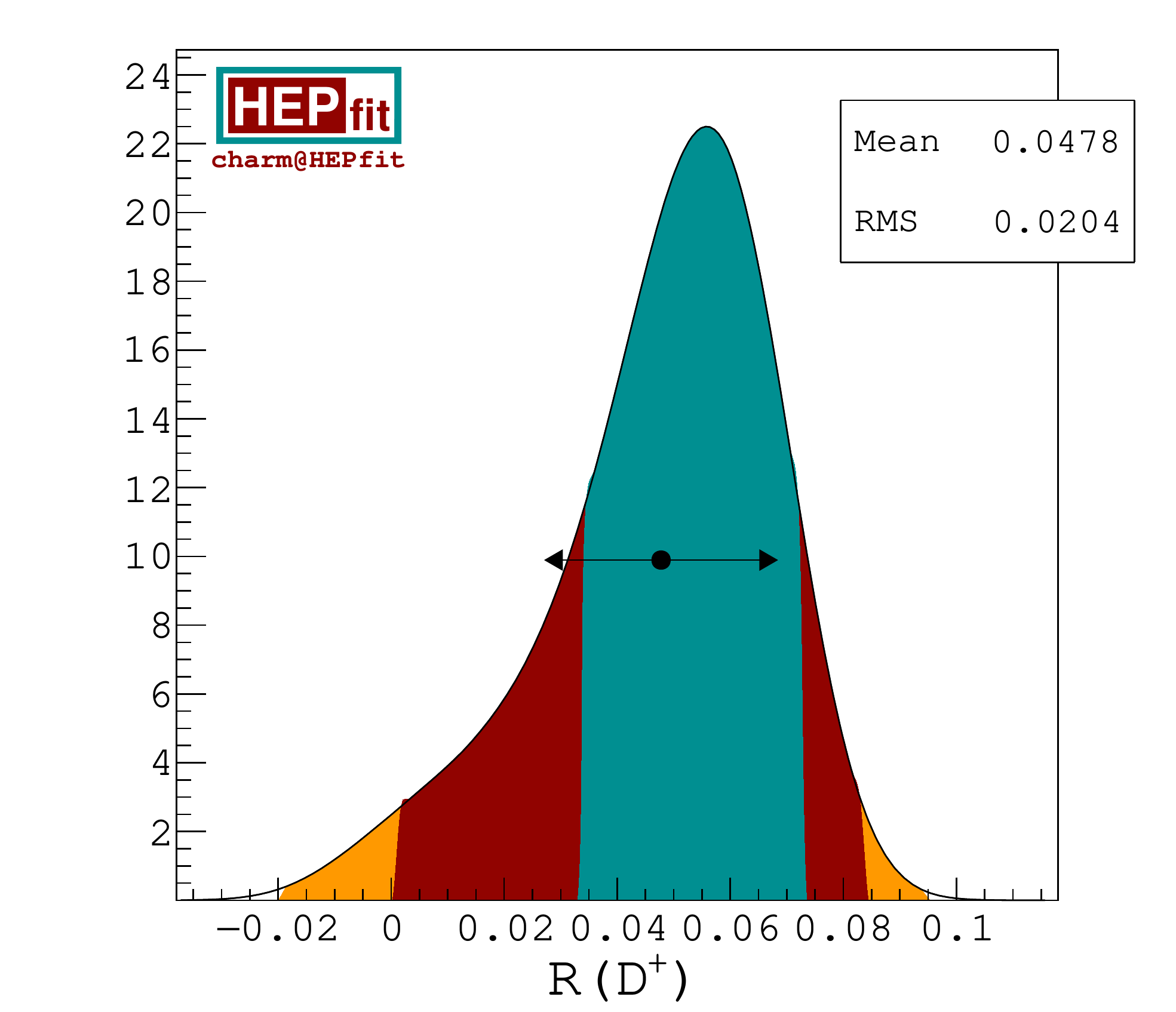}}
\subfloat{\includegraphics[width=.25\textwidth]{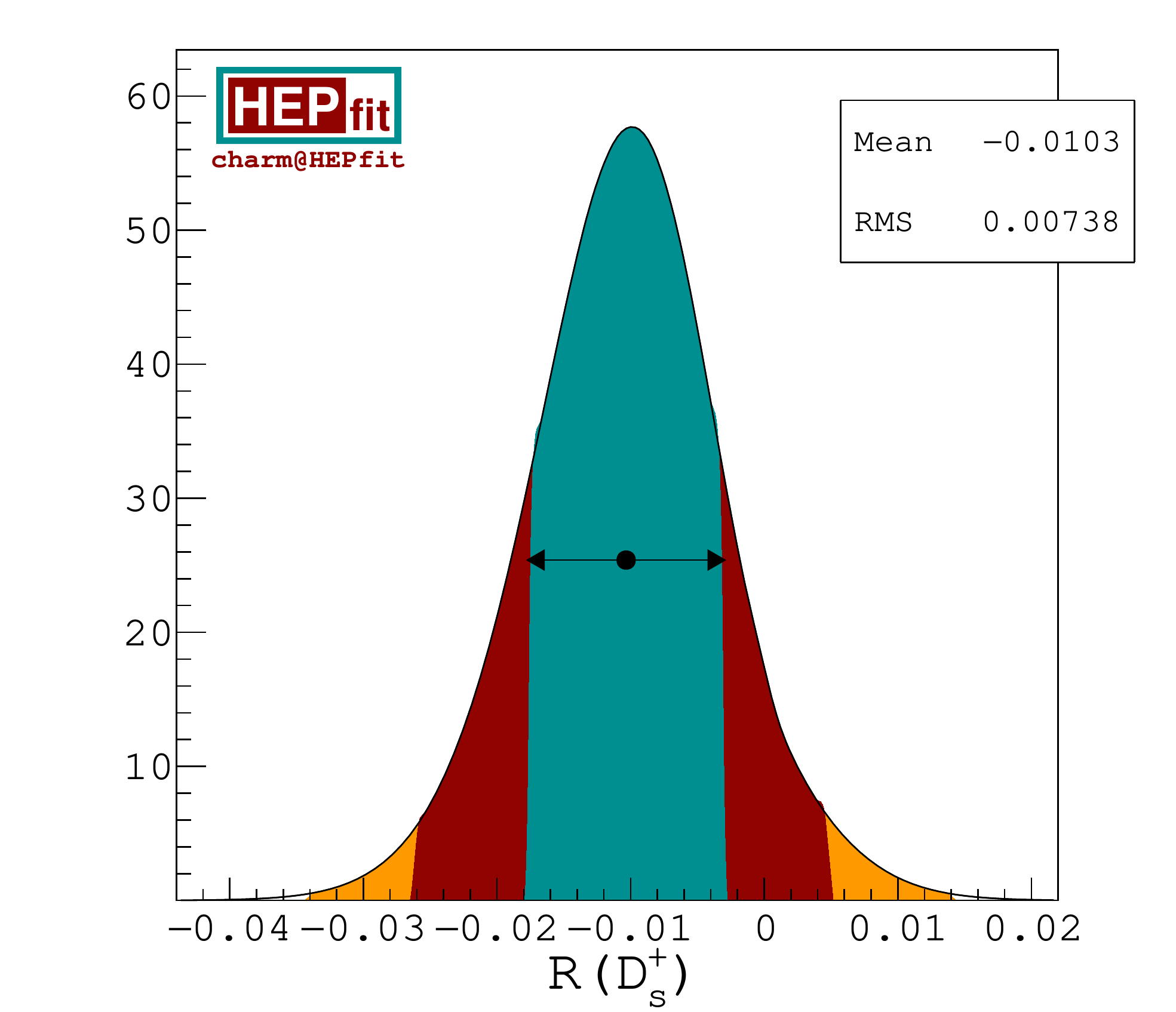}}\\
\caption{Fit results for $R(D^0,\pi^0)$, $R(D^+,\pi^+)$ and $R(D^+_s,K^+)$ using the branching fraction data in table~\ref{tab:BR-fit} and the HFLAV world average of $\Delta{\rm A}_{\rm CP}$. The green, red and orange regions are the 68\%, 95\% and 99\% probability regions respectively.}
\label{fig:D0Dp}
\end{center}
\end{figure*}

 One can also define rate asymmetries involving interference of CA and DCS decays of the neutral $D^0$ meson to the neutral $K\pi$ final state. A method for measuring this was first proposed in~\cite{Bigi:1994aw}. The rate asymmetry for the neutral $D^0$ initial state is defined as
 \begin{equation}
 R(D^0,\pi^0)\equiv \frac{\Gamma\left(D^0\to K_S\pi^0\right) - \Gamma\left(D^0\to K_L\pi^0\right)}{\Gamma\left(D^0\to K_S\pi^0\right) + \Gamma\left(D^0\to K_L\pi^0\right)},
 \label{eq:RD0}
 \end{equation}
For the charged $D^+$ in the initial state, the rate asymmetry is defined as  $R(D^+,\pi^+)$ with the substitutions $D^0\to D^+$ and $\pi^0\to \pi^+$. For $D_s^+$ the rate asymmetry is defined as $R(D_s^+,K^+)$ with the substitution $D^0\to D_s^+$ and $\pi^0\to K^+$  in the above relation.
The rate asymmetry in eq.~(\ref{eq:RD0}) leads us to another U-spin breaking parameter $\epsilon_0^\prime$, the real part of which is can be shown to be given by
\begin{equation}
{\rm Re}\left(\epsilon_0^\prime\right)=\frac{R(D^0,\pi^0)}{4\tan^2\theta_C}-\frac{1}{2}\,.
\end{equation}
From a CLEO Collaboration measurement of $K_S-K_L$ asymmetry~\cite{He:2007aj}, the measured values for  $R(D^0,\pi^0)$, $R(D^+,\pi^+)$ are
\begin{eqnarray}
R(D^0,\pi^0)^{\rm CLEO}&=&0.108\pm0.025\pm0.024,\nonumber\\
R(D^+,\pi^+)^{\rm CLEO}&=&0.022\pm0.016\pm0.018,
\end{eqnarray}
leading to a value of ${\rm Re}(\epsilon^\prime_0)=0.00\pm0.16$~\cite{Gronau:2015rda}. We compute $R(D^0,\pi^0)$, $R(D^+,\pi^+)$ and ${\rm Re}({\epsilon^\prime_0})$ from our fit to the branching fractions and CP asymmetries and get (c.f. figure~\ref{fig:D0Dp})\footnote{The posterior distributions of all three observables are non-Gaussian and hence, the error bars have been interpreted as the RMS of the distributions.}
\begin{eqnarray}
R(D^0,\pi^0)&=&0.1166\pm0.0061,\nonumber\\
R(D^+,\pi^+)&=&0.048\pm0.020,\nonumber\\
{\rm Re}(\epsilon^\prime_0)&=&0.045\pm0.029.
\end{eqnarray}
which is in fair agreement with the CLEO measurements. The various predictions for $R(D^0,\pi^0)$ and $R(D^+,\pi^+)$ that have been made previously are listed in table~\ref{tab:RR}.

\begin{table*}[t!]
\begin{center}
\begin{tabular}{|l|c||l|c||l|c|}
\hline
$R(D^0,\pi^0)$			 &							&$R(D^+,\pi^+)$			&						&$R(D_s^+,K^+)$			        &							        \\
\hline
$\sim10\%$ 			 &\cite{Bigi:1994aw}				&$-0.010\pm0.026$ 			&\cite{Gao:2006nb}			&$-0.003^{+0.019}_{-0.017}$ 		&\cite{Bhattacharya:2008ss}		    	\\
$\sim0.106$ 			&\cite{Gao:2006nb}				&$-0.006^{+0.033}_{-0.028}$ 	&\cite{Bhattacharya:2008ss}	&$-0.0022\pm0.0087$ 		    	&\cite{Bhattacharya:2009ps}		    	\\
$0.107$ 				&\cite{Bhattacharya:2009ps}		&$-0.005\pm0.013$ 			&\cite{Bhattacharya:2009ps}	&$-0.008\pm0.007$ 			    	&\cite{Cheng:2010ry,Gao:2014ena}		\\
$0.09^{+0.04}_{-0.02}$ 	&\cite{Muller:2015lua}			&$-0.019\pm0.016$ 			&\cite{Cheng:2010ry}		&\;\;\;$0.11^{+0.04}_{-0.14}$ 		&\cite{Muller:2015lua}			    	\\
$0.113\pm0.001$ 		&\cite{Wang:2017ksn}			&\;\;\;$0.025\pm0.008$ 		&\cite{Wang:2017ksn}		&\;\;\;$0.012\pm0.006$ 		    	&\cite{Wang:2017ksn}			    	\\
\hline
\end{tabular}
\end{center}
\caption{Estimates of $R(D^0,\pi^0)$, $R(D^+,\pi^+)$ and $R(D_s^+,K^+)$ from the existing literature.}
\label{tab:RR}
\end{table*}%

We present a prediction of $R(D_s^+,K^+)$ (c.f. figure~\ref{fig:D0Dp})\footnote{The predicted value is the same for both the solutions of the strong phases, negative and positive.}:
\begin{equation}
R(D_s^+,K^+) = -0.0103\pm 0.0074.
\end{equation}
which can be compared with other predictions made in the past as listed in table~\ref{tab:RR}.
All the results are reasonably compatible. However, these imply that ${\rm BR}(D_s^+\to K^+K_S)>{\rm BR}(D_s^+\to K^+K_L)$ in~\cite{Muller:2015lua,Wang:2017ksn}, whereas for the rest the contrary is true if one considers the central values of the ratio. Since this rate asymmetry depends on the estimate of the strong phases, a measurement of the latter can be used to test our predictions of the strong phases.

\subsection[Amplitude relations and \texorpdfstring{{$\mathrm{SU}(3)_F$}}{SU(3)\_F} breaking]{Amplitude relations and {\boldmath $\mathrm{SU}(3)_F$} breaking}
\label{sec:exam}
\begin{figure*}
\begin{center}
\subfloat{\includegraphics[width=.25\textwidth]{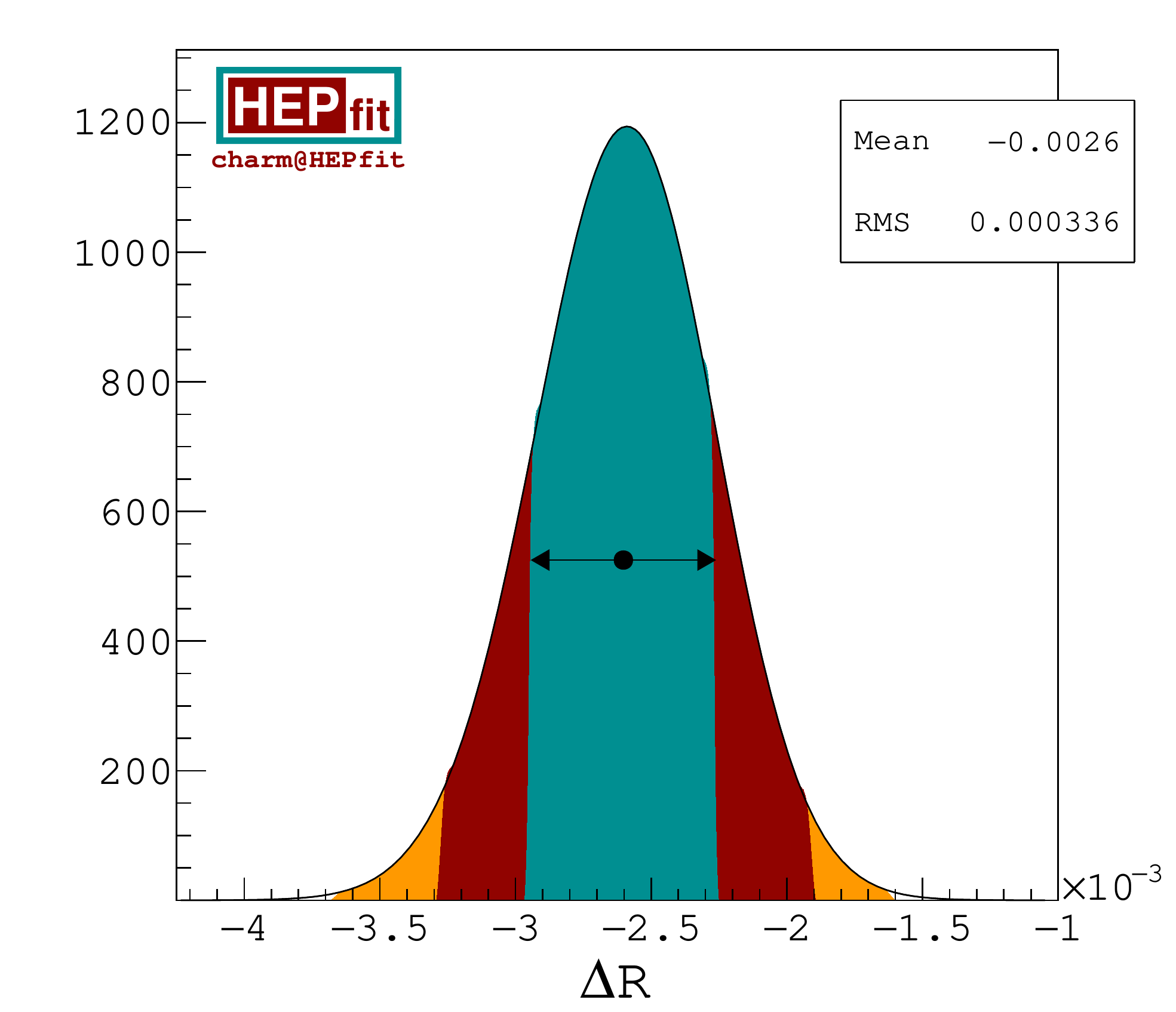}}
\subfloat{\includegraphics[width=.25\textwidth]{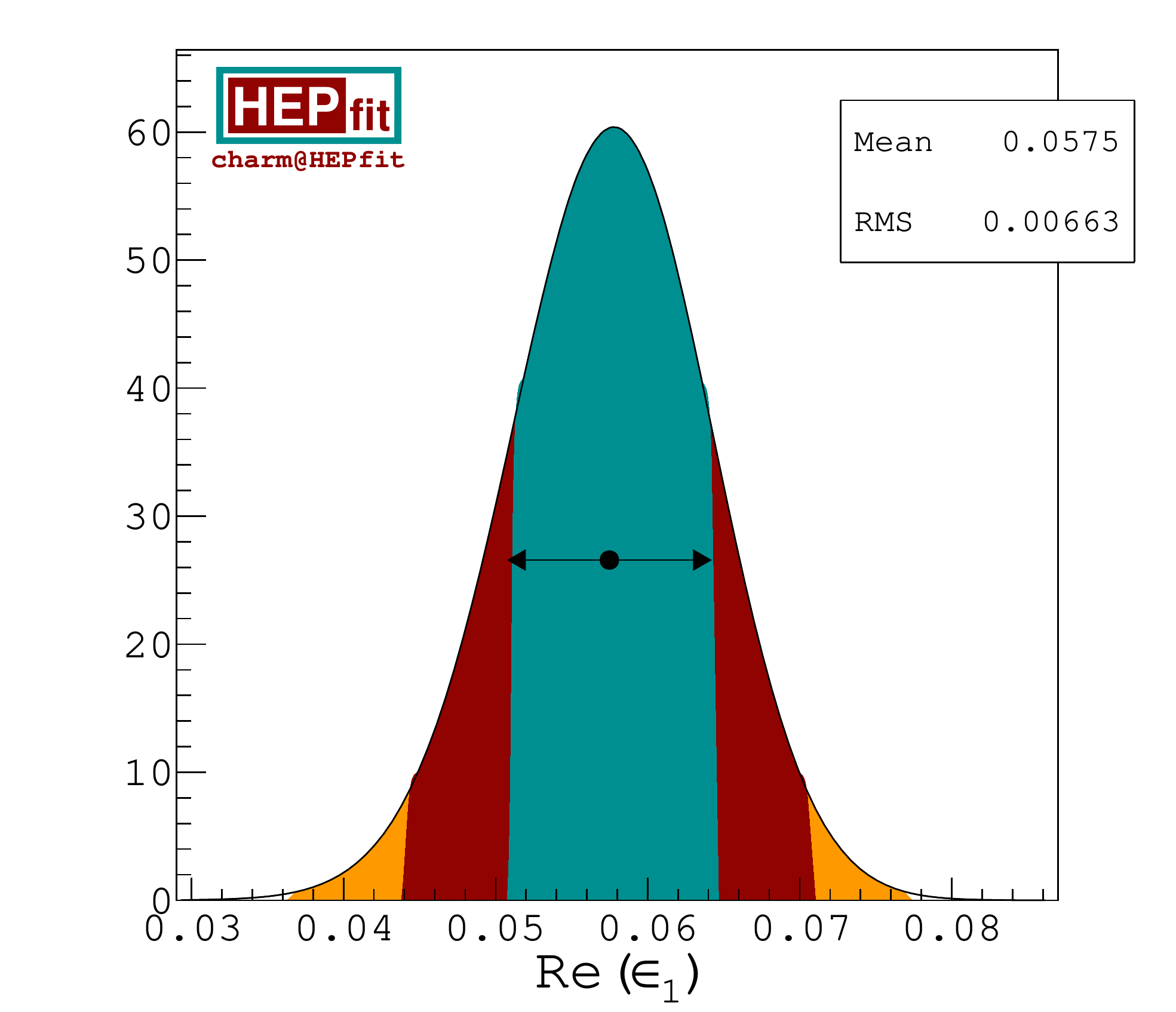}}
\subfloat{\includegraphics[width=.25\textwidth]{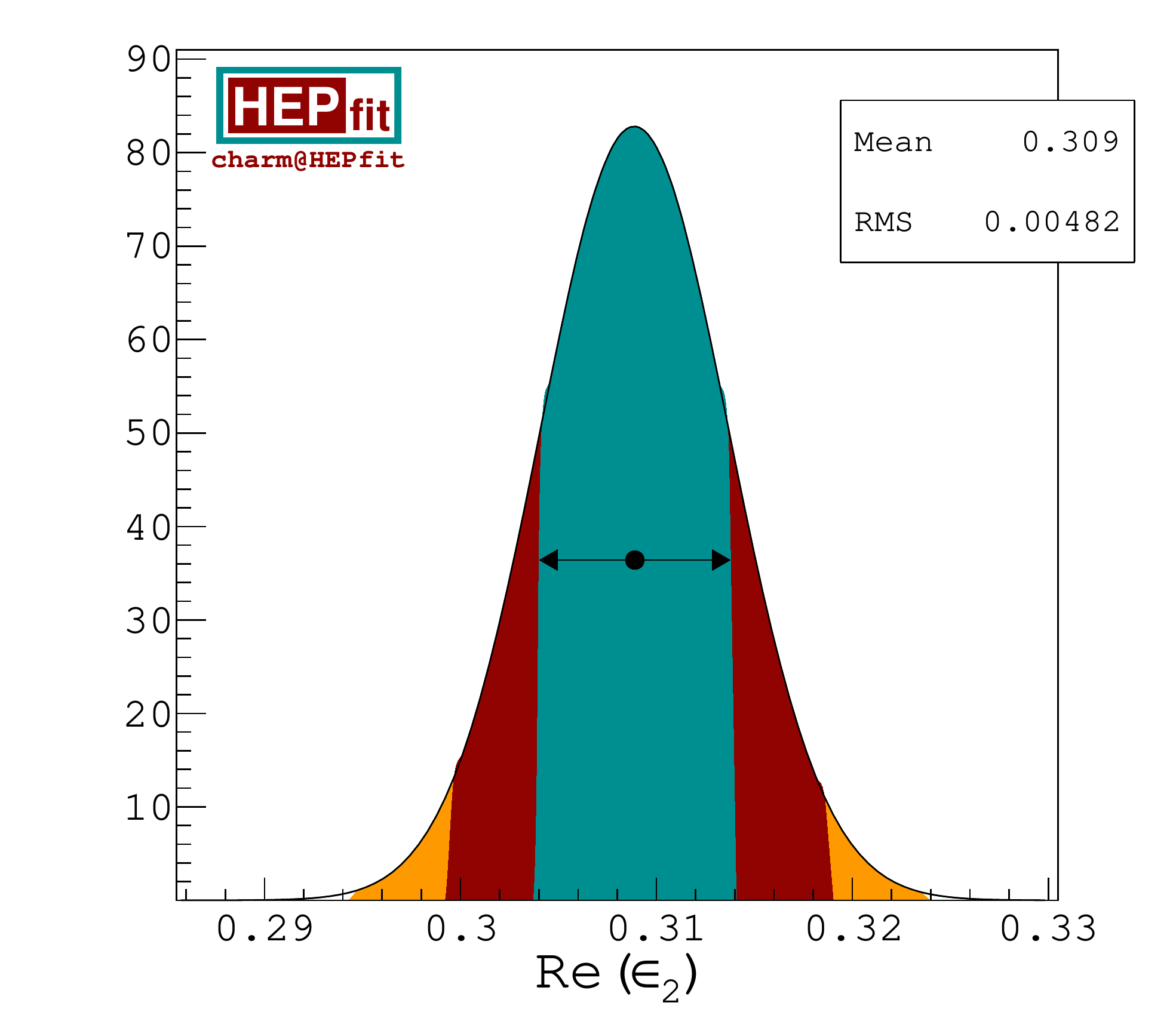}}
\caption{Fit results for $\Delta R$, $\epsilon_1$ and $\epsilon_2$ using the branching fraction data in table~\ref{tab:BR-fit} and the HFLAV world average of $\Delta{\rm A}_{\rm CP}$. The green, red and orange regions are the 68\%, 95\% and 99\% probability regions respectively.}
\label{fig:Gronau}
\end{center}
\end{figure*}
While our parametrization is well motivated by  $\mathrm{SU}(3)_{F}$
arguments, it is also good to check if there are some ways of
validating it. Here we follow a more general
theoretical construction of $\mathrm{SU}(3)_{F}$ arguments put forward by Gronau
in \cite{Gronau:2015rda} which also allows for a measure of the degree at which
$\mathrm{SU}(3)_{F}$ is broken by applying a higher order perturbation expansion
in $\mathrm{SU}(3)_{F}$ breaking. The amplitude relations for $D^0$ decays to
pairs of neutral pseudoscalar mesons can be written as:

\begin{widetext}
 \begin{eqnarray}
 R_1  & \equiv & \frac{|A(D^0 \to K^+\pi^-)|}{|A(D^0 \to \pi^+K^-)|\tan^2\theta_C}~,\;\;
 R_2  \equiv  \frac{|A(D^0 \to K^+K^-)|}{|A(D^0\to \pi^+\pi^-)|}~,
 \nonumber\\
 R_3 & \equiv & \frac{|A(D^0 \to K^+K^-)| + |A(D^0\to \pi^+\pi^-)|}
{|A(D^0 \to \pi^+K^-)|\tan\theta_C + |A(D^0\to K^+\pi^-)|\tan^{-1}\theta_C}~,\;\;
R_4  \equiv  \sqrt{\frac{|A(D^0 \to K^+K^-)||A(D^0\to \pi^+\pi^-)|}
{|A(D^0\to \pi^+K^-)||A(D^0\to K^+\pi^-)|}}~.
\end{eqnarray}
\end{widetext}
These four ratios are not mutually independent.
They obey a trivial identity
\begin{equation}
R_4 = R_3 \sqrt{\frac{1 - [(R_2-1)/(R_2+1)]^2}{1 - [(R_1-1)/(R_1+1)^2}}~.
\end{equation}

It can be shown that $R_i=1$ in the limit of $\mathrm{SU}(3)_F$ and the relation
\begin{eqnarray}
\Delta R &\equiv& R_3 - R_4 \nonumber\\
&+& \frac{1}{8}\left[\left(\sqrt{2R_1-1} - 1\right)^2 -
\left(\sqrt{2R_2 - 1} -1\right )^2\right] \nonumber\\
&=&{\cal O}(\epsilon_1^4, \epsilon_2^4) + {\cal O}(\hat{\delta}_1\epsilon_1^2, \hat{\delta}_2\epsilon_2^2)~.
\end{eqnarray}
differs from zero by terms of the order ${\cal O}(\epsilon_1^4, \epsilon_2^4) + {\cal O}(\hat{\delta}_1\epsilon_1^2, \hat{\delta}_2\epsilon_2^2)$, where $\epsilon_i$ and $\hat{\delta}_i$ are U-spin and Isospin breaking terms, respectively. One can then write the real parts of the $\mathrm{SU}(3)_{F}$ breaking parameters $\epsilon_1$ and $\epsilon_2$ as
\begin{eqnarray}
{\rm Re}(\epsilon_i)&=&\frac{1}{2}\left(\sqrt{2R_i-1}-1\right) -{\rm Re}(\hat{\delta}_i) - 2{\rm Re}(\hat{\delta}_i){\rm Re}(\epsilon_i) \nonumber\\
&+& \mathcal{O}(\hat{\delta}_i\epsilon_i) + \mathcal{O}(\epsilon_i^3)
\end{eqnarray}
 with $i=1,2$. The U-spin breaking in $D^0\to K^+\pi^-$ is denoted by
 $\epsilon_1$ and that in $D^0\to K^+K^-$ is denoted by $\epsilon_2$. It is
 expected~\cite{Gronau:2015rda} that $\epsilon_2$ quantifies breaking in both
 the tree and penguin amplitudes while $\epsilon_1$ quantifies the breaking in only
 tree amplitudes. Hence, the former is expected to be somewhat larger than the
 latter. In our work we do not consider isospin breaking and hence $\hat{\delta}_i=0$.
 We test these relations in our paramterization of the amplitudes and use the
 parameters extracted from the branching fractions as inputs. We find a fair
 agreement with the results quoted in~\cite{Gronau:2015rda} for $\Delta R$,
 ${\rm Re}(\epsilon_1)$ and ${\rm Re}(\epsilon_2)$ as is evident from
 figure~\ref{fig:Gronau}.
\subsection{Correlations between CP asymmetries}
\label{sec:CPCorr}
\begin{figure*}
\begin{center}
\subfloat{\includegraphics[width=.25\textwidth]{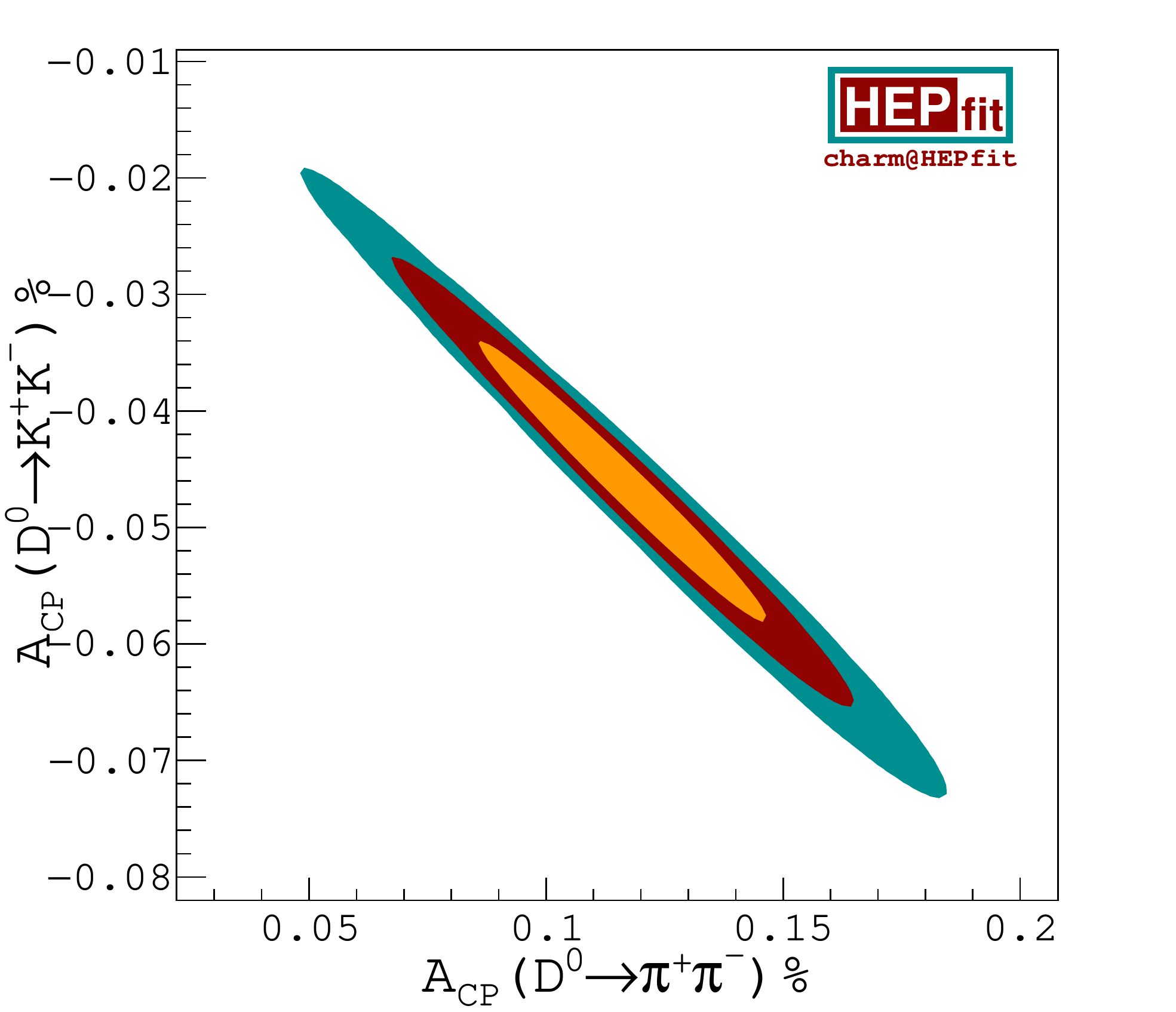}}
\subfloat{\includegraphics[width=.25\textwidth]{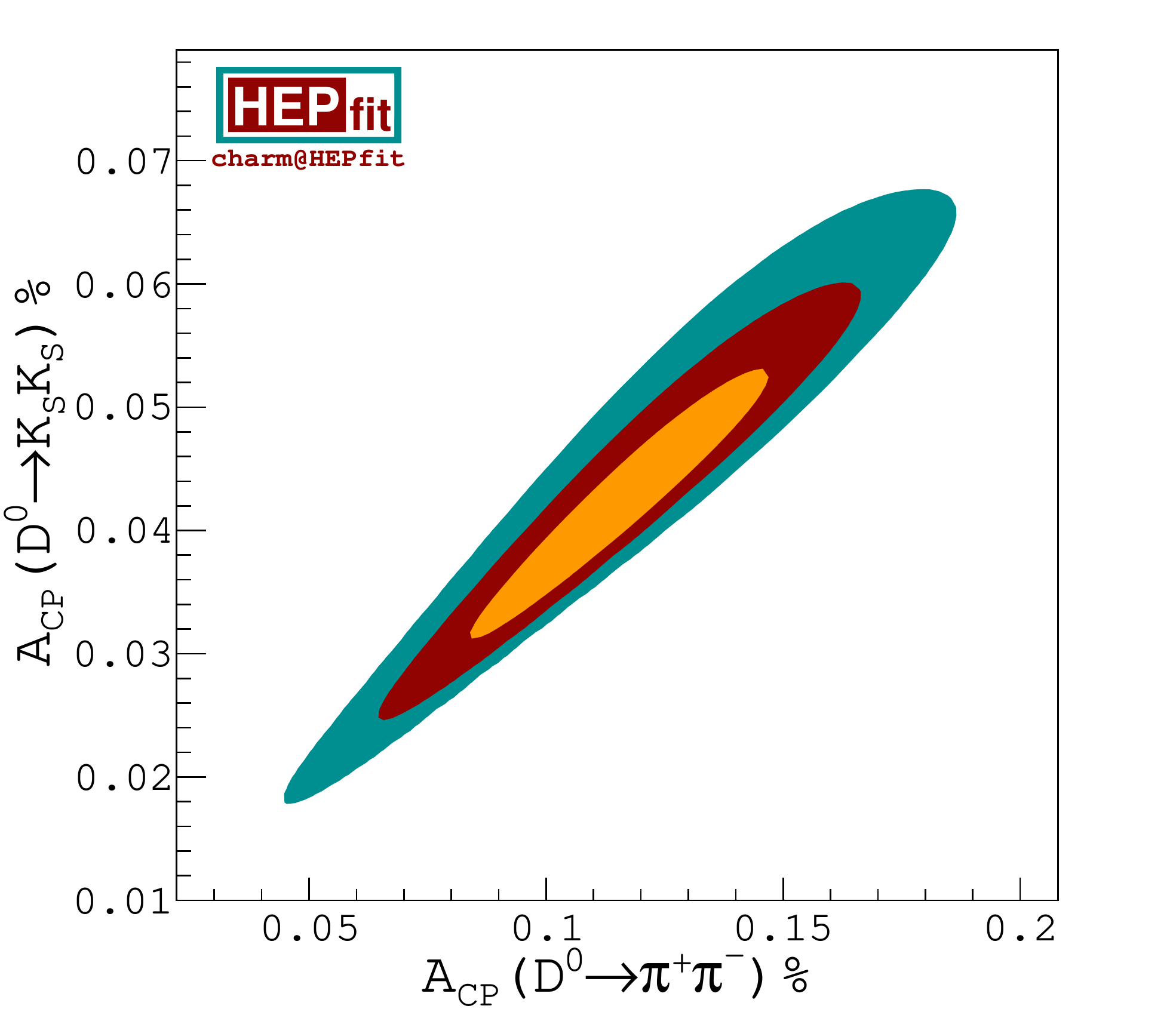}}
\subfloat{\includegraphics[width=.25\textwidth]{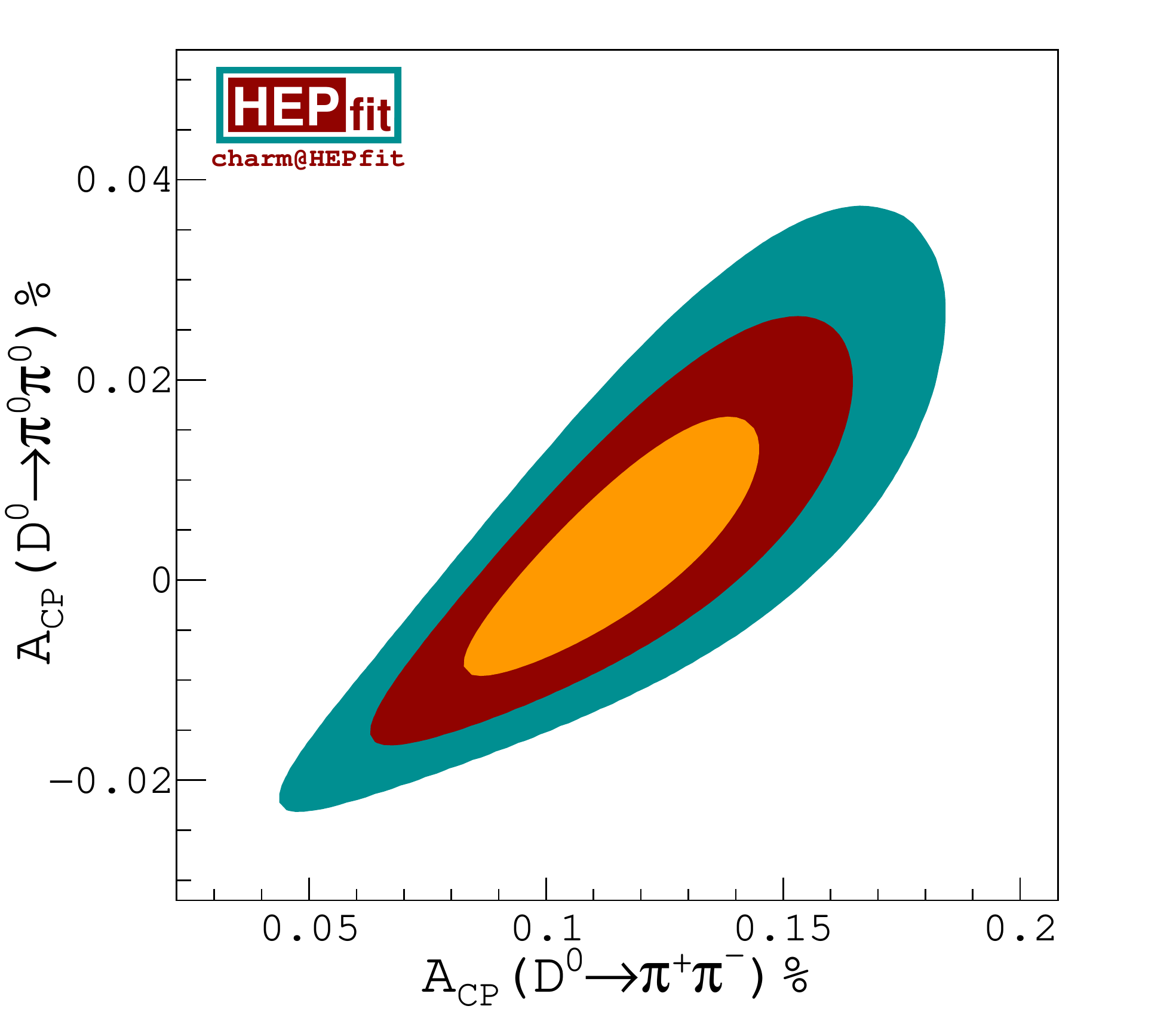}}\\
\subfloat{\includegraphics[width=.25\textwidth]{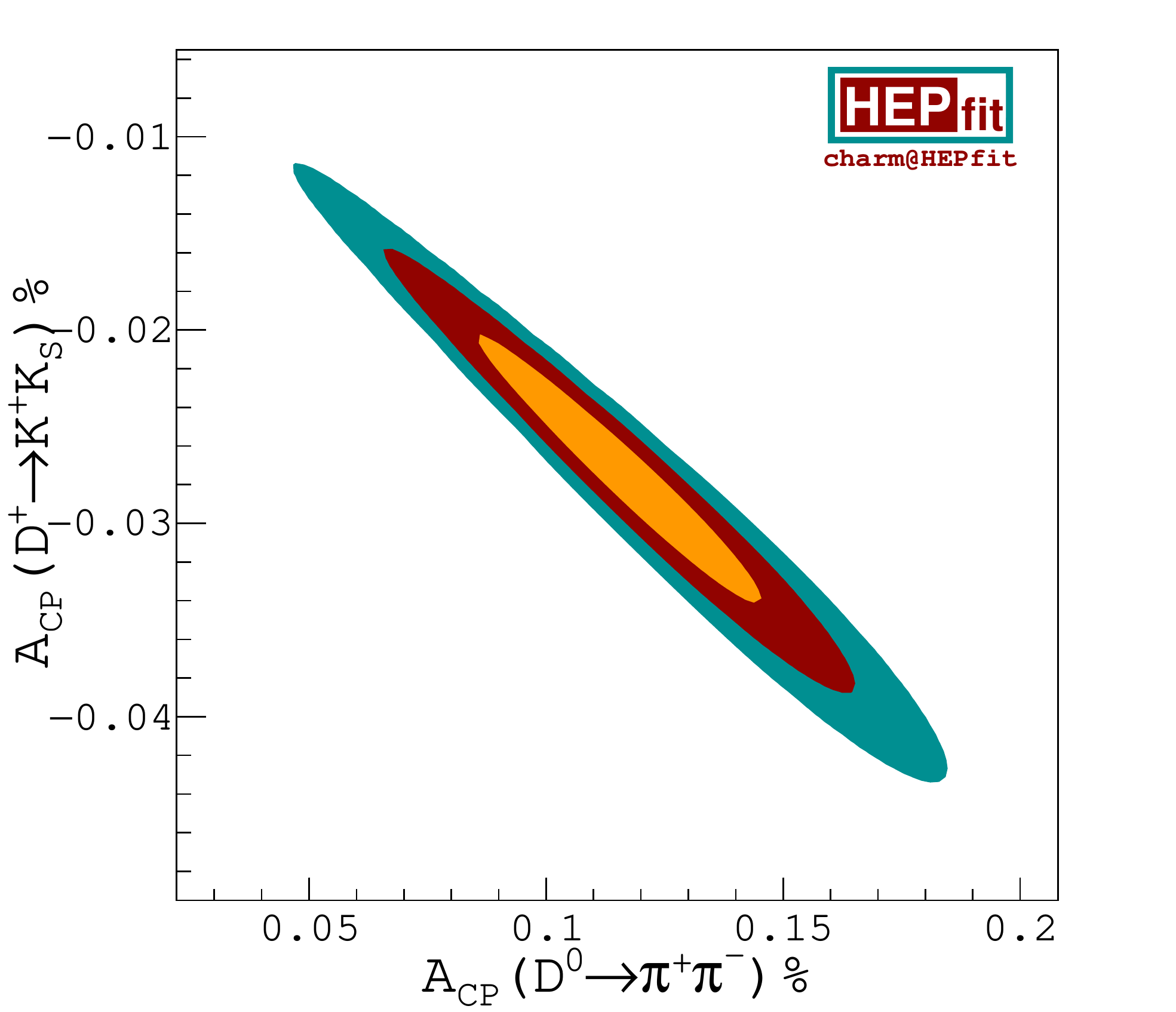}}
\subfloat{\includegraphics[width=.25\textwidth]{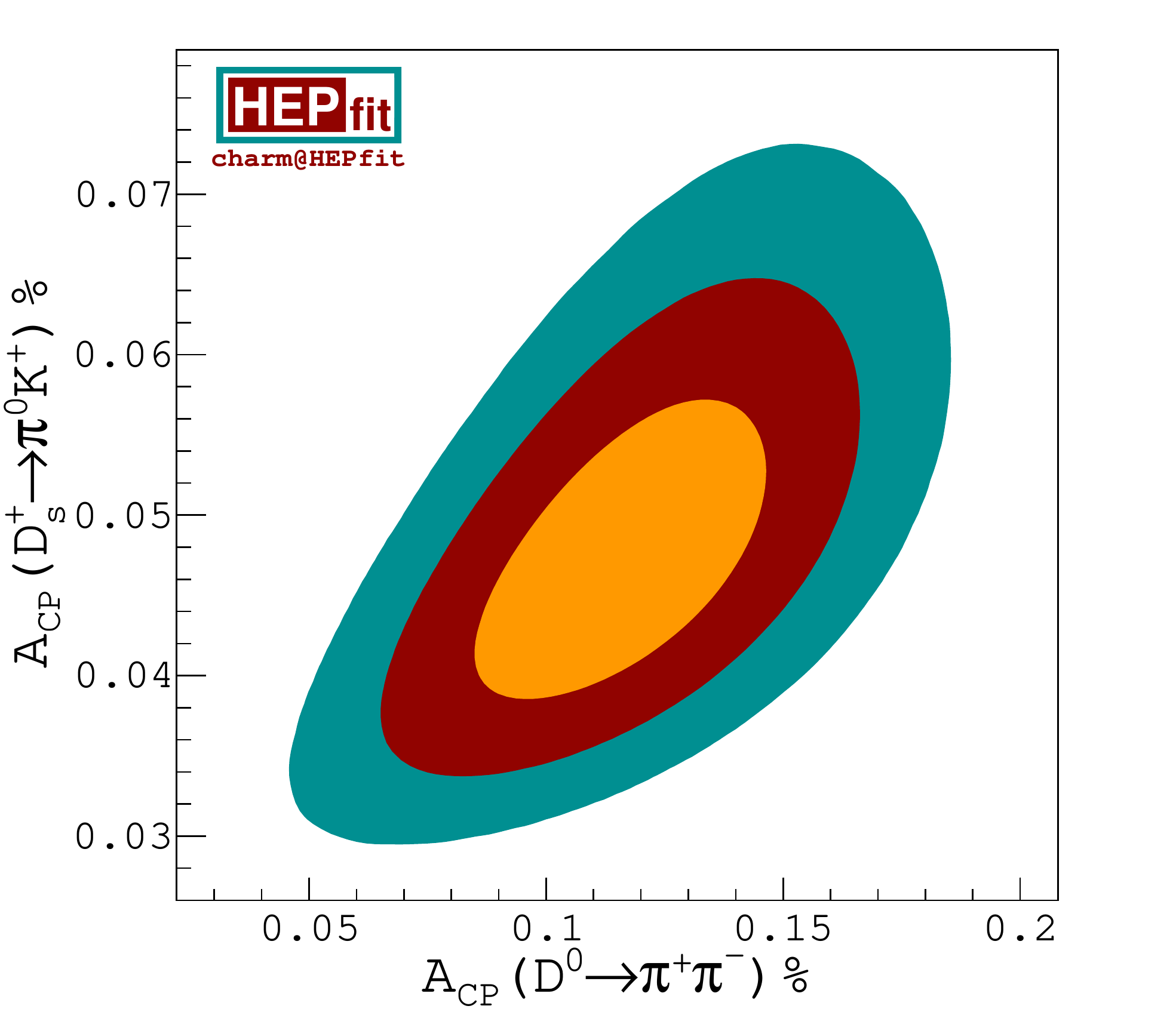}}
\subfloat{\includegraphics[width=.25\textwidth]{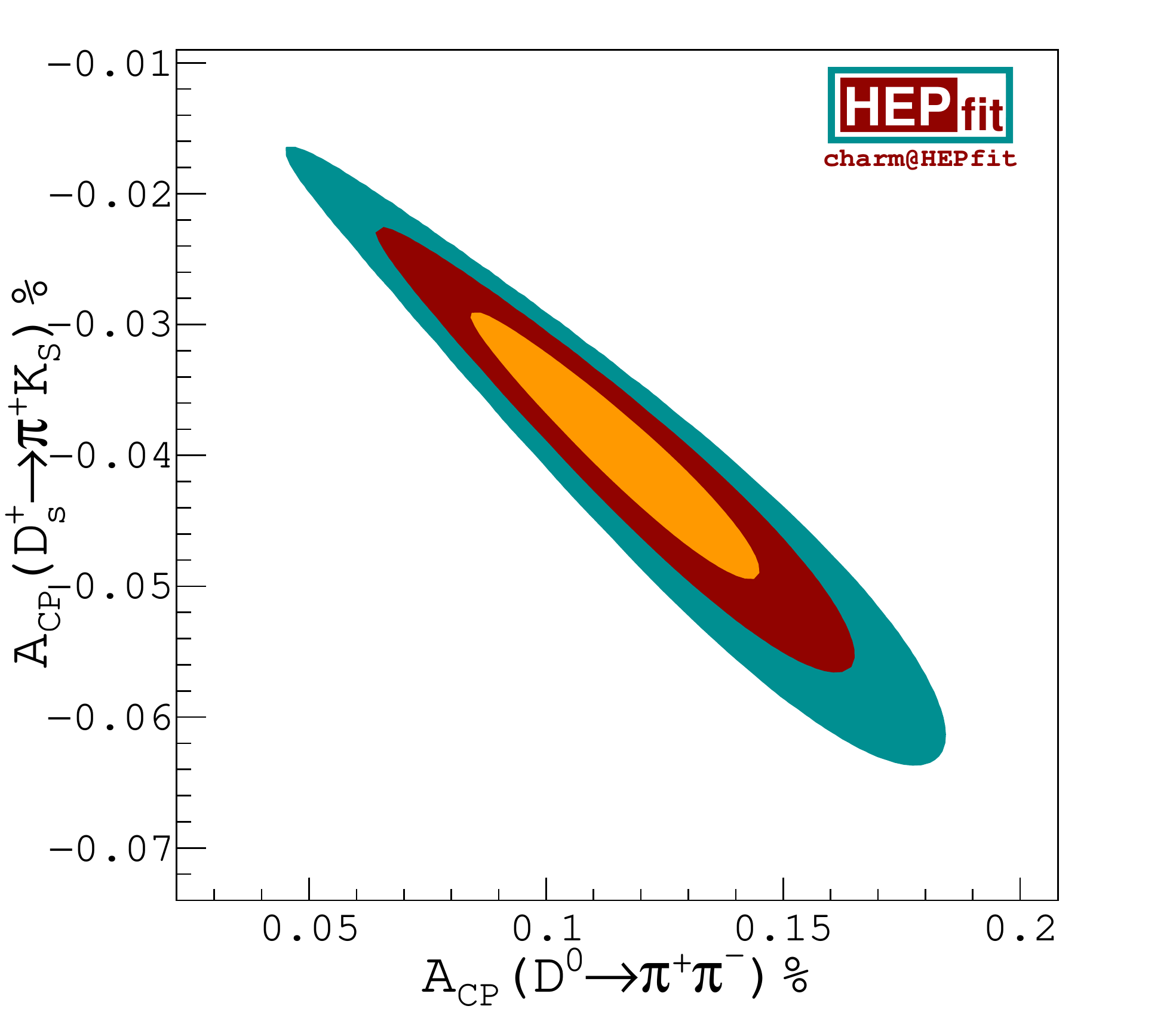}}
\caption{Correlations between asymmetries (in \%) as given in equation~(\ref{eq:asym_corr}) using the branching fraction data in table~\ref{tab:BR-fit} and the HFLAV world average of $\Delta{\rm A}_{\rm CP}$ quoted in section~\ref{sec:CPamp}. The orange, red and green regions are the 68\%, 95\% and 99\% probability regions respectively.}
\label{fig:asym_corr}
\end{center}
\end{figure*}
As a second test of our parameterization we propose the correlation between the CP asymmetries that we have earlier explained are parametrically correlated. Since the asymmetries are correlated through the combination of parameters, $\left(P + \Delta_3\right)/T$, it is possible to combine the expression for the asymmetries to obtain relations between them. By considering the $\pi\pi$ and $KK$ final states we have, symbolically,
\begin{widetext}
\begin{eqnarray}
{\rm A}_{\rm CP}(D\to KK) = f_{KK}\left(\vec{p}\right)A_{CP}\left(D^0\to \pi^+\pi^-\right)
+ g_{KK}\left(\vec{p}\right)A_{CP}\left(D^0\to \pi^0\pi^0\right) + h_{KK}\left(\vec{p}\right),
\end{eqnarray}
\end{widetext}
where $f_{KK}\left(\vec{p}\right)$, $g_{KK}\left(\vec{p}\right)$ and $h_{KK}\left(\vec{p}\right)$ are functions of $\vec p=\left\{T,C,\kappa,\kappa^\prime,K,K^\prime,\Delta,\phi,\epsilon_\delta, \delta_0,\delta_0^\prime,\delta_\frac{1}{2},\delta_1\right\}$ and depend on the final $KK$ pair. With the central values for the parameters from our fits in table~\ref{tab:fit} we get for negative phases
\begin{widetext}
\begin{eqnarray}
{\rm A}_{\rm CP}\left(D^0\to K^+K^-\right) &=& -0.657 A_{CP}\left(D^0\to\pi^+\pi^-\right) + 0.750 A_{CP}\left(D^0\to\pi^0\pi^0\right) + 2.78 \times 10^{-4},\nonumber\\
{\rm A}_{\rm CP}\left(D^0\to K_SK_S\right) &=& 3.47 A_{CP}\left(D^0\to \pi^+\pi^-\right) - 8.88 A_{CP}\left(D^0\to \pi^0\pi^0\right) - 3.28 \times 10^{-3}.\nonumber\\
\end{eqnarray}
\end{widetext}
when we consider the fit with positive phases the constant terms change their signs. In the case in which
we consider the limit $\Delta_4\to0$ we have,
\begin{eqnarray}
{\rm A}_{\rm CP}\left(D^0\to K^+K^-\right) &=& -0.394 {\rm A}_{\rm CP}\left(D^0\to \pi^+\pi^-\right)\nonumber\\
&&-1.05 \times 10^{-6},\nonumber\\
{\rm A}_{\rm CP}\left(D^0\to K_S K_S\right) &=& 0.342{\rm A}_{\rm CP}\left(D^0\to \pi^+\pi^-\right) \nonumber\\
&&+ 2.75  \times 10^{-5},\nonumber\\
{\rm A}_{\rm CP}\left(D^0\to \pi^0\pi^0\right) &=& 0.352{\rm A}_{\rm CP}\left(D^0\to \pi^+\pi^-\right) \nonumber\\
&&- 3.72 \times 10^{-4}.
\label{eq:asym_corr}
\end{eqnarray}
Likewise, the CP asymmetries in the other channels can also be correlated. These correlations between the predicted asymmetries are plotted in figure~\ref{fig:asym_corr} and includes several SCS decay modes in which CP violation is possible. A deviation from these correlations would indicate a breakdown of our parameterization. The correlations have been derived by using only the branching fraction data and the measurement of $\Delta{\rm A}_{\rm CP}$ from LHCb. Most notably, the formalism we use renders the CP asymmetry in $D^0\to K_SK_S$ completely correlated to $\Delta{\rm A}_{\rm CP}$ since the weak exchange diagram present in the $\Delta U=0$ part of the amplitude of the former decay mode, and absent in the latter, is generated by rescattering and is not an independent contribution.

\subsection{Constraints on penguin amplitudes from future measurements}
\label{sec:future}
\begin{figure*}
\begin{center}
\subfloat[BII-50]{\includegraphics[width=.3\textwidth]{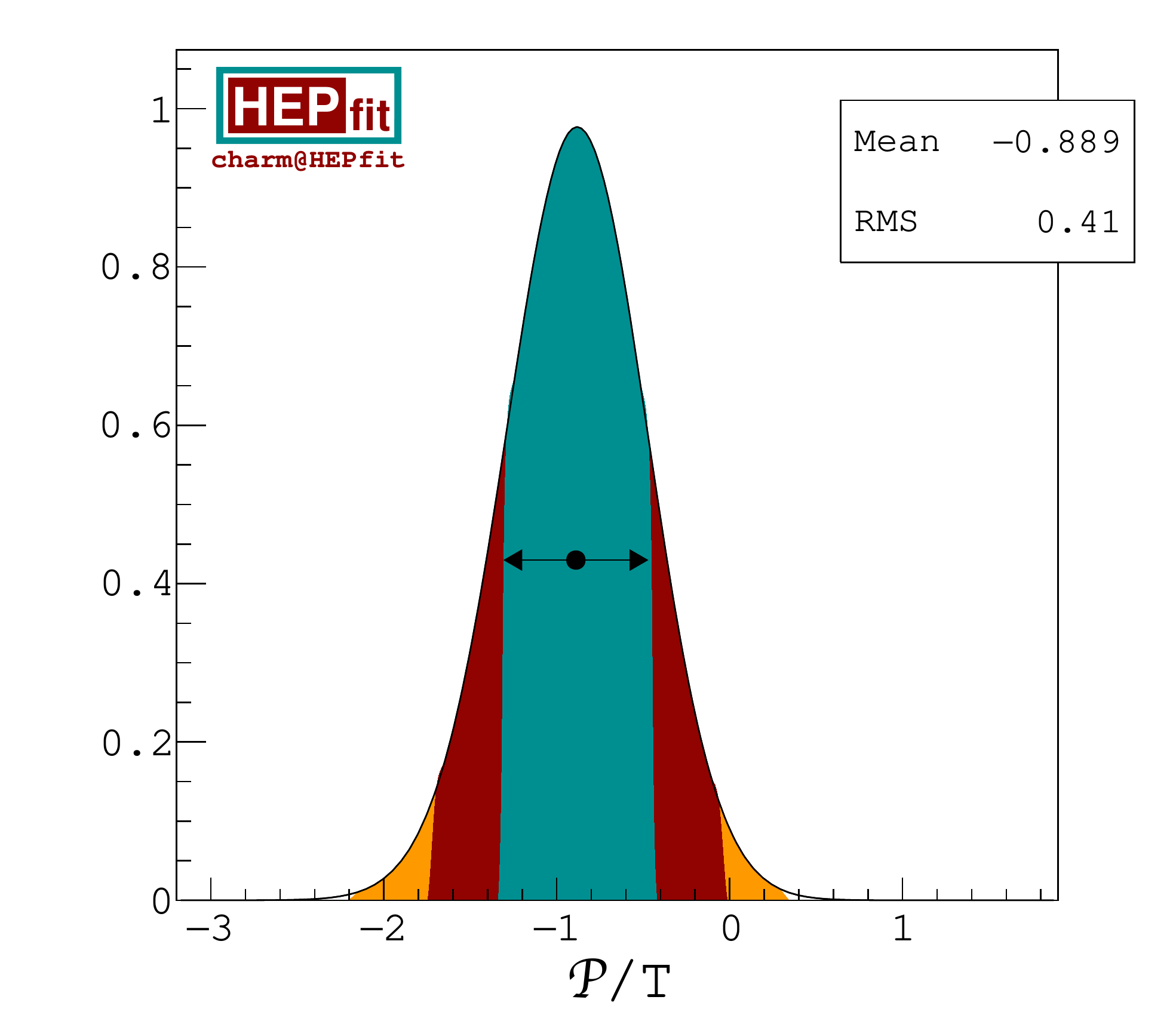}}
\subfloat[BII-50 + $\Delta{\rm A}_{\rm CP}$]{\includegraphics[width=.3\textwidth]{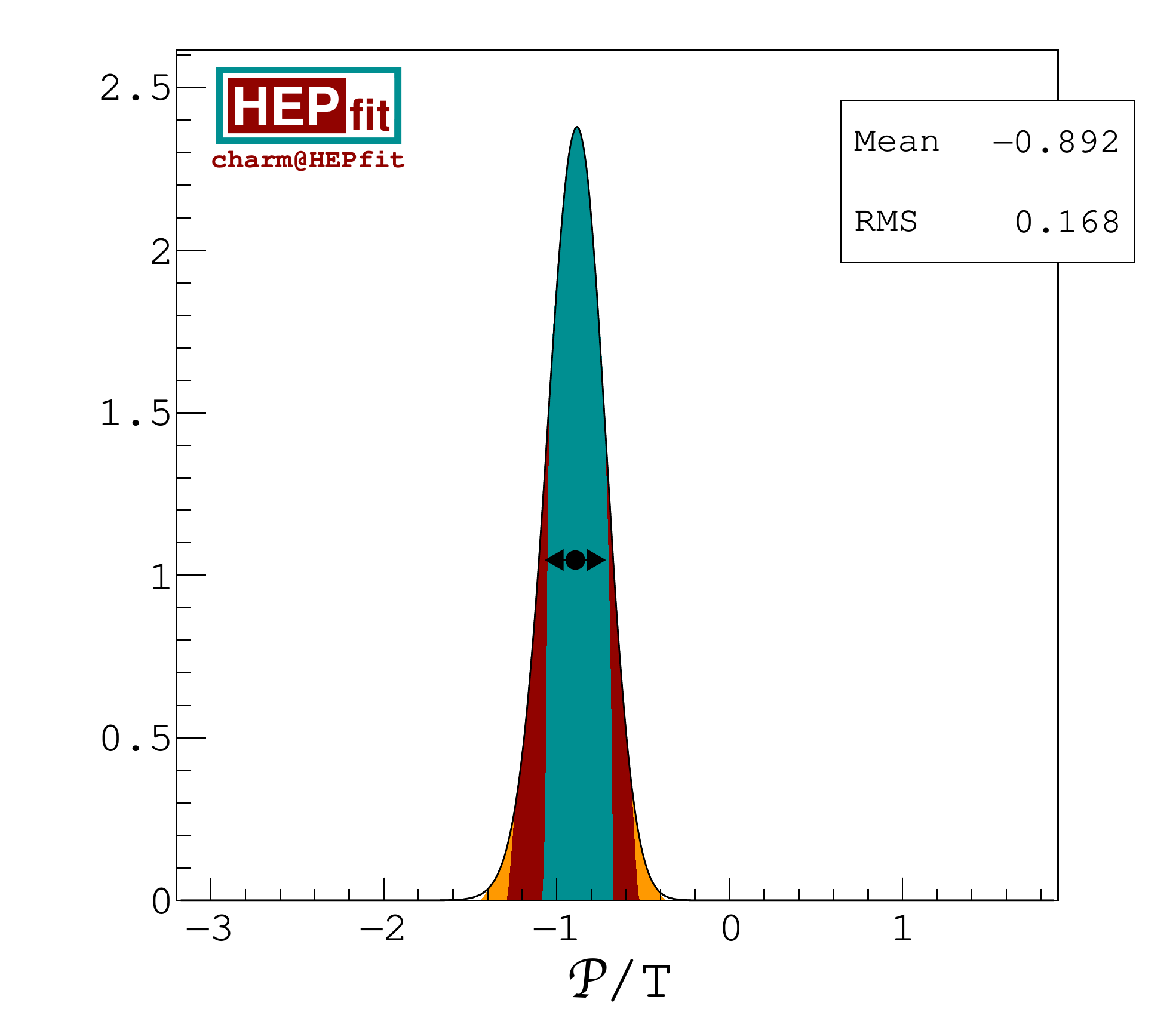}}
\subfloat[BII-50 + LHCb-50]{\includegraphics[width=.3\textwidth]{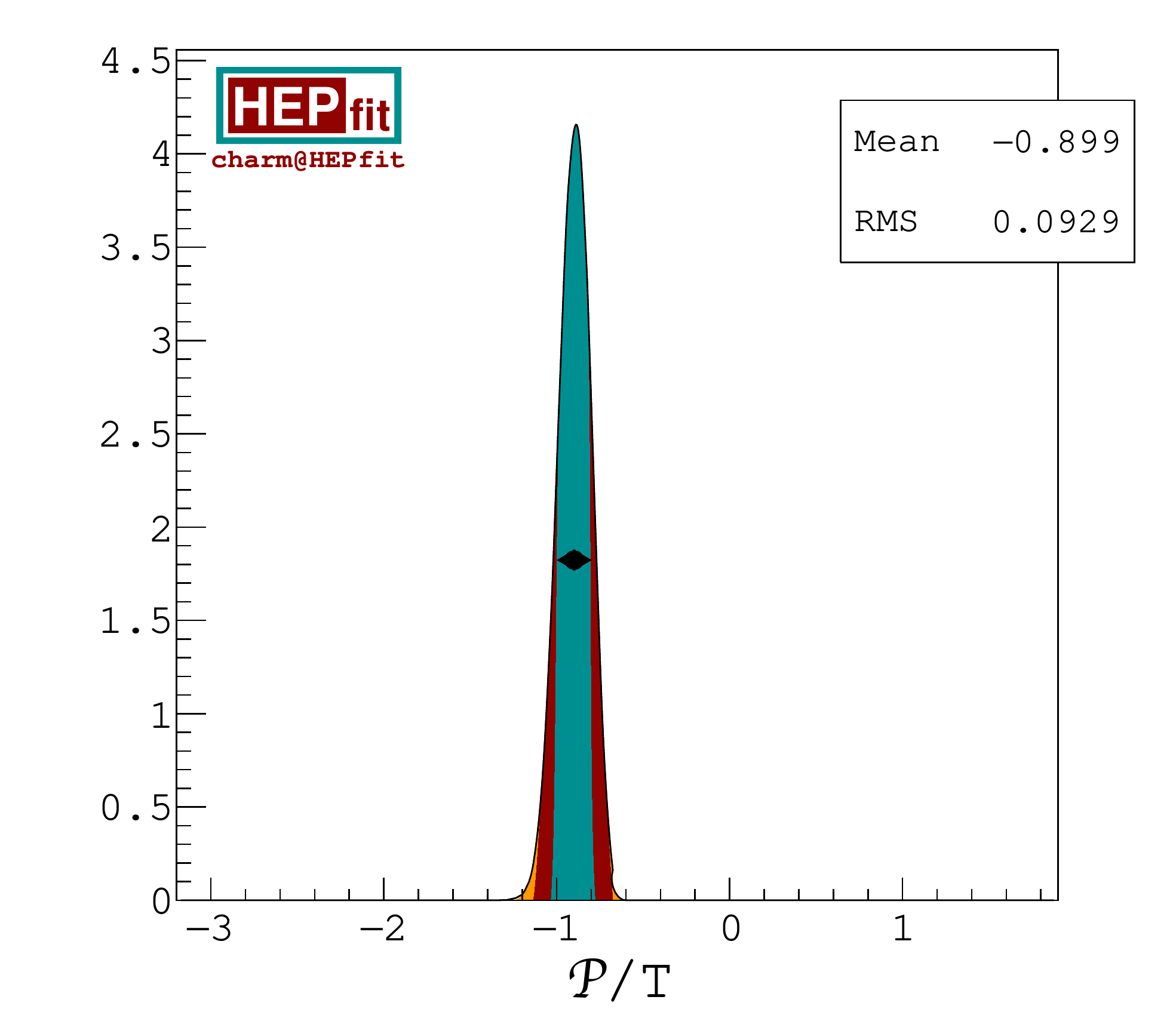}}
\caption{\it Fit results for $\mathpzcB{P}/T$ using the branching fraction data in table~\ref{tab:BR-fit} and CP asymmetries listed in table~\ref{tab:ACPF}. The green, red and orange regions are the 68\%, 95\% and 99\% probability regions respectively.(BII-50: Belle II 50 ab$^{-1}$, LHCb-50: LHCb 50 fb$^{-1}$)}
\label{fig:BELLE}
\end{center}
\end{figure*}

\begin{table*}
\begin{center}
\begin{tabular}{|l|c|c|c|c|}
\hline
\multirow{3}{*}{${\rm A}_{\rm CP}$(channel)}				& \multirow{3}{*}{mode (\%)} & \multicolumn{3}{c|}{RMS (\%)}\\
\cline{3-5}
&											& \multirow{2}{*}{Current Fit}	& Belle II 							&LHCb \\
&											&			& 50 ab$^{-1}$~\cite{Schwartz:2017gni}	&50 fb$^{-1}$~\cite{LHCb-PUB-2014-040}					\\
\hline
$D^0\to\pi^+\pi^-$		    	&\;0.1174 		&0.020		& 0.05		&--					\\
$D^0\to\pi^0\pi^0$		    	&-0.0034 		&0.009		& 0.09		&--					\\
$D^0\to K^+K^-$		        &-0.0465 		&0.008	  	& 0.03		&-- 					\\
$D^0\to K_SK_S$		        &\;0.0431 		&0.007		& 0.17		&--					\\
$D^+\to K^+K_S$		        &-0.0276 		&0.005		& 0.05		&--					\\
$D^+_s\to \pi^+K_S$		    	&-0.0403 		&0.007		& 0.29		&--					\\
$\Delta{\rm A}_{\rm CP}$    	&-0.164\;\; 	& -- 			& -- 	    				&0.01	\\
\hline
\end{tabular}
\caption{\it Numbers used to generate the constraints on $\mathpzcB{P}/T$ from future experiments. The column marked ``Current Fit'' shows the RMS from our prediction of the asymmetries using the branching fraction data and the LHCb measurement of $\Delta{\rm A}_{\rm CP}$ only and for the negative solution for the phases.}
\label{tab:ACPF}
\end{center}
\end{table*}

With Belle II starting up and LHCb having built a very strong charm program over the past few years, it is instructive to see what these measurements will mean in terms of constraining the penguin amplitudes. While the measurements of the branching fractions are expected to improve significantly too, this will not additionally constrain the penguin amplitudes directly. However, the ratio $(P + \Delta_3)/T$ would certainly benefit from an improved determination of $T$. Considering $T$ is already extracted at a precision of less than $O(1\%)$, improvements in this parameter will leave a negligible effect. On the other hand, not all the phases appearing in the SCS decays are very well constrained. An improvement in these would certainly improve the constraints on the penguin amplitudes. In particular, an improved measurement of $D^0\to K_S K_S$, which is non-vanishing only when $\mathrm{SU}(3)_F$ is broken, is quite important for further constraining the parameters that arise from this breaking specially because the branching ratio of this channel is not well measured currently.

To keep the analysis simple and on the more conservative side we do not take into account any improvement in the measurement of the branching fractions. We project the central values of the CP asymmetries using their value at the global mode of the current fit and use the errors projected by the experiments. We use projections for Belle II at 50 ab$^{-1}$ for various asymmetries. We also use the projected measurement of $\Delta{\rm A}_{\rm CP}$ at LHCb 50 fb$^{-1}$ data. Finally, we combine all these projected measurements. These are tabulated in table~\ref{tab:ACPF}.

In figure~\ref{fig:BELLE} we show how the constraints on $\mathpzcB{P}/T$ will change with additional data. As is evident, the constraints are not much better than what we see in figure~\ref{fig:ACP_corr} with only the full Belle II data. The reason for this is that the projected precision of measurement of these asymmetries from the full Belle II data of 50 fb$^{-1}$ is comparable or worse than the precision of the prediction of the asymmetries from the current measurement of $\Delta{\rm A}_{\rm CP}$ as can be seen from table~\ref{tab:ACPF}. Once the measurement of $\Delta{\rm A}_{\rm CP}$ improves, the constraint on $\mathpzcB{P}/T$ gets much better, but significantly so only after 50 fb$^{-1}$ of data from LHCb.

\section{Summary}
\label{sec:sum}
The main purpose of this work is to take advantage of the high precision reached by the measurements of the branching ratios in two particle final states consisting of kaons and (or) pions of the pseudoscalar charmed particles to deduce the predictions of the Standard Model for the CP violating asymmetries in their decays. To this extent we have constructed amplitudes in agreement with the measured branching ratios, where the $\mathrm{SU}(3)_F$ violations come mainly from the final state interaction and from the non-conservation of the strangeness changing vector currents.

So in this work we extend the formalism presented in~\cite{Buccella:2013tya} with a larger menu of branching fractions for $D\to P P$ with $P=K,\pi$ but excluding the branching fractions which have $\eta/\eta^\prime$ in the final state. We extend the old parameterization with the parameters $K^{(\prime)}$ and $\kappa^{(\prime)}$ to address $\mathrm{SU}(3)_F$ breaking effects both in the tree and colour suppressed amplitudes. We introduce $\epsilon_\delta$ to address the splitting of the phases due to mass splitting between the $D^{0,+}$ and the $D_s^+$. Another parameter $\Delta$ is included to address the decays of $D^+_{(s)}$ mesons. To accommodate for CP asymmetry in the SCS decays we introduce three parameters $P$, $\Delta_3$ and $\Delta_4$. The latter is parametrically suppressed due to an approximate selection rule. The former two cannot be resolved from CP asymmetries of the SCS decays we consider and hence only the sum can be extracted from data and we deem its ratio with $T$ as the parameter relevant for the fit.

We perform a fit of the parameters to the branching fractions and $\Delta{\rm A}_{\rm CP}$ using \HEPfit and predict several CP asymmetries using our parametrization. In our framework, ignoring very small effects, the CP asymmetries show distinct correlations which can serve as a test of our framework. We also explore $\mathrm{SU}(3)_F$ breaking effects as advocated by Gronau~\cite{Gronau:2015rda} and find a good agreement with the results from that work. The rate asymmetries extracted from the branching fraction data agrees well with the CLEO collaboration data. As a future extension of this work, we will extend the parameterization to final states with $\eta/\eta^\prime$.

Within the ambit of our work we find reasonable success in trying to parameterize $D\to PP$ decays within a $\mathrm{SU}(3)_F$ framework. The important conclusions of our work are:

\begin{itemize}
\item We succeed in describing the measured branching fractions by invoking $\mathrm{SU}(3)_F$ breaking using large phases from FSI, non-conservation of the
strangeness changing vector current and slight shifts in the reduced matrix elements for CA and the DCS vs. the SCS decay amplitudes. This does not require the introduction of the parameter $P$, $\Delta_3$ or $\Delta_4$.

\item The values of the FSI phases, when considering the negative solutions, fall nicely along the pattern of the expected mass ordering of the resonance from the presence of which these FSI phases are generated. This also fixes the imaginary parts, which are relevant for the CP violating asymmetries. The negative solution is motivated by considering the masses of the resonances to be arranged according to the Gell-Man-Ne'eman-Okubo mass formula which requires the strong phase in the $I=1$ channel to be larger than that in the $I=1/2$ channel.

\item Once we relate the 15 in the $\Delta U = 0$ part of the amplitude to that in the $\Delta U = 1$ part of the amplitude the asymmetries depend on three new parameters. Of these, the combination  $P + T + \Delta_3$ incorporates the uncertain strength of the penguin contributions. Then we apply an approximate selection rule that forbids the simultaneous creation of a $d \bar{d}$ and a $s \bar{s}$ pair, similar to the OZI rule, to the penguin annihilation contribution coming from the 3.  Hence, the third parameter, $\Delta_4$, is expected to be small by this approximate selection rule and contributes mainly to the asymmetry in $D^0\to K_SK_S$ as can be seen from the $\mathrm{SU}(3)_F$ limit. Moreover, the terms proportional to $T + C$ are constrained by the branching fraction data. The combination $P+\Delta_3$ cannot be disentangled from measurements. Hence all the CP asymmetries depend on this combination of parameter and are thus correlated.

\item We show that amongst the current measurement of CP asymmetries, $\Delta{\rm A}_{\rm CP}^{\rm dir}$ is by far the strongest constraint on the combination $(P+\Delta_3)/T$. We use this fact and the parametric correlation between the $\Delta U =0$ part of the amplitudes to predict several asymmetries which are listed in table~\ref{tab:ACP_HFLAV}. Since $\Delta{\rm A}_{\rm CP}^{\rm dir}$ constrains the penguin amplitude to $\mathpzcB{P}\sim \mathcal{O}(T)$, the part proportional to it no longer dominates the CP asymmetries. Indeed the part proportional to $(T + C)$ becomes sizable in comparison. Hence the penguin amplitudes can no longer be expected to bring about an order of magnitude enhancement beyond the 1\% level in the CP asymmetries of several channel and can enhance them by only a factor of few. This lies in contrast with what was previously expected as the effect of penguin amplitudes in the CP asymmetries of SCS $D\to PP$ decays.

\item With the correlations between the asymmetries and the direction pointed at by the data we can propose methods for validating our $\mathrm{SU}(3)_F$ framework by looking at rate asymmetries between several $K_S-K_L$ final states and the correlation between CP asymmetries in different channels. In particular, as a consequence of the strong phases determined by the fit, we predict the yet unmeasured rate asymmetry: $$R(D_s^+,K^+) = -0.0103\pm 0.0074.$$

\item When we choose the negative solution for the phases, we also predict $$\delta_{K\pi}=\delta_{K^-\pi^+}-\delta_{K^+\pi^-} = 3.14^\circ \pm 5.69^\circ.$$
\end{itemize}

In this framework of $\mathrm{SU}(3)_F$ breaking that is driven by large phase from FSI due to rescattering through scalar resonances, it can be shown that CP asymmetries in all SCS modes are constrained to the per mille level by the current measurement of $\Delta{\rm A}_{\rm CP}^{\rm dir}$.

\acknowledgments
A. P. would like to acknowledge partial support from ERC Ideas Starting Grant n.~279972 ``NPFlavour''  while most of this work was being done. We would like to thank Alessandra Pugliese and Maurizio Lusignoli for fruitful discussions during the initial stages of the work. We would like to thank Luca Silvestrini and Enrico Franco for their help and support for several aspects of this work.

\appendix

\section{Posterior distributions of the parameters for the full fit}
\label{app:post}
\begin{figure*}[t]
\begin{center}
\subfloat{\includegraphics[width=.23\textwidth]{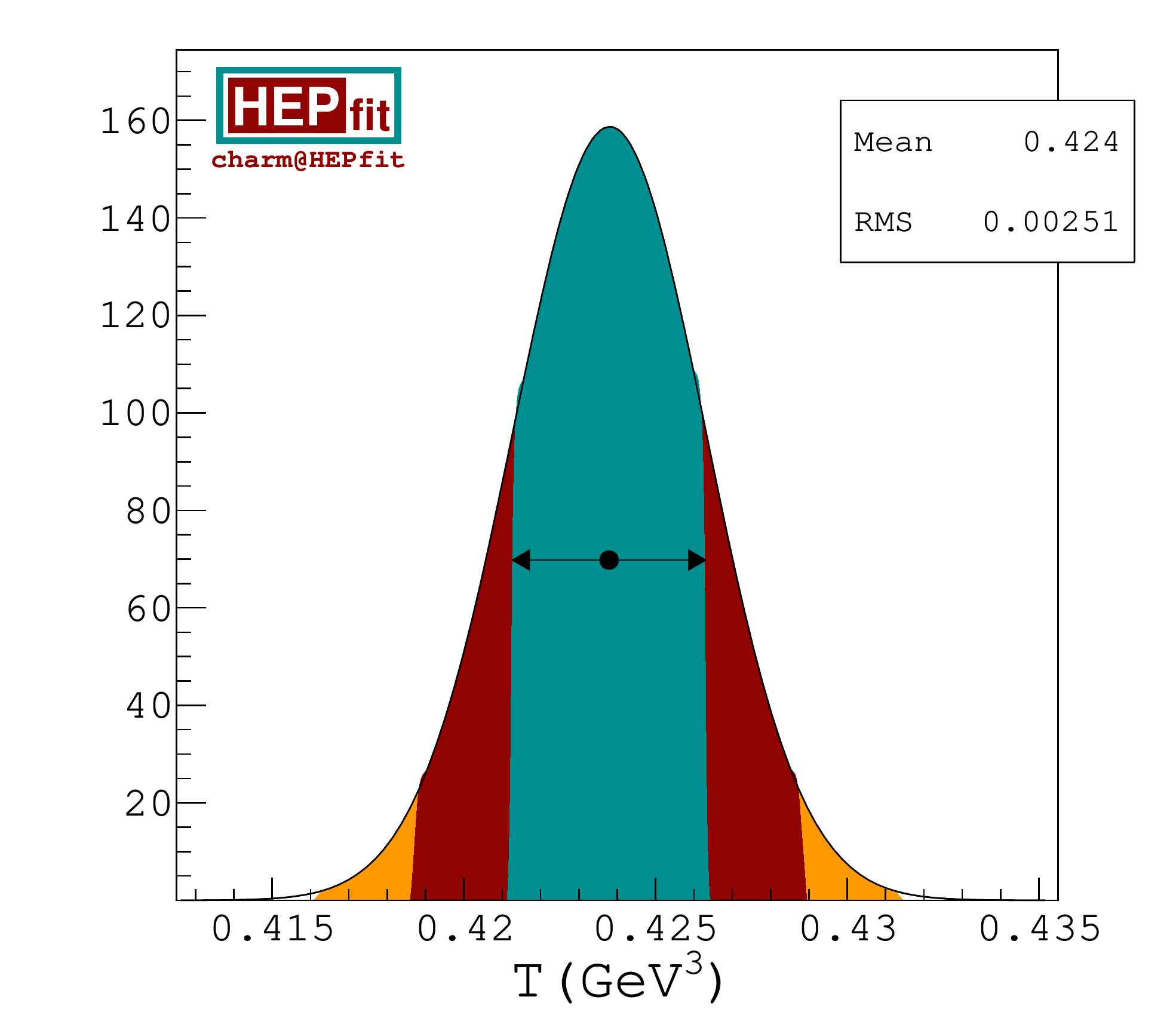}}
\subfloat{\includegraphics[width=.23\textwidth]{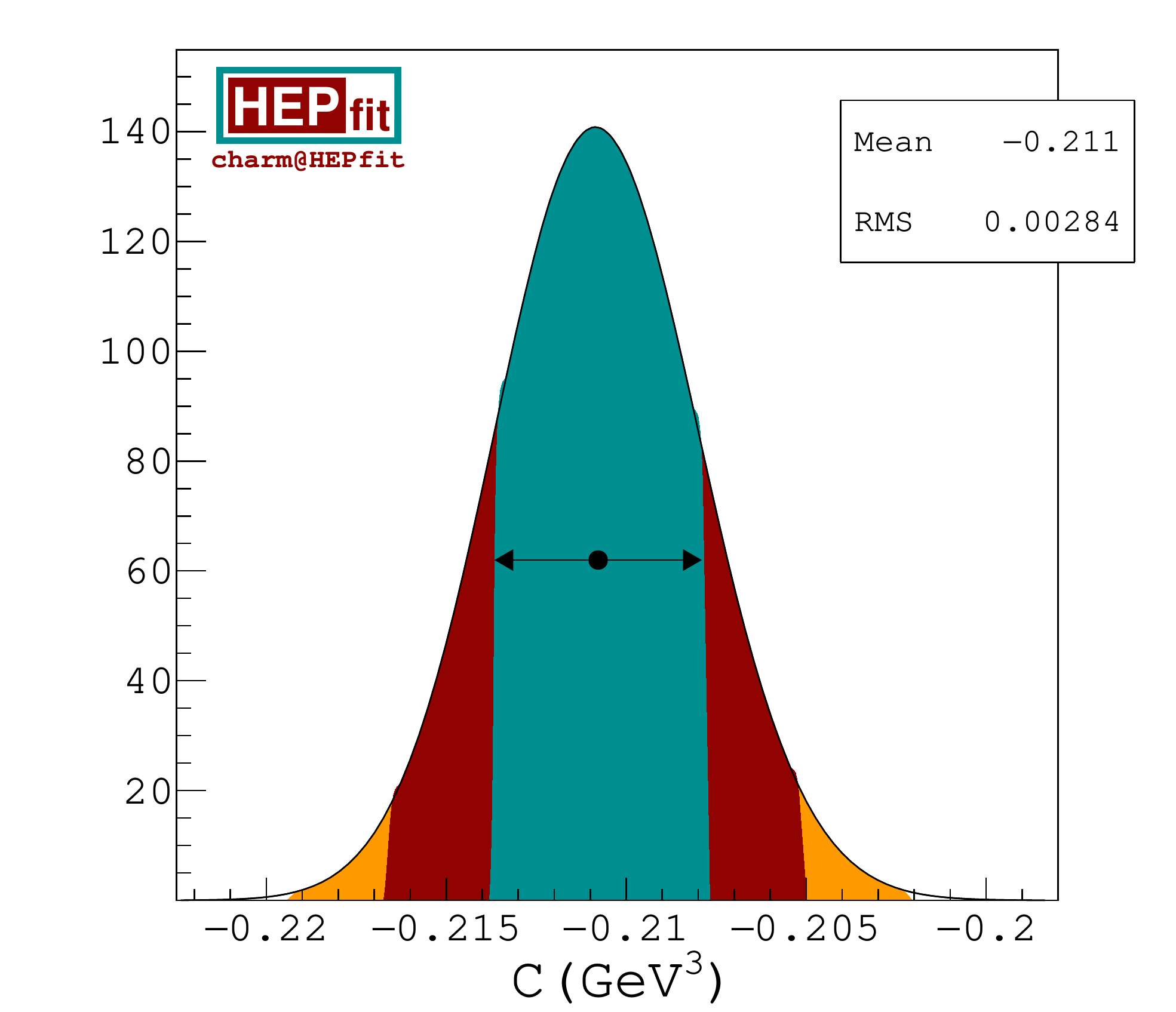}}
\subfloat{\includegraphics[width=.23\textwidth]{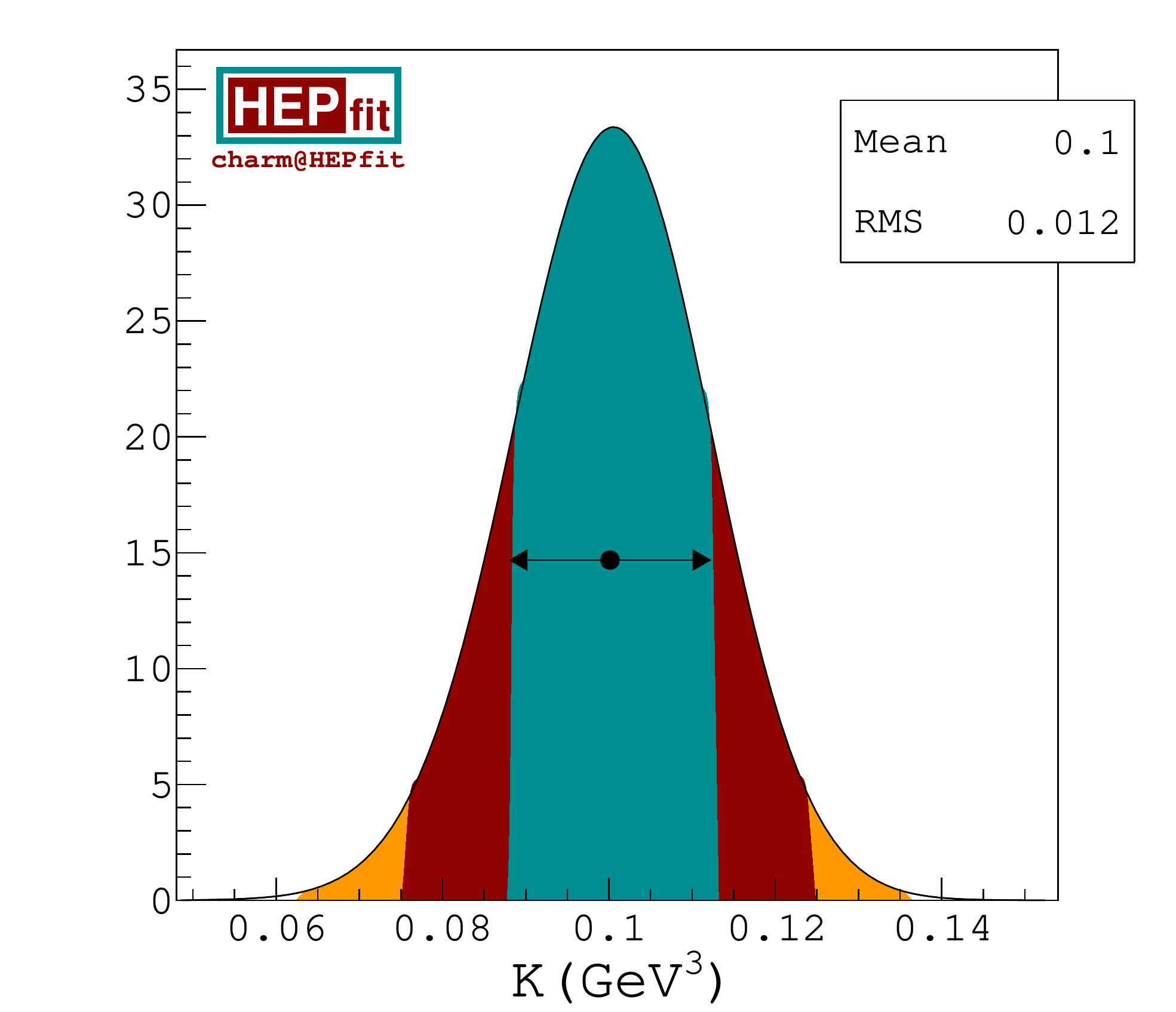}}
\subfloat{\includegraphics[width=.23\textwidth]{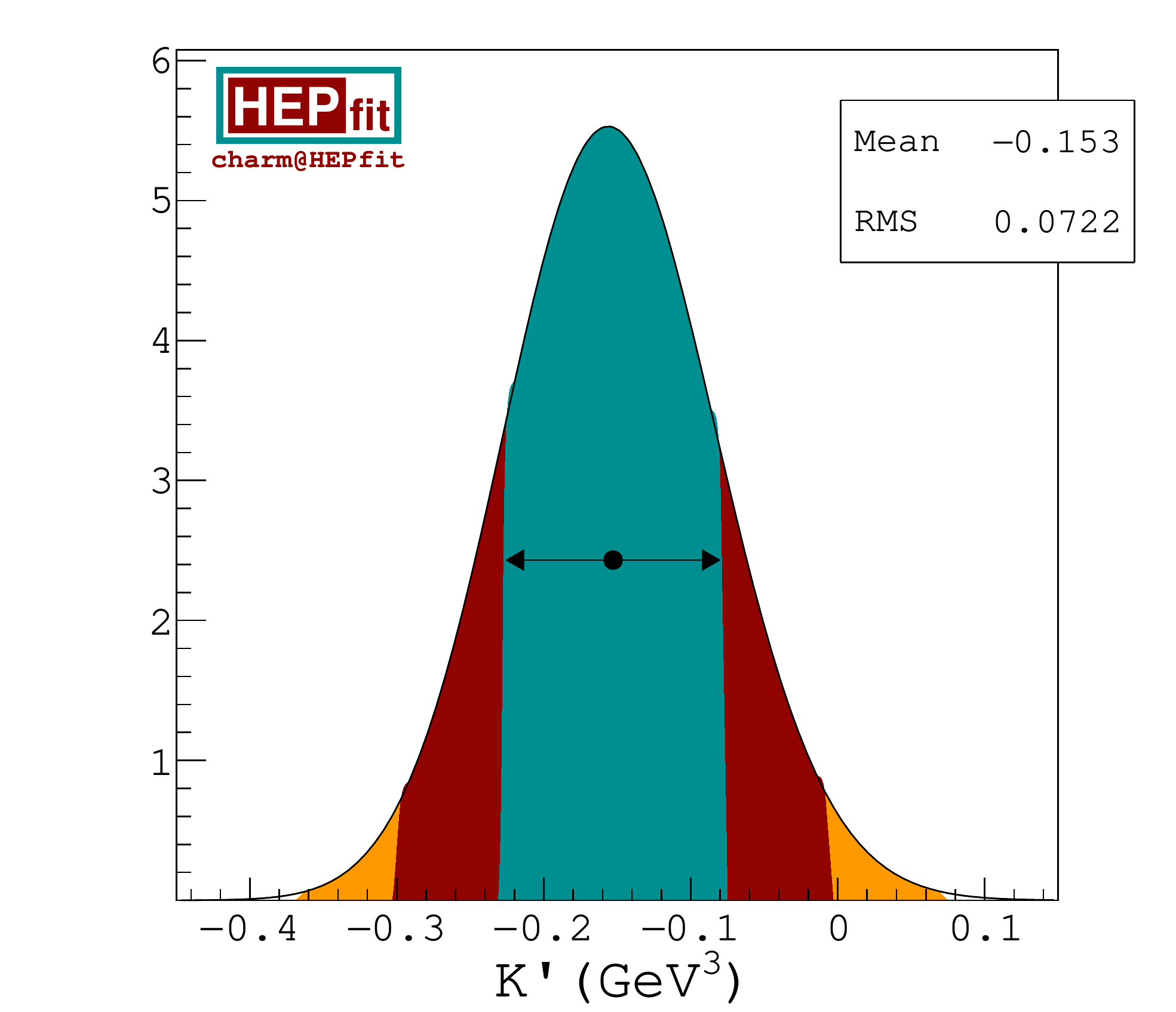}}\\
\subfloat{\includegraphics[width=.23\textwidth]{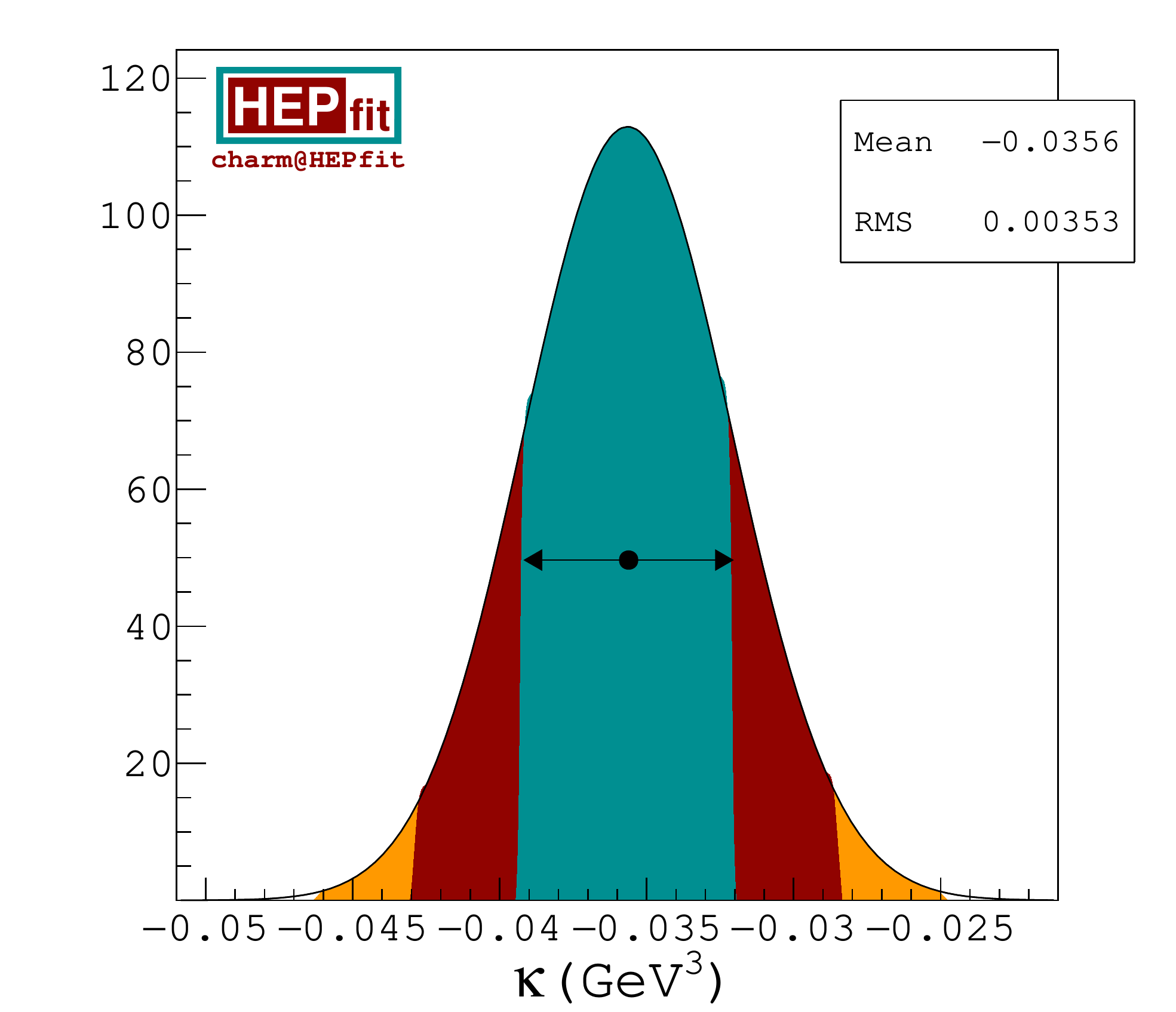}}
\subfloat{\includegraphics[width=.23\textwidth]{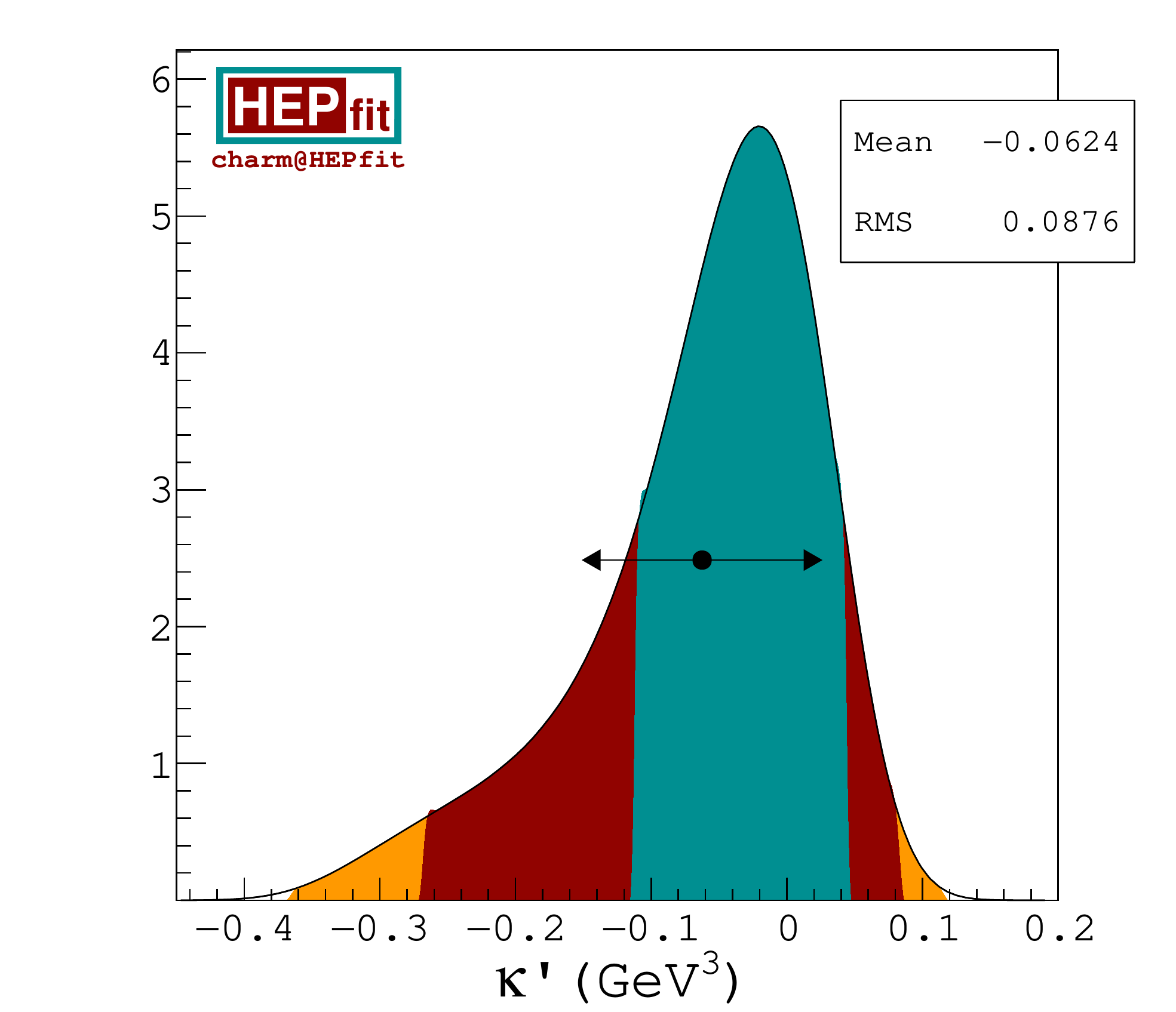}}
\subfloat{\includegraphics[width=.23\textwidth]{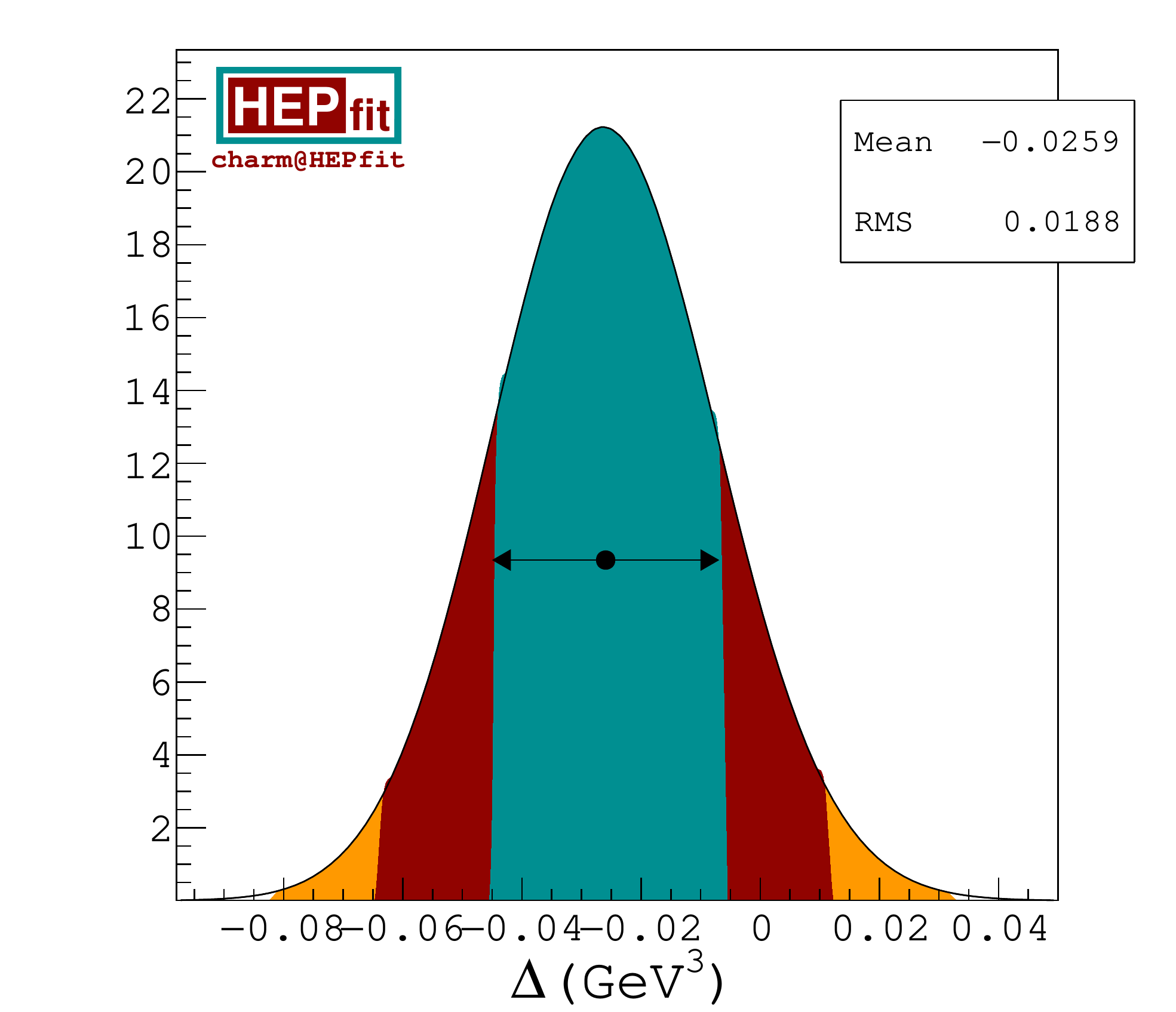}}
\subfloat{\includegraphics[width=.23\textwidth]{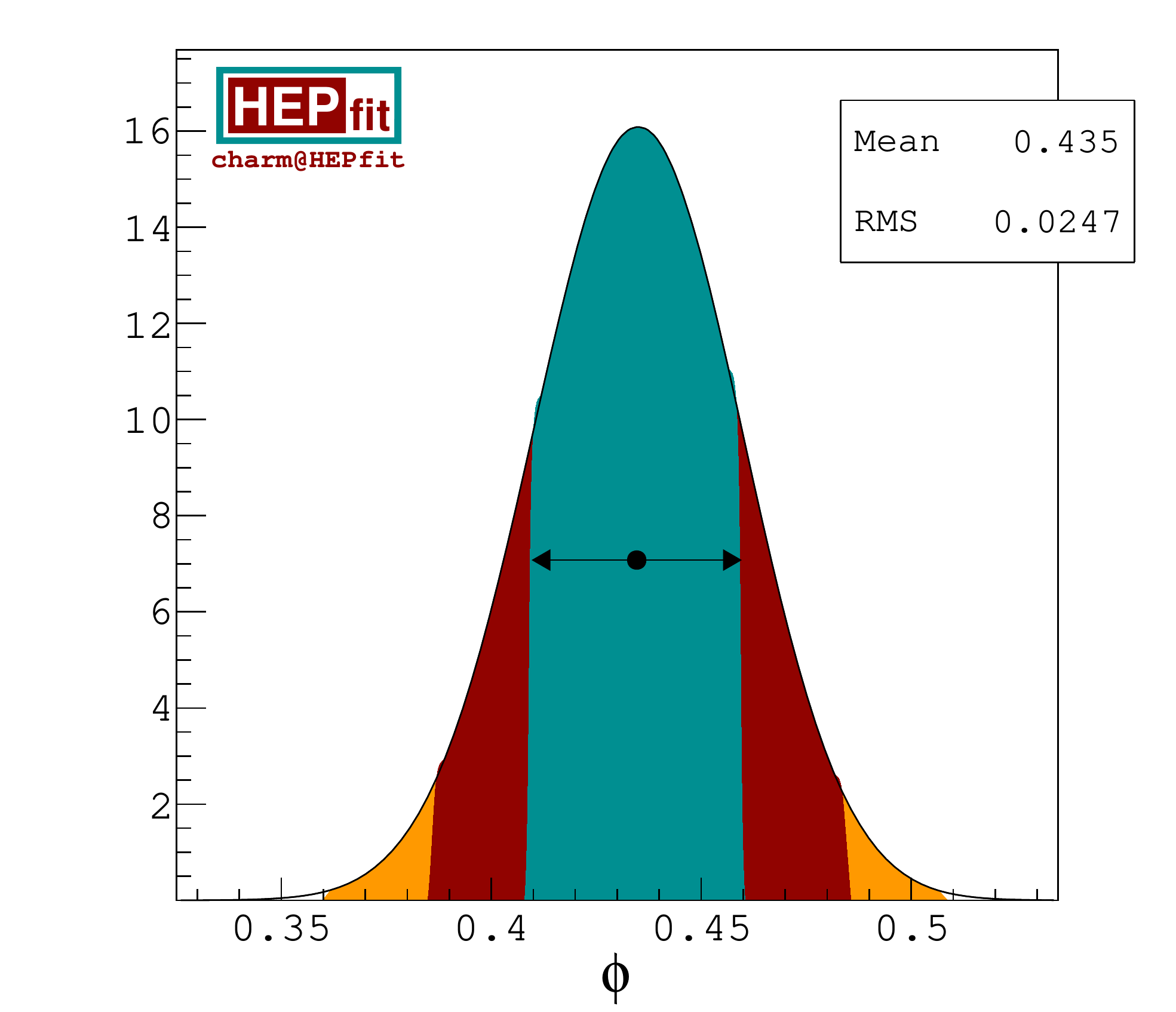}}\\
\subfloat{\includegraphics[width=.23\textwidth]{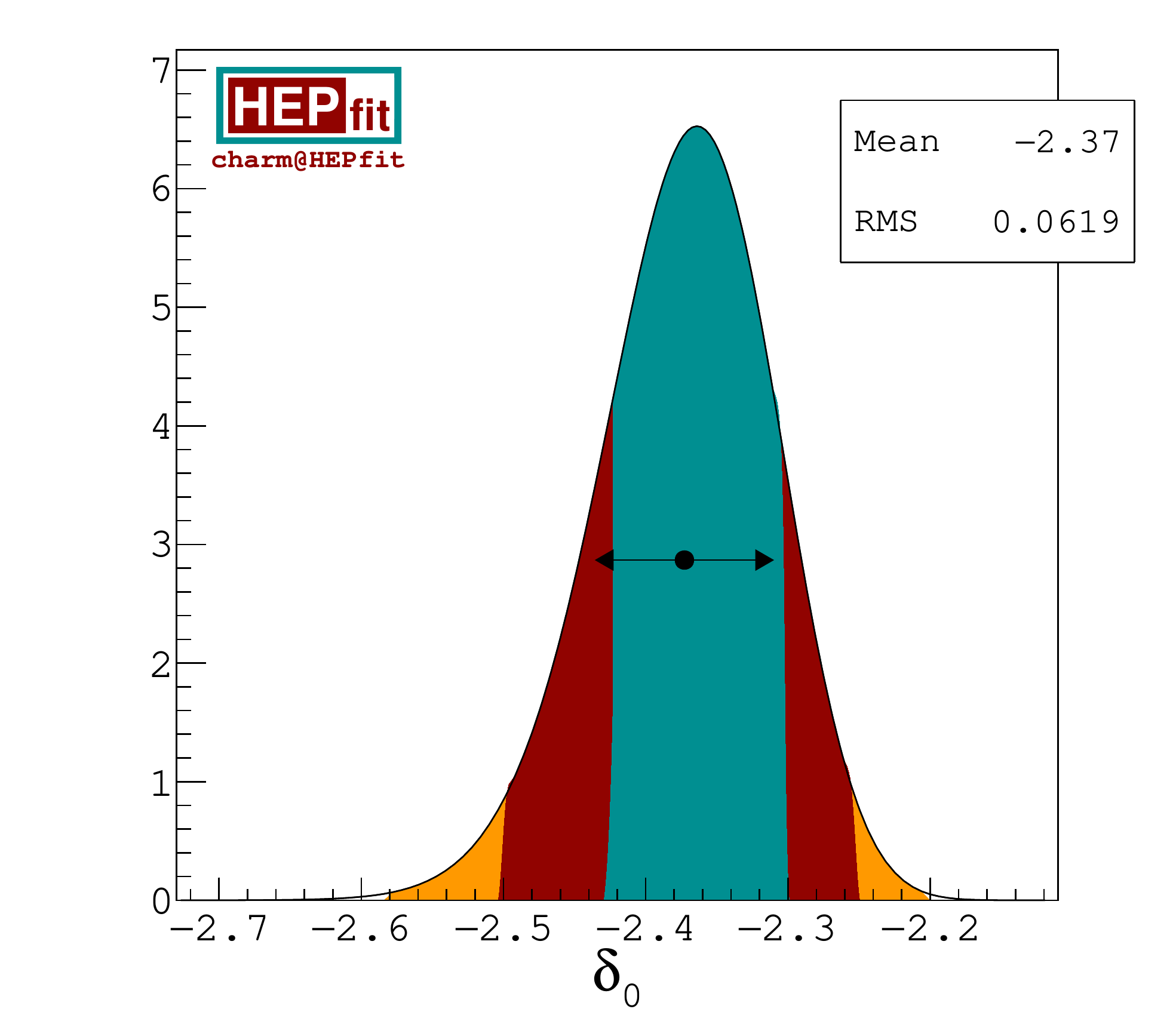}}
\subfloat{\includegraphics[width=.23\textwidth]{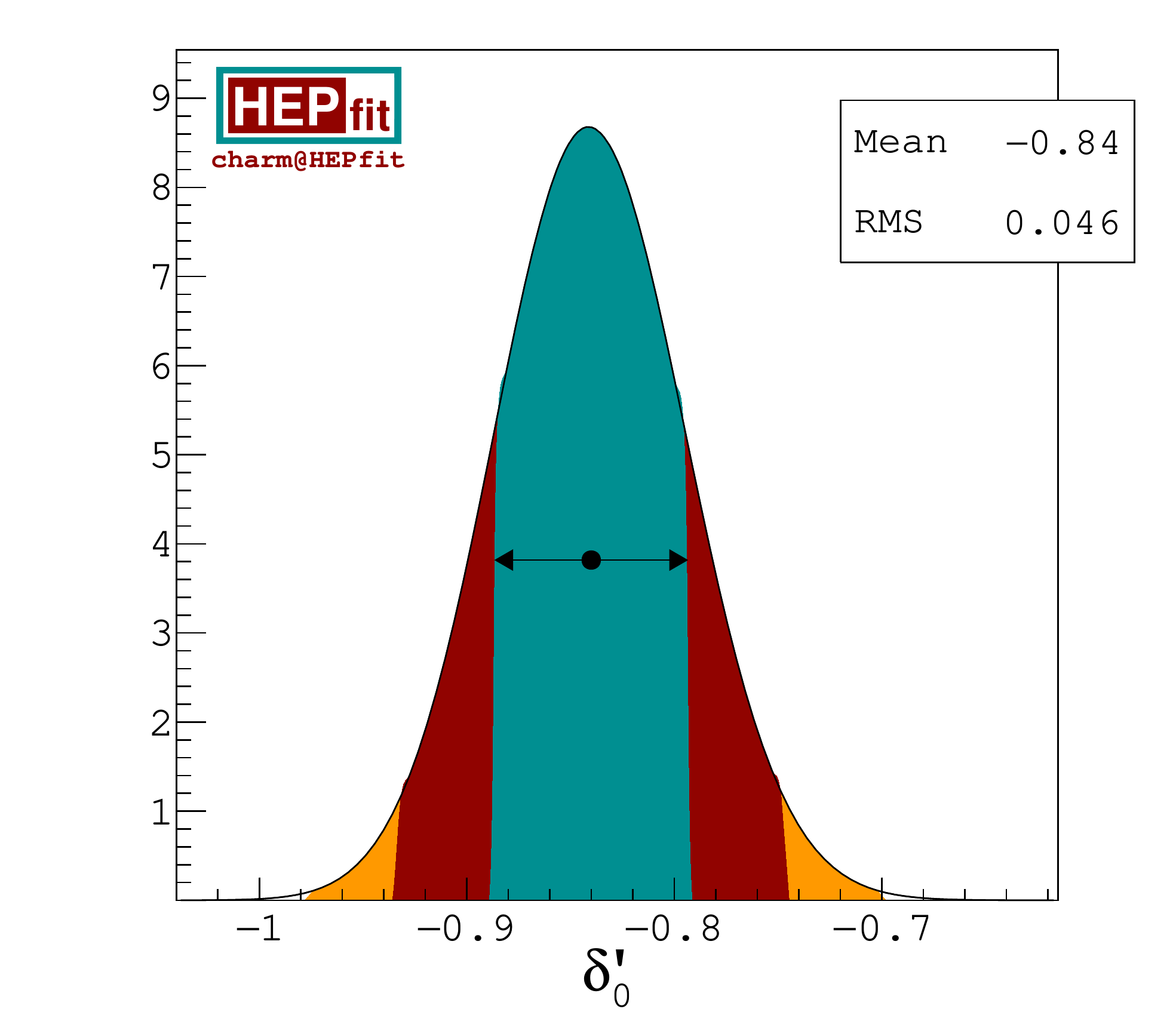}}
\subfloat{\includegraphics[width=.23\textwidth]{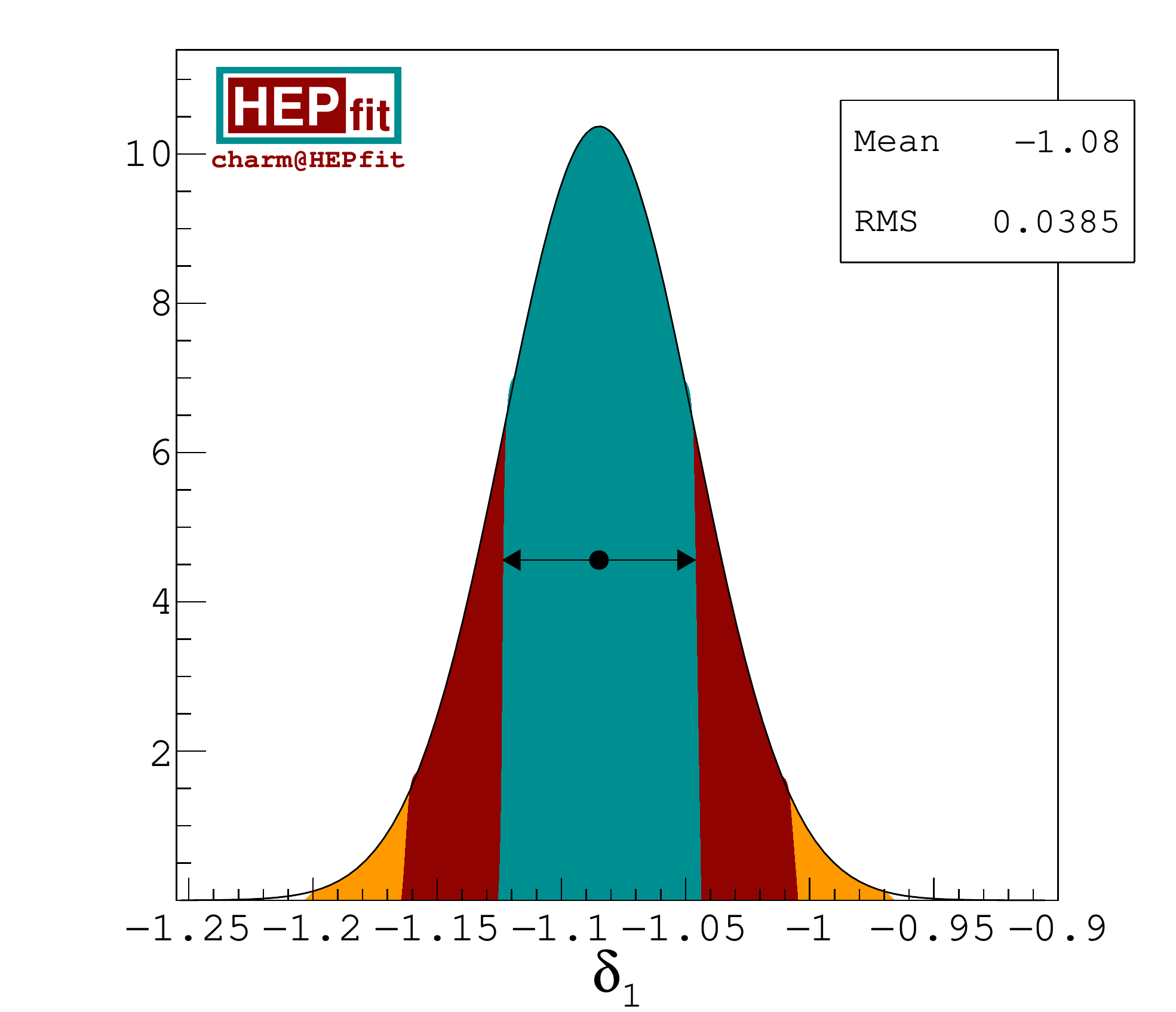}}
\subfloat{\includegraphics[width=.23\textwidth]{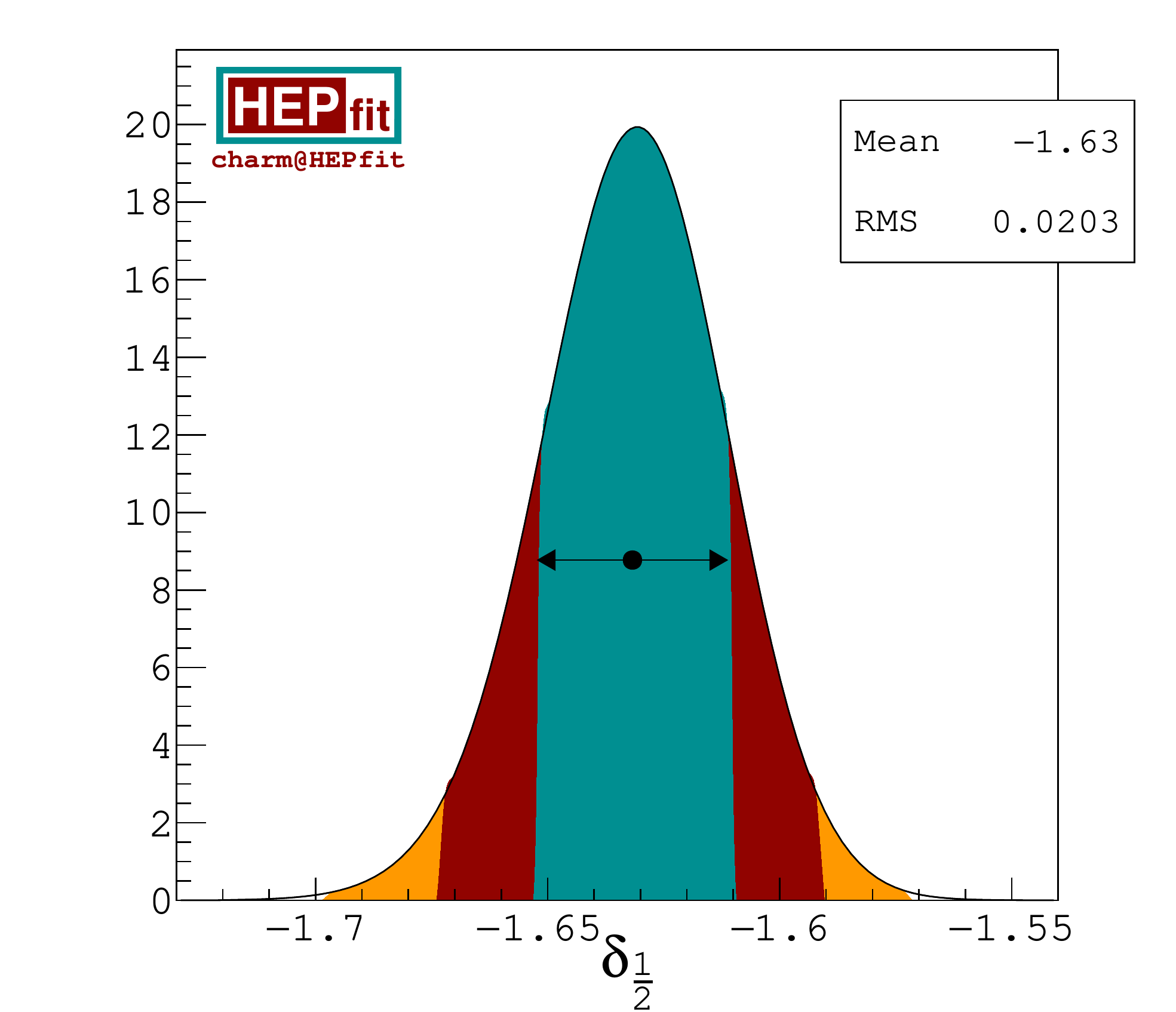}}\\
\subfloat{\includegraphics[width=.23\textwidth]{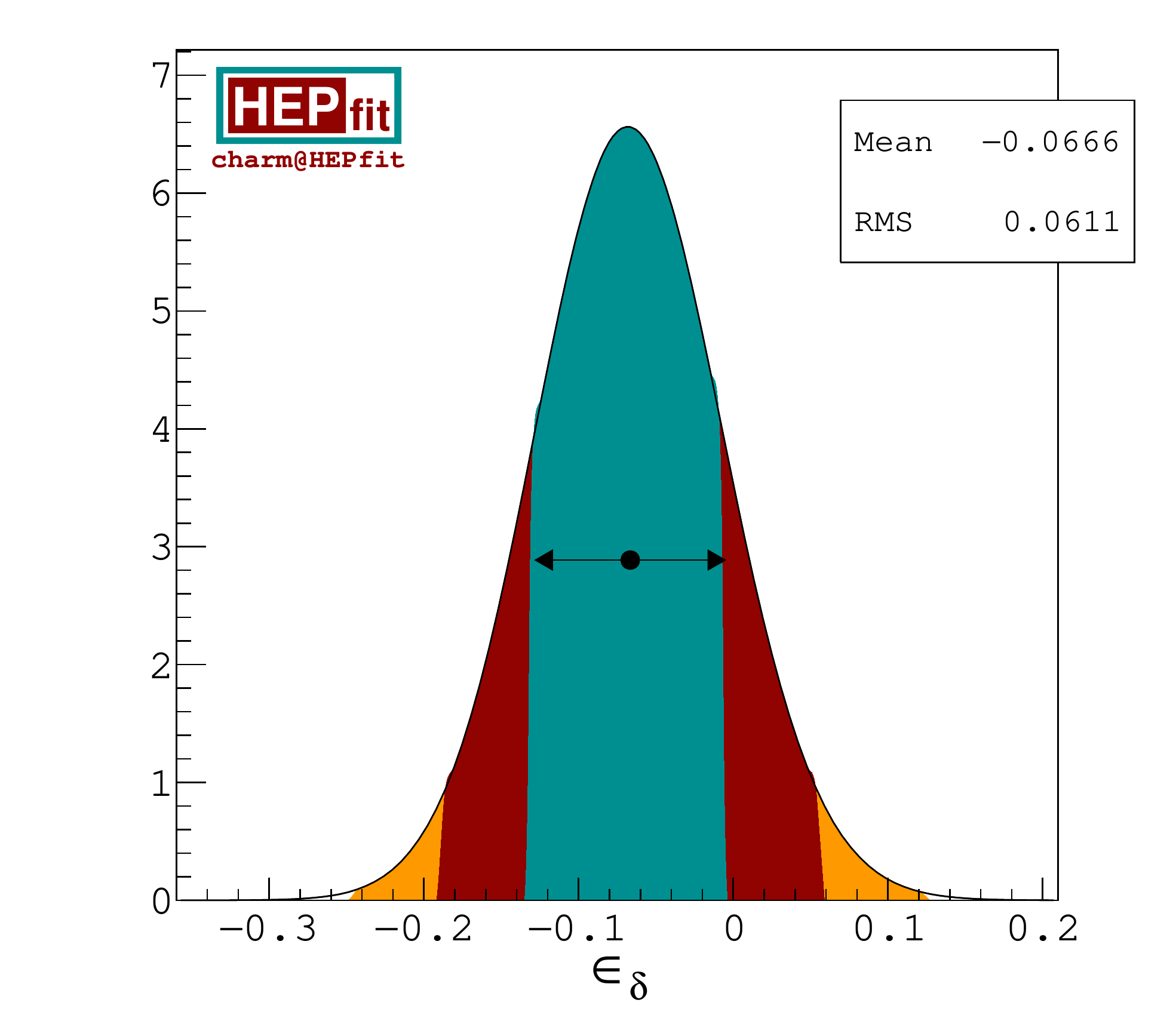}}
\subfloat{\includegraphics[width=.23\textwidth]{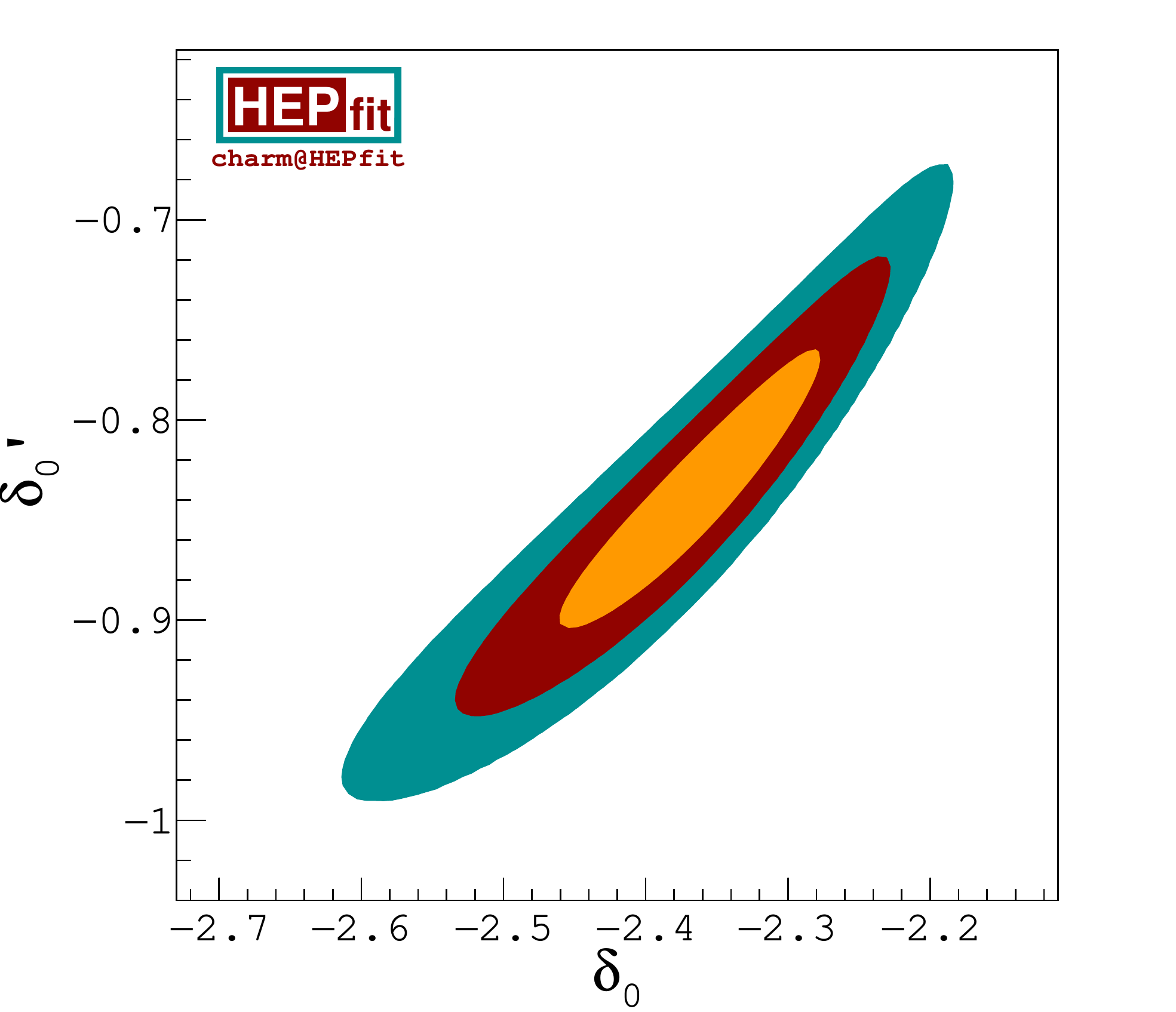}}
\subfloat{\includegraphics[width=.23\textwidth]{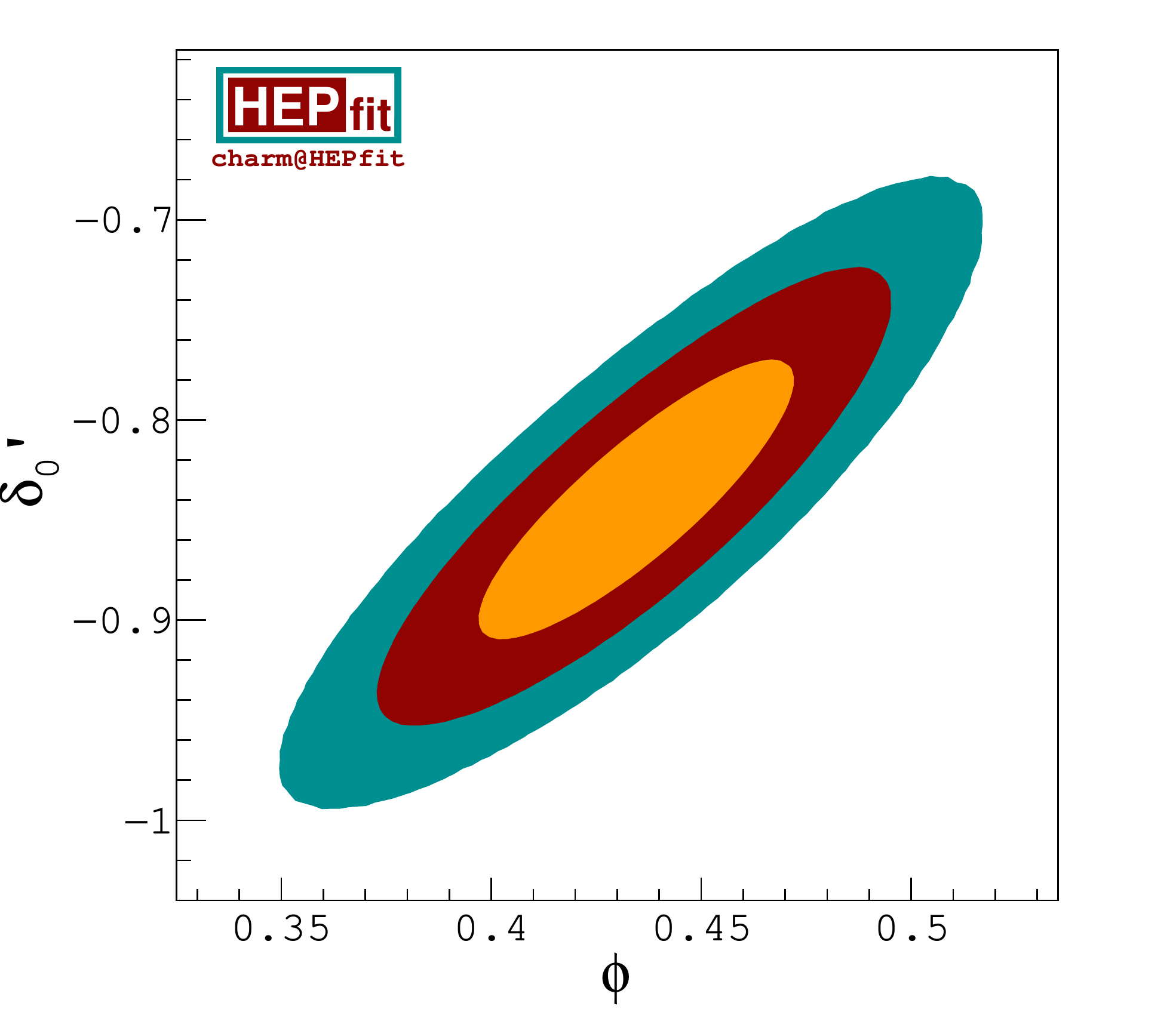}}
\subfloat{\includegraphics[width=.23\textwidth]{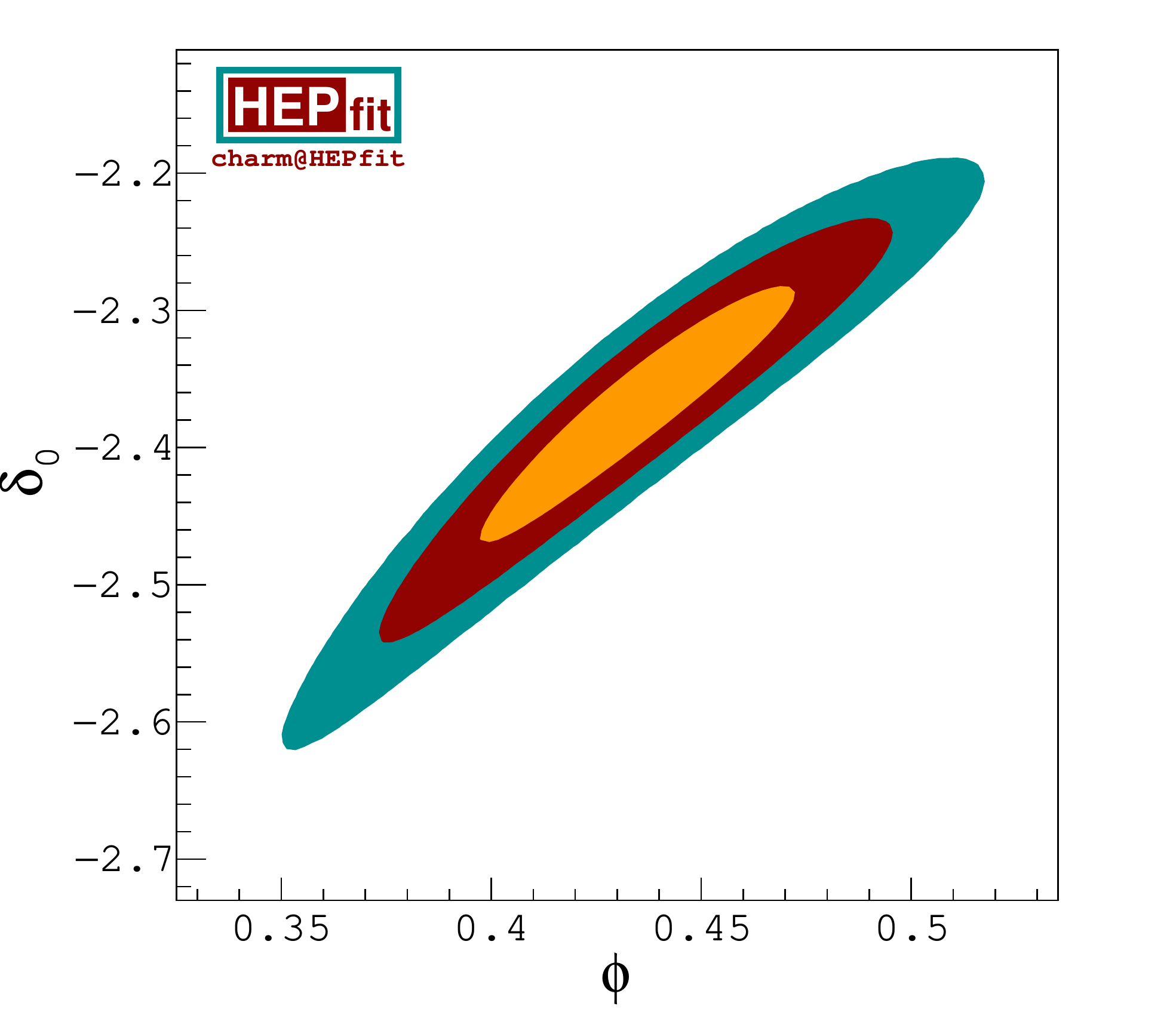}}
\caption{The marginalized posterior distributions of the parameters from the fit as given in table~\ref{tab:fit}. The green, red and orange regions are the 68\%, 95\% and 99\% probability regions respectively. The bottom three 2D marginalized plots show the correlations between the parameters $\delta_0$, $\delta_0^\prime$ and $\phi$, The orange, red and green regions are the 68\%, 95\% and 99\% probability regions respectively.}
\label{fig:post}
\end{center}
\end{figure*}
The fit of the parameters to the branching fractions and $\Delta{\rm A}_{\rm CP}^{\rm dir}$ and the predictions for the CP asymmetries was done with \HEPfit. A model was built specifically for this purpose. The code necessary for replicating this analysis can be made available on request. In figure~\ref{fig:post} we show the posterior distributions of the parameters from the fit with the mean and RMS listed in table~\ref{tab:BR-fit} and $\Delta{\rm A}_{\rm CP}^{\rm dir}$ quoted in section~\ref{sec:CPamp}. Only the posteriors for $\delta_0$ and $\kappa^\prime$ show some deviation from being Gaussian distributions. The two phases $\delta_0$ and $\delta_0^\prime$ and the angle $\phi$ that appear in the SCS decays are highly correlated. We show the correlation plots for these parameters in the bottom three plots of figure~\ref{fig:post}. The values of the CKM parameters used in these fits are from the UTfit average~\cite{Bona:2006ah}:
\begin{eqnarray}
\lambda=0.22534 \pm 0.00089\qquad A=0.833\pm0.012\\
\bar{\rho}=0.153\pm0.013\qquad \bar{\eta}=0.343\pm0.011
\end{eqnarray}

\nocite{*}

\bibliography{CharmDecays}

\end{document}